\documentclass[10pt,paper]{JHEP3}
\usepackage[dvips]{graphicx}
\usepackage{amsmath}

\skip\footins = 1\bigskipamount plus 2pt minus 4pt

\newcommand{\tr}{\mathop{\rm tr}}

\title{5-field terms in the open superstring effective action}

\author{Luiz Antonio Barreiro$^{1,2}$ and  Ricardo Medina$^{1}$\\
$^{1}$Instituto de Ci{\^e}ncias Exatas, Universidade Federal de Itajub\'{a}\\
\hphantom{$^{1}$}Itajub\'a, Minas Gerais, Brazil\\
$^{2}$Instituto de Ci{\^e}ncias Exatas, Funda\c{c}\~{a}o de Ensino e Pesquisa de Itajub\'{a}\\
\hphantom{$^{1}$}Itajub\'a, Minas Gerais, Brazil\\
\hphantom{$^{1}$}E-mails: \email{barreiro@unifei.edu.br,
rmedina@unifei.edu.br}.}

\abstract{Some time ago the bosonic and fermionic 4-field terms of
the non-abelian low energy effective action of the open
superstring were obtained, to all order in $\alpha'$. This was
done at tree level by directly generalizing the abelian case,
treated some time before, and considering the known expressions of
all massless superstring 4-point amplitudes (at tree level). In
the present work we obtain the bosonic 5-field terms of this
effective action, to all order in $\alpha'$. This is done by
considering the simplified expression of the superstring 5-point
amplitude for massless bosons, obtained some time ago.}

\begin{document}

\section{Introduction}

It has been known for a long time that String Theory modifies the
Yang-Mills and Einstein lagrangians  by adding $\alpha'$ terms to
them \cite{Scherk1}. Since the middle eighties many results of
this type were found, among which it was seen that the low energy
dynamics of abelian open strings (corresponding to photons) was
described by the Born-Infeld lagrangian \cite{Fradkin1, Tseytlin1,
Abouelsaood1, Bergshoeff1, Leigh1}, as long as the field strength
is kept constant. This is an all order in $\alpha'$ result. Some
non-abelian $\alpha'$ corrections of the effective lagrangian were
also determined by that time using the scattering amplitude
approach \cite{Tseytlin1, Gross1, Kitazawa1}.\\
\noindent After the discovery of D-branes as Ramond-Ramond charged
states \cite{Polchinski0} and that the low energy dynamics of
non-abelian open superstrings captured an equivalent description
in terms of them \cite{Witten1}, a supersymmetric generalization
of the abelian Born-Infeld lagrangian, in this context, was
constructed \cite{Cederwall0, Cederwall0p, Bergshoeff0,
Aganagic1}. Besides these results, abelian corrections were
obtained in
\cite{Wyllard2, Andreev2, DeRoo1}, to all order in $\alpha'$.\\
\noindent A non-abelian generalization of the Born-Infeld
lagrangian, in the context of Superstring Theory, has been
proposed in \cite{Tseytlin2} by means of a symmetrized trace
prescription. The complete ${\cal O}({\alpha'}^4)$ terms of
\cite{Koerber2, Sevrin1, Nagaoka1} have been written following
this prescription for the $F^6$ terms. It has been seen that, in
the open superstring effective lagrangian, the covariant
derivative terms are as important as the $F^n$ ones
\cite{Hashimoto1, Bilal1}. Also, an interesting proposal about the
general structure of the non-abelian Born-Infeld action and its
covariant derivative terms has been made in \cite{Cornalba1} by
means of the Seiberg-Witten map \cite{Seiberg1}, but this result
has unknown coefficients and explicit expressions in $D=10$ are
not given. So, in this sense, in the non-abelian case, the usually
known exact results for the lagrangian terms are strictly
perturbative in $\alpha'$. This has been done by different methods
which include the 1-loop effective action for $N=4$ SYM
\cite{Refolli1, Grasso1}; deformations of the Yang-Mills
lagrangian considering either BPS solutions to it \cite{Fosse1,
Koerber1, Koerber2} or supersymmetry requirements \cite{DeRoo2,
Cederwall1, Cederwall2, Drummond1}; and the scattering amplitude
approach \cite{Tseytlin1, Gross1, Kitazawa1,
Brandt1}. \\
\noindent It is generally believed that scattering
amplitudes are only used in String Theory to find the $first$
$\alpha'$ corrections in the effective lagrangians, but this is
not true: it was first seen in \cite{DeRoo1} that all the
$\alpha'$ information of the open superstring 4-point amplitudes,
of massless bosons and fermions, can be taken to the effective
lagrangian, at least in the abelian case. This was soon
generalized to the non-abelian case and to the 4-point effective
actions of the NS-NS sector of Closed Superstring Theory
\cite{Chandia1}. In the present work we go further and find the
$D^{2n}F^5$ terms of the open superstring non-abelian effective
lagrangian, to all order in $\alpha'$. This is done by using the
5-point amplitude, first completely calculated in
\cite{Brandt1}\footnote{In \cite{Kitazawa1} a partial computation
of this 5-point amplitude was done.} and afterwards simplified in
\cite{Machado1}. In contrast to the 4-point case, due to the
presence of poles in the $\alpha'$ terms of the scattering
amplitude, this is a very
much more complicated problem to solve.\\
\noindent We have organized this paper in a main body and
appendices. These last ones contain, besides conventions,
identities and tensors, some lengthy formulas and derivations,
which are important, but which would otherwise have turned the
main body too long. The structure of the main body of the paper is
as follows. In section \ref{345-point} we give a very brief review
about scattering amplitudes and the low energy effective
lagrangian inferred from them. In section \ref{34-point} we review
the known structure of the effective lagrangian, as far as 3 and
4-point amplitudes are concerned. In section \ref{5-point} we
review the derivation of the 5-point amplitude (at tree level),
since this is the starting point for the present paper,  and we
confirm that this formula satisfies the usual properties of
scattering amplitudes in Open Superstring Theory. We have placed
the main result of this work, namely, the explicit $D^{2n}F^5$
terms of the effective lagrangian (to all order in $\alpha'$) in
section \ref{lagrangian}. This section also contains the
scattering subamplitude that leads to that lagrangian and some
explicit examples of $\alpha'$ terms up to ${\cal O}({\alpha'}^4)$
order\footnote{In principle, with the $\alpha'$ expansions that we
give in appendix \ref{expansions} we could explicitly write the
$D^{2n}F^5$ terms up to ${\cal O}({\alpha'}^6)$ order.}. Finally,
section \ref{final} contains a summary and final remarks,
including future directions. Throughout this work we have treated
the very involved calculations that arise in scattering
amplitudes using $Mathematica$'s $FeynCalc \ 3.5$ package.\\

\section{Review of scattering amplitudes and low energy effective lagrangian}
\label{345-point}

The $M$-point tree level scattering amplitude for massless bosons,
in superstring theory, is given by \cite{Schwarz1}
\begin{eqnarray}
{\cal A}^{(M)} = i \ (2 \pi)^{10} \delta(k_1 + k_2 + \ldots + k_M)
\cdot {\sum_{j_1, j_2, \ldots , j_M}}' \tr(\lambda^{a_{j_1}}
\lambda^{a_{j_2}} \ldots \lambda^{a_{j_M}}) \ A(j_1, j_2, \ldots,
j_M) \ , \label{general-amplitude}
\end{eqnarray}
where the sum $\sum'$ in the indices $\{j_1, j_2, \ldots, j_M\}$
is done over $non-cyclic$ equivalent permutations of the group
$\{1, 2, \ldots, M\}$. The matrices $\lambda^{a_{j}}$ are in the
adjoint representation of the Lie group\footnote{At tree level,
the Lie groups $SO(N)$ and $USp(N)$ have been shown to be
consistent with the description of superstring
interactions\cite{Schwarz1}. It has also been seen that $U(N)$ is
the appropriate gauge group when describing the low energy
dynamics of $N$ coincident D-branes \cite{Witten1}. In the present
work it will not be necessary to make any especial reference to
any of these Lie groups.} and $A(j_1,
j_2, \ldots, j_M)$, called $subamplitude$, corresponds to the
$M$-point amplitude of open superstrings which do not carry color
indices and which are placed in the ordering $\{j_1, j_2, \ldots,
j_M\}$ (modulo cyclic permutations).

Since the present work is based in the determination of the
effective lagrangian by means of the known expressions of the
scattering amplitudes\footnote{To the knowledge of the authors,
this method was first introduced in \cite{Scherk1}.}, in the
following subsections we briefly review the 3, 4 and 5-point (tree
level) subamplitudes of massless bosons in Open Superstring
Theory. From them, an on-shell effective lagrangian of the form
\begin{eqnarray}
{\cal L}_{\rm ef} = {\cal L}_{YM}  + {\cal L}_{D^{2n}F^4} + {\cal
L}_{D^{2n}F^5} \label{effective1}
\end{eqnarray}
emerges, where
\begin{eqnarray}
\label{LYM}
{\cal L}_{YM} &=& -\frac{1}{4} \tr(F^2) \\
\label{LD2nF4} {\cal L}_{D^{2n}F^4} &=& {\alpha'}^2 g^2 \tr(F^4) +
{\alpha'}^3 g^2 \tr(D^2F^4) + {\alpha'}^4 g^2 \tr(D^4F^4) + \ldots \\
\label{LD2nF5} {\cal L}_{D^{2n}F^5} &=& {\alpha'}^3 g^3 \tr(F^5) +
{\alpha'}^4 g^3 \tr(D^2F^5) + {\alpha'}^5 g^3 \tr(D^4F^5) + \ldots
\ .
\end{eqnarray}
The lagrangian in (\ref{effective1}) agrees with the open
superstring effective lagrangian up to 5-field terms: any
difference between them is sensible only to 6 or higher-point
scattering amplitudes. Also, there is no unique way in writing its
terms since some freedom arises from the $[D,D]F=-ig[F,F]$
relation, the Bianchi identity and integration by parts (see
\cite{Tseytlin1},\cite{Koerber1} and \cite{Grasso1}, for example),
thus, allowing to interchange some terms from ${\cal
L}_{D^{2n}F^4}$ and ${\cal L}_{D^{2n}F^5}$. In the next section we
will specify the convention that we will use in writing the
$D^{2n}F^4$ terms.

\section{3 and 4-point subamplitudes and their effective
lagrangian in Open Superstring Theory}
\label{34-point}

The tree level 3 and 4-point subamplitudes for massless bosons, in
Open Superstring Theory, have been known for a long time. Their
on-shell expressions are the following \cite{Schwarz1},
respectively:
\begin{eqnarray}
\label{A123}
A(1,2,3) & = & 2 g \left[\frac{}{} (\zeta_1 \cdot
k_2)(\zeta_2 \cdot \zeta_3) + (\zeta_2 \cdot k_3)(\zeta_3 \cdot
\zeta_1) + (\zeta_3 \cdot k_1)(\zeta_1 \cdot \zeta_2) \frac{}{}
\right] \ ,
\\
\label{A1234} A(1,2,3,4) & = & 8 \ g^2 \ {\alpha'}^2 \frac{\Gamma(-
\alpha' s) \Gamma(- \alpha' t)}{\Gamma(1- \alpha' s - \alpha' t)}
K(\zeta_1, k_1; \zeta_2, k_2; \zeta_3, k_3; \zeta_4, k_4) \ ,
\end{eqnarray}
where
\begin{eqnarray}
K(\zeta_1, k_1; \zeta_2, k_2; \zeta_3, k_3; \zeta_4, k_4) =
t_{(8)}^{\mu_1 \nu_1 \mu_2 \nu_2 \mu_3 \nu_3 \mu_4 \nu_4}
\zeta_{\mu_1}^1 k_{\nu_1}^1 \zeta_{\mu_2}^2 k_{\nu_2}^2
\zeta_{\mu_3}^3 k_{\nu_3}^3 \zeta_{\mu_4}^4 k_{\nu_4}^4
\label{kinematicfactor}
\end{eqnarray}
is a kinematic factor, $t_{(8)}$ being a known tensor
\cite{Schwarz1}. The $s$ and $t$ variables in (\ref{A1234}) are part
of the three Mandelstam variables, which are defined as
\begin{eqnarray}
s = -(k_1 + k_2)^2, \ \ \, t = -(k_1 + k_4)^2, \ \ \, u = -(k_1 +
k_3)^2 \ \ \ .
\label{Mandelstam}
\end{eqnarray}

The Gamma factor in (\ref{A1234}) has a completely known $\alpha'$
expansion (see appendix \ref{expansion-Gamma}) which begins like
\begin{eqnarray}
{\alpha'}^2 \frac{\Gamma(- \alpha' s) \Gamma(- \alpha' t)}
{\Gamma(1- \alpha' s - \alpha' t)} = \frac{1}{s t} - \frac{\
\pi^2}{6}{\alpha'}^2 + {\cal O}({\alpha'}^3)\,. \label{gammas}
\end{eqnarray}
Since the leading term in the $\alpha'$ expansion of $A(1,2,3,4)$ in
(\ref{A1234}) is the Yang-Mills 4-point subamplitude,
$A_{YM}(1,2,3,4)$, then it may also be written as
\begin{eqnarray}
A(1,2,3,4) & = &  {\alpha'}^2 st \ \frac{\Gamma(- \alpha' s)
 \Gamma(- \alpha' t)}{\Gamma(1- \alpha' s - \alpha' t)}
A_{YM}(1,2,3,4) \ . \label{A1234II}
\end{eqnarray}
This last formula will be useful in the next section, when making
a comparison between the 4 and 5-point subamplitudes.\\
\noindent From the expression of $A(1,2,3)$ in (\ref{A123}) it is
immediate that the effective lagrangian may be written with no
$D^{2n}F^3$ terms (as it has already been done in
(\ref{effective1})) since this amplitude contains no $\alpha'$
corrections and agrees with
the one from the Yang-Mills lagrangian.\\
\noindent Using the expression of $A(1,2,3,4)$ in (\ref{A1234})
the lagrangian ${\cal L}_ {D^{2n}F^4}$ has been determined some
time ago\cite{Chandia1}\footnote{In fact, not only the $D^{2n}F^4$
terms were determined in \cite{Chandia1}, but also all
boson-fermion and fermion-fermion terms which are sensible to
4-point amplitudes.}, to all order in $\alpha'$. This was done by
directly generalizing the procedure considered in \cite{DeRoo1}
for the non-abelian case. The final result is the
following\footnote{There are three differences between eq.
(\ref{Seff-nonabelian}) and the corresponding one in section 2.2
of \cite{Chandia1}:
\begin{enumerate}
\item We have now written the $F^4$ terms of the integrand in an
abbreviated manner, by means of the $t_{(8)}$ tensor and, as a
consequence of the symmetries of this tensor, it has been possible
to write ${\cal L}_{D^{2n}F^4}$ in terms of only one function
$f_{\rm sym}$, instead of the three ones contained in the ${\cal
G}^{\ a_1 a_2 a_3 a_4}_{\rm \alpha' corr.}(s, t, u)$ function of
\cite{Chandia1}. \item We have introduced a symmetrized
prescription in the $\alpha'$ expansion of function $f$, explained
in eq. (\ref{fsym}), and that is why we have now called it $f_{\rm
sym}$. \item We have included derivative terms which were not
considered in \cite{Chandia1}, namely, the ones with
$D^2=D^{\alpha}D_{\alpha}$ operating on a same field strength.
\end{enumerate}
Neither item 2 nor item 3 invalidate the result of
\cite{Chandia1}, since there were considered results sensible up
to 4-point amplitudes only.}:
\begin{multline}
{\cal L}_{D^{2n}F^4}  = - \frac{1}{8} g^2 {\alpha'}^2 \
 \int \int \int \int \
\left\{ \frac{}{}\prod_{j=1}^{4} d^{10} x_j \ \delta^{(10)}(x-x_j)
\frac{}{} \right\} \times \\
\begin{split}
& \times f_{\rm sym} \left( \frac{}{} \frac{ (D_{1} + D_{2})^2 +
(D_{3} + D_{4})^2}{2} ,
\frac{(D_{1} + D_{4})^2 + (D_{2} + D_{3})^2}{2} \frac{}{} \right) \times \\
& \times t_{(8)}^{\mu_1 \nu_1 \mu_2 \nu_2 \mu_3 \nu_3 \mu_4 \nu_4}
\tr \biggl( F_{\mu_1 \nu_1}(x_1)F_{\mu_2 \nu_2}(x_2)F_{\mu_3
\nu_3}(x_3)F_{\mu_4 \nu_4}(x_4) \biggr) \ ,
\label{Seff-nonabelian}
\end{split}
\end{multline}
where the function $f$ is given by
\begin{eqnarray}
f(s, t) & = & \frac{\Gamma(- \alpha' s)\Gamma(- \alpha' t)}
{\Gamma(1 - \alpha' s - \alpha' t)} - \frac{1}{{\alpha'}^2 st} \ .
\label{f}
\end{eqnarray}
In (\ref{Seff-nonabelian}) `$f_{\rm sym}$' denotes that in the
power series of it, which involves powers of $s$ and $t$, we are
using a symmetrized convention:
\begin{eqnarray}
(s^k t^l)_{\rm sym} = \frac{( \ \mbox{Sum of all different
permutations of k powers of $s$ and l powers of $t$} \
)}{\left(\begin{array}{c}
                k+l\\
                k
               \end{array}\right)} \ .
\label{fsym}
\end{eqnarray}
\noindent In the case of ordinary numbers the formula in
(\ref{fsym}) coincides with $s^k t^l$, but this does not happen in
the case of covariant derivative operators, as it is in
(\ref{Seff-nonabelian}).\\
\noindent The terms of ${\cal L}_{D^{2n}F^4}$ were written
explicitly up to ${\cal O}({\alpha'}^5)$ order in \cite{Chandia1} .\\
\noindent The fact that $A(1,2,3,4)$ has no poles at any (non zero)
$\alpha'$ order made the construction of ${\cal L}_{D^{2n}F^4}$, in
(\ref{Seff-nonabelian}), quite direct: besides some trace factors,
the lagrangian is constructed in terms of the 4-point Gamma factor
in which the momenta are substituted by covariant derivatives
(appropriately symmetrized). All this implies an enormous
simplification with respect to other methods of obtaining the
$\alpha'$ correction terms in the effective lagrangian, since all of
them require, order by order in $\alpha'$, of an explicit
construction of gauge invariant terms with unknown coefficients
which are afterwards determined by some matching with Superstring
Theory 4-point amplitudes (see \cite{Tseytlin1}, \cite{Bilal1},
\cite{Koerber1} and \cite{Grasso1}, for example).\\

\section{The 5-point amplitude of massless bosons in Open Superstring Theory}
\label{5-point}

\noindent A first calculation of the 5-point amplitude for
massless bosons in Open Superstring Theory was done a long time
ago in \cite{Kitazawa1}. This was a partial calculation with
enough information to calculate the $F^5$ and the $D^2F^4$ terms
that appear at ${\cal O}({\alpha'}^3)$ order in the effective
lagrangian (\ref{effective1}). Long after this result, in
\cite{Koerber1} it was seen that, of all these terms, only the
$D^2F^4$ ones were correctly determined. The corrected $F^5$ terms
of \cite{Koerber1} were also confirmed subsequently by other
methods \cite{DeRoo2,Brandt1,Grasso1, Drummond1}.\\
\noindent Since the main result of the present work, namely, the
determination of the $D^{2n}F^5$ terms of the effective
lagrangian, lies on the computation of the 5-point subamplitude,
in the next subsection we will review the main steps of it in some
detail. In the subsection after the next one we will check that
our 5-point formula satisfies the usual properties of massless
open string subamplitudes \cite{Schwarz1, Mangano1}: cyclicity,
(on-shell) gauge invariance, world-sheet parity and
factorizability.

\subsection{Review of the 5-point subamplitude derivation}
\label{review-5}

\noindent A complete expression for the 5-point amplitude was
first obtained in \cite{Brandt1}:
\begin{eqnarray}
\lefteqn{ A(1, 2, 3, 4, 5)  =} && \nonumber\\&& 2 g^3 (2
\alpha')^2 \biggl[
  (\zeta_1 \cdot \zeta_2)(\zeta_3 \cdot \zeta_4)
      \Bigl\{   (\zeta_5 \cdot k_1)(k_2 \cdot k_3) L_2
     - (\zeta_5 \cdot k_2)(k_1 \cdot k_3) L_3 +
\nonumber \\ & & \hphantom{2 g^3 (2 \alpha')^2 \Bigl[ (\zeta_1
\cdot \zeta_2)(\zeta_3
  \cdot \zeta_4) \Bigl\{}
+ (\zeta_5 \cdot k_3) \left( (k_2 \cdot k_4){L_3}'+(k_2 \cdot
k_3)L_2 \right)
  \Bigr\}+
\nonumber \\
& & \hphantom{2 g^3 (2 \alpha')^2 \Bigl[} + (\zeta_1 \cdot
\zeta_3)(\zeta_2 \cdot \zeta_4) \Bigl\{
    -(\zeta_5 \cdot k_1)(k_2 \cdot k_3) L_7 +
\nonumber
\\
&& \hphantom{2 g^3 (2 \alpha')^2 \Bigl[+ (\zeta_1 \cdot
\zeta_3)(\zeta_2
    \cdot \zeta_4) \Bigl\{}
+  (\zeta_5 \cdot k_2) \left( (k_3 \cdot k_4){L_1}'-(k_2\cdot
k_3)L_7
  \right) -
\nonumber \\
& & \hphantom{2 g^3 (2 \alpha')^2 \Bigl[+ (\zeta_1 \cdot
\zeta_3)(\zeta_2
    \cdot \zeta_4) \Bigl\{}
- (\zeta_5 \cdot k_3) (k_1 \cdot k_2)L_1   \Bigr\}+ \nonumber\\
& & \hphantom{2 g^3 (2 \alpha')^2 \Bigl[} + (\zeta_1 \cdot
\zeta_4)(\zeta_2 \cdot \zeta_3) \Bigl\{   (\zeta_5 \cdot k_1)
\left(   (k_3 \cdot k_4){K_4}'-(k_1 \cdot k_3)K_5 \right) -
\nonumber\\
 & & \hphantom{2 g^3 (2 \alpha')^2 \Bigl[+ (\zeta_1
\cdot \zeta_4)(\zeta_2
    \cdot \zeta_3) \Bigl\{ }
- (\zeta_5 \cdot k_2)(k_1 \cdot k_3) K_5 + (\zeta_5 \cdot k_3)(k_1
\cdot k_2) K_4   \Bigr\}+ \nonumber
\end{eqnarray}
\begin{eqnarray}
& & \hphantom{2 g^3 (2
\alpha')^2 \Bigl[} + (\zeta_2 \cdot \zeta_3) \Bigl\{   (\zeta_5
\cdot k_1)  \left( (\zeta_1 \cdot k_2)(\zeta_4 \cdot k_3) L_2 -
(\zeta_1 \cdot k_3)(\zeta_4 \cdot k_2) L_7   \right)+
\nonumber \\
& & \hphantom{2 g^3 (2 \alpha')^2 \Bigl[+ (\zeta_2 \cdot
\zeta_3)\Bigl\{ } + (\zeta_5 \cdot k_2) \Bigl(   (\zeta_1 \cdot
k_3)(\zeta_4 \cdot k_1) K_5 + (\zeta_1 \cdot k_3)(\zeta_4 \cdot
k_3) {L_4}' +
\nonumber\\
&& \hphantom{2 g^3 (2 \alpha')^2 \Bigl[+ (\zeta_2 \cdot
\zeta_3)\Bigl\{ +
    (\zeta_5 \cdot k_2)\Bigl(} +(\zeta_1 \cdot k_2)(\zeta_4 \cdot k_3)
    L_2 + (\zeta_1 \cdot k_4)(\zeta_4 \cdot k_3) {K_4}' \Bigr)-
\nonumber \\
& & \hphantom{2 g^3 (2 \alpha')^2 \Bigl[+ (\zeta_2 \cdot
\zeta_3)\Bigl\{} - (\zeta_5 \cdot k_3) \Bigl(   (\zeta_1 \cdot
k_2)(\zeta_4 \cdot k_1) K_4 + (\zeta_1
 \cdot k_3)(\zeta_4 \cdot k_2) L_7 +
\nonumber\\
 && \hphantom{2 g^3 (2 \alpha')^2 \Bigl[+ (\zeta_2
\cdot \zeta_3)\Bigl\{-
    (\zeta_5 \cdot k_3)\Bigl(}
+(\zeta_1 \cdot k_2)(\zeta_4 \cdot
 k_2) L_4 +
(\zeta_1 \cdot k_4)(\zeta_4 \cdot k_2) {K_5}'
  \Bigr)     \Bigr\}+ \nonumber\\
 & & \hphantom{2 g^3 (2 \alpha')^2 \Bigl[} + (\zeta_1
\cdot \zeta_4) \Bigl\{   (\zeta_5 \cdot k_2)  \left( (\zeta_2
\cdot k_1)(\zeta_3 \cdot k_4) K_2 - (\zeta_2 \cdot k_4)(\zeta_3
\cdot k_1) K_3   \right)- \nonumber \\ & & \hphantom{2 g^3 (2
\alpha')^2 \Bigl[+ (\zeta_1 \cdot \zeta_4) \Bigl\{} - (\zeta_5
\cdot k_1) \Bigl(   (\zeta_2 \cdot k_3)(\zeta_3 \cdot k_4) {K_4}'
 - (\zeta_2 \cdot k_4)(\zeta_3 \cdot k_2) {K_5}' +
\nonumber
\\
&& \hphantom{2 g^3 (2 \alpha')^2 \Bigl[+ (\zeta_1 \cdot \zeta_4)
\Bigl\{- (\zeta_5 \cdot k_1)\Bigl(} +(\zeta_2 \cdot
 k_4)(\zeta_3 \cdot k_4) {K_1}' -
(\zeta_2
 \cdot k_1)(\zeta_3 \cdot k_4) {K_2}
  \Bigr)+
\nonumber
\\ & & \hphantom{2 g^3 (2 \alpha')^2 \Bigl[} + (\zeta_5
\cdot k_4) \Bigl(   (\zeta_2 \cdot k_1)(\zeta_3 \cdot k_2) K_4   +
(\zeta_2 \cdot k_1)(\zeta_3 \cdot k_1) K_1- \nonumber\\ &&
\hphantom{2 g^3 (2 \alpha')^2 \Bigl[+ (\zeta_5 \cdot k_4) \Bigl(}
- (\zeta_2 \cdot k_3)(\zeta_3 \cdot k_1) K_5 - (\zeta_2 \cdot
k_4)(\zeta_3 \cdot k_1) {K_3}
  \Bigr)     \Bigr\} \biggr] +
\nonumber
\\ & & \hphantom{2 g^3 (2 \alpha')^2 \Bigl[} +  \left( \
\mbox{cyclic permutations of indexes (1,2,3,4,5)} \ \right) \ .
\label{amplitude}
\end{eqnarray}
In this formula, $K_i$, ${K_i}'$, $L_i$ and ${L_i}'$ are $\alpha'$
dependent factors\footnote{In \cite{Brandt1} they were called
`kinematic factors' but we will not use this terminology any longer
since the word `kinematic' is usually reserved for expressions which
depend on both, momenta $and$ polarizations. In the present work we
will use the terminology `$\alpha'$ dependent factor' to denote an
expression which depends on $\alpha'$ $and$ the momenta $k_i$.},
defined by a double integral of the form
\begin{eqnarray}
\int_0^1 d x_3 \int_0^{x_3} d x_2 \ x_3^{2 \alpha' \alpha_{13}}
(1-x_3)^{2 \alpha'
  \alpha_{34}} {x_2}^{2 \alpha' \alpha_{12}} (1-x_2)^{2 \alpha'
  \alpha_{24}} (x_3-x_2)^{2 \alpha' \alpha_{23}} \cdot
\varphi(x_2, x_3) \ , \label{double-int}
\end{eqnarray}
where the function $\varphi(x_2, x_3)$ has a specific expression
for each of them (see appendix A.1 of \cite{Brandt1} for further
details). They can all be calculated as a product of a Beta and a
Hypergeometric function \cite{Kitazawa1} and they have a well
defined Laurent expansion in $\alpha'$ (after regularizing in some
cases). For example, in the case of the factors $K_2$ and $K_3$ we
have that \cite{Brandt1}\footnote{In \cite{Brandt1}, the
expansions of $K_2$ and $K_3$ contained the variables $\rho$ and
$\alpha_{24}$ which we have already substituted using the
relations (\ref{rho}), (\ref{alpha13}) and (\ref{alpha24}) of
appendix \ref{expansion-K3K}.}:
\begin{eqnarray}
K_2 & = & \frac{1}{(2 \alpha')^2} \left\{ \frac{1}{\alpha_{12} \
\alpha_{34}} \right\} - \frac{ \ \pi^2}{6} \left\{\frac{
\alpha_{51} \ \alpha_{12} - \alpha_{12} \ \alpha_{34} +
\alpha_{34} \ \alpha_{45}}{\alpha_{12} \ \alpha_{34}} \right\} +
\nonumber
\\&+& \zeta(3) \ (2 \alpha') \left\{  \frac{\alpha_{12}^2 \ \alpha_{51}
 - \alpha_{34}^2 \ \alpha_{12} +
\alpha_{45}^2 \ \alpha_{34} + \alpha_{51}^2 \ \alpha_{12} -
\alpha_{12}^2 \ \alpha_{34} + \alpha_{34}^2 \ \alpha_{45} - 2
\alpha_{12} \ \alpha_{23} \ \alpha_{34} }{\alpha_{12} \
\alpha_{34}} \right\}
+ \nonumber\\
&+& {\cal O}((2\alpha')^2) \,, \label{K2}
\\
K_3 & = & \frac{ \ \pi^2}{6} - \zeta(3) \ (2 \alpha') \left\{
\alpha_{12} + \alpha_{23} + \alpha_{34} + \alpha_{45} +
\alpha_{51} \right\} + {\cal O}((2\alpha')^2) \, .
\label{K3}
\end{eqnarray}
\noindent In (\ref{K2}) and (\ref{K3}) we are using the notation
\begin{eqnarray}
\alpha_{ij} = k_i \cdot k_j \ \ \ \ (i,j = 1,2,3,4,5; \ i \neq j) \
. \label{alphaij}
\end{eqnarray}

\noindent In appendix A.2 of \cite{Brandt1}, among some relations,
the following were found:
\begin{eqnarray}
\alpha_{34} K_2 & = & \alpha_{13} K_1 + \alpha_{23} K_4
\nonumber \\
\alpha_{24} K_3 & = & \alpha_{12} K_1 - \alpha_{23} K_5
\nonumber \\
\alpha_{13} L_1 & = & \alpha_{34} {L_3}' - \alpha_{23} {L_4}'
\nonumber \\
\alpha_{12} K_2 & = & \alpha_{24} {K_1}' + \alpha_{23} {K_4}'
\nonumber \\
\alpha_{13} K_3 & = & \alpha_{34} {K_1}' - \alpha_{23} {K_5}'
\nonumber \\
\alpha_{24} {L_1}' & = & \alpha_{12} L_3 - \alpha_{23} L_4
\nonumber \\
\alpha_{34}{K_4}'-\alpha_{13}K_5 & = &
\alpha_{12}K_4-\alpha_{24}{K_5}' \,.
\label{relations1}
\end{eqnarray}
\noindent The first three of them were explicitly written in eq.
(A.6) of \cite{Brandt1}. The following three relations can be
obtained from the first ones by the $duality$ operation mentioned in
that appendix\footnote{From the point of view of the string
world-sheet, this $duality$ operation is nothing else than a $twist$
of a disk with five insertions, with respect to the fifth vertex.
This will be seen in the third item of subsection
\ref{properties}.}. All of these expressions can be derived using
the definition of each $K_i$, ${K_i}'$, $L_i$ and ${L_i}'$, as a
double integral (see appendix A.1 of \cite{Brandt1}), and
integration by parts. The last relation in (\ref{relations1}) comes
from demanding invariance of $K_6$ under the mentioned $duality$
operation and using its expression given in appendix A.2 of the same
reference.

\noindent Now, in \cite{Machado1} there were found additional
relations, independent of the ones in (\ref{relations1}):
\begin{eqnarray}
K_2 + (K_1-L_3+{K_1}'-{L_3}')/2 = 0,\nonumber \\
K_1-K_4+K_5 = 0, \nonumber \\
{K_1}'-{K_4}'+{K_5}' = 0, \nonumber \\
K_5+{K_5}'+(K_1-L_4+{K_1}'-{L_4}')/2 = 0, \nonumber \\
L_2-L_3-{L_4}'=0, \nonumber \\
L_2-{L_3}'- L_4=0, \nonumber \\
K_1-{K_1}'-(L_4 - {L_4}') = 0.
\label{relations2}
\end{eqnarray}
These independent relations come from the very definition of each
factor, since the function $\varphi(x_2, x_3)$ in every case is a
fraction. For example, for the factors $K_1$, $K_4$ and $K_5$, the
corresponding $\varphi(x_2, x_3)$ function is given by
\begin{eqnarray}
\varphi_{K_1}(x_2,x_3) = \frac{1}{x_2 \ x_3} , \ \ \ \
\varphi_{K_4}(x_2,x_3) = \frac{1}{x_2 \ (x_3-x_2)} , \ \ \ \
\varphi_{K_5}(x_2,x_3) = \frac{1}{x_3 \ (x_3-x_2)} \ \ ,
\label{fractions}
\end{eqnarray}
which can easily be seen to satisfy
\begin{eqnarray}
\varphi_{K_1}(x_2,x_3) - \varphi_{K_4}(x_2,x_3) +
\varphi_{K_5}(x_2,x_3) = 0 \ .
\label{sumoffractions}
\end{eqnarray}
The integrated version of eq. (\ref{sumoffractions}) is precisely
the second of the equations in (\ref{relations2}).

\noindent Besides all these relations, we now introduce a new
factor, $T$, defined as
\begin{eqnarray}
T = (2 \alpha')^2 \biggl[ \ \alpha_{12} \ \alpha_{34} \ K_2 + (
\alpha_{51} \ \alpha_{12} - \alpha_{12} \ \alpha_{34} +
\alpha_{34} \ \alpha_{45} ) \ K_3 \ \biggr]. \label{K2-K}
\end{eqnarray}
The same as $K_3$, this factor remains invariant under cyclic
permutations of indexes $(1,2,3,4,5)$. We prove this in appendix
\ref{expansion-K3K}.

\noindent The factors $K_2$ and $K_3$ have been written in terms of
Beta and Hypergeometric functions in equations (\ref{K2hyperg}) and
(\ref{K3hyperg}).

\noindent So, summarizing, the 5-point amplitude in
(\ref{amplitude}) is given in terms of sixteen factors $K_i$,
${K_i}'$, $L_i$ and ${L_i}'$, which are related with the
additional factor $T$ by fifteen independent linear relations
given in (\ref{relations1}), (\ref{relations2}) and (\ref{K2-K}).
This allows to write the 5-point amplitude in (\ref{amplitude}) in
terms of only two factors, which we choose to be $T$ and
$K_3$\footnote{Eq. (\ref{amp2}) is the same as the one in Table 2
of \cite{Machado1}, but with not exactly the same choice of
momentum dependent factors and kinematic expressions.}
\begin{eqnarray}
A(1,2,3,4,5) =  T \cdot A(\zeta,k) \ + (2  \alpha')^2 \ K_3 \cdot
B(\zeta,k) \ . \ \label{amp2}
\end{eqnarray}
Here, $A(\zeta,k)$ and $B(\zeta,k)$ are two kinematical expressions
which are known explicitly, after all the substitutions of the
factors have been done in (\ref{amplitude}). They can be identified
with subamplitudes of specific terms of the effective lagrangian
(\ref{effective1}), as we will see in the next lines.

\noindent In appendix \ref{expansion-K3K} we have that the $\alpha'$
expansion of $T$ begins as
\begin{eqnarray}
T & = & 1 + {\cal O}((2  \alpha')^3) \ . \label{K}
\end{eqnarray}
Substituting the leading terms of the $\alpha'$ expansions of $T$
and $(2  \alpha')^2  K_3$ in (\ref{amp2}), we have that $A(\zeta,k)$
should agree with the Yang-Mills 5-point subamplitude and that
$B(\zeta,k)$ should agree with the $F^4$ terms 5-point subamplitude.
We have checked (computationally) that this really happens on-shell,
after using momentum conservation and physical state conditions.

\noindent So our final formula for the 5-point subamplitude is
\begin{eqnarray}
A(1,2,3,4,5) = T \cdot A_{YM}(1,2,3,4,5) + (2  \alpha')^2 K_3
\cdot A_{F^4}(1,2,3,4,5) \ .
\label{A12345final}
\end{eqnarray}
In (\ref{AYM5}) and (\ref{AF412345}) we give the expressions for
$A_{YM}(1,2,3,4,5)$ and $A_{F^4}(1,2,3,4,5)$, respectively.\\
\noindent The formula in (\ref{A12345final}) has exactly the same
structure of the corresponding one for the 4-point subamplitude,
written in (\ref{A1234II}), but with two kinematic expressions and
two factors which contain the $\alpha'$ dependence.

\subsection{Properties satisfied by the 5-point amplitude}
\label{properties}

\noindent On this subsection we confirm that $A(1,2,3,4,5)$, given
in (\ref{A12345final}), satisfies the usual properties of massless
open string subamplitudes, namely, cyclic invariance, on-shell gauge
invariance, world-sheet parity and factorizability.

\begin{enumerate}
\item{\underline{Cyclic invariance}:}\\
\noindent It was mentioned in the previous subsection, and it is
proved in appendix \ref{expansion-K3K}, that the factors $T$ and
$K_3$ are invariant under cyclic permutations of indexes
$(1,2,3,4,5)$. Now, once $A_{YM}(1,2,3,4,5)$ and
$A_{F^4}(1,2,3,4,5)$ are the Yang-Mills and the $F^4$ terms 5-point
subamplitudes, which are non-abelian field theory amplitudes, by
construction they are cyclic invariant. So, in this sense, the
cyclic invariance of $A(1,2,3,4,5)$ is already manifestly written in
formula (\ref{A12345final}).

\item{\underline{On-shell gauge invariance}:}\\
\noindent This property consists in that the subamplitude should
become zero if any of the polarizations $\zeta_i$ is substituted by
the corresponding momentum $k_i$, after using physical state
($\zeta_j \cdot k_j = 0$) and on-shell ($k_j^2=0$) conditions for
all external string states, together with momentum
conservation\cite{Schwarz1}. That this indeed happens in
(\ref{A12345final}) can be understood from the fact that
$A_{YM}(1,2,3,4,5)$ and $A_{F^4}(1,2,3,4,5)$ are 5-point
subamplitudes that come from gauge invariant terms, so both of them
should independently become zero when doing $\zeta_i \rightarrow
k_i$ for any $i=1,2,3,4,5$. So the on-shell gauge invariance of
$A(1,2,3,4,5)$ is also manifestly
written in (\ref{A12345final}).\\
\noindent In any case, the explicit check of the on-shell gauge
invariance of $A_{F^4}(1,2,3,4,5)$ can be seen as follows. This
subamplitude is given by
\begin{eqnarray}
A_{F^4}(1,2,3,4,5)  =  \ 2 \  g^3
\left\{ \frac{}{} \left[ \frac{}{} K(\zeta_1, \zeta_2; \zeta_3,
k_3; \zeta_4, k_4; \zeta_5, k_5) \ + \right. \right. \ \ \ \ \ \
\ \ \ \ \ \ \ \ \ \ \ \ \ \ \ \ \ \ \ \ \ \ \ \ \ \ \ \ \ \
\ \ \ \ \ \nonumber \\
+ \frac{1}{\alpha_{12}} \left( \frac{}{} (\zeta_1 \cdot \zeta_2)
K(k_1, k_2; \zeta_3, k_3; \zeta_4, k_4; \zeta_5, k_5) + (\zeta_1
\cdot k_2) K(\zeta_2, k_1 + k_2; \zeta_3, k_3; \zeta_4, k_4;
\zeta_5, k_5) \ - \right. \nonumber \\
\left. \left. \left. - (\zeta_2 \cdot k_1) K(\zeta_1, k_1 + k_2;
\zeta_3, k_3; \zeta_4, k_4; \zeta_5, k_5) \frac{}{} \right)
\frac{}{} \right]+ \mbox{(cyclic permutations)} \frac{}{} \right\}
\ . \ \ \ \ \
 \ \ \ \ \
\label{AF412345}
\end{eqnarray}
It has been obtained as the coefficient of $\pi^2/6 \ (2 \alpha')^2$
in the $\alpha'$ series of the subamplitude
$A_{D^{2n}F^4}(1,2,3,4,5)$, given in (\ref{AD2nF4}). As mentioned in
appendix \ref{D2nF4}, in this formula the expression $K(A, a; B, b;
C, c; D, d)$ denotes the same kinematic construction of
(\ref{kinematicfactor}), evaluated in the corresponding variables.
Due to the symmetries of the $t_{(8)}$ tensor (see appendix
\ref{t8tensor}), it is not difficult to see that this last
expression becomes zero whenever any of the $\zeta_i$ is substituted
by $k_i$.

\item{\underline{World-sheet parity}:}\\
\noindent The tree level interaction of $n$ open strings is
described by a conformal field theory on a disk with $n$
insertions on its boundary\cite{Polchinski1}. A world-sheet parity
transformation, $\sigma \rightarrow l - \sigma$ (where $\sigma \
\varepsilon \ [0, \ l]$ is the string internal coordinate),
corresponds to a twisting of this disk with respect to any of
those insertions (in figure \ref{twist5}, for example, it
is shown a twisting with respect to the fifth insertion).\\
\begin{figure}[h]
\centerline{\includegraphics*[scale=0.8,angle=0]{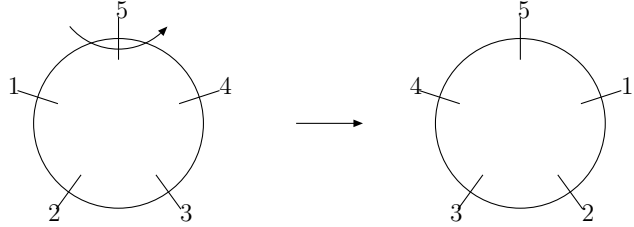}}
\caption{Twisting of a disk with five insertions, with respect to
the fifth one. Under this operation the indexes $(1,2,3,4,5)$
transform as indicated in equation (\ref{twisting5}). }
\label{twist5}
\end{figure}
\noindent When there is an interaction of an odd number of open
strings, the subamplitude changes sign under a world-sheet parity
transformation \cite{Schwarz1, Polchinski1} . So, in the case of
five interacting superstrings, a twisting with respect to, say the
fifth one,
\begin{eqnarray}
(1,2,3,4,5) \rightarrow (4,3,2,1,5) \ , \label{twisting5}
\end{eqnarray}
should imply that
\begin{eqnarray}
A(1,2,3,4,5) = - A(4,3,2,1,5) \ . \label{twisting}
\end{eqnarray}
To see that the amplitude in (\ref{A12345final}) indeed satisfies
this condition it can be argued as follows. First, in appendix
\ref{K2K3Ktwisting} it is proved that the factors $T$ and $K_3$
are invariant under a twisting transformation. Second, the
Yang-Mills subamplitude $A_{YM}(1,2,3,4,5)$ already satisfies
(\ref{twisting}) \cite{Mangano1}\footnote{This twisting
transformation is also called `inversion' in \cite{Mangano1} since
due to the cyclic symmetry (\ref{twisting}) it can also be written
as $A(1,2,3,4,5) = - A(5,4,3,2,1)$ \ .}. And third, using the
symmetries of the $t_{(8)}$ tensor it is not difficult to prove
that $A_{F^4}(1,2,3,4,5)$, in (\ref{AF412345}), does also satisfy
the twisting condition (\ref{twisting}).

\item{\underline{Factorizability}:}\\
\noindent There are two non trivial tests that confirm the right
structure of the poles of $A(1,2,3,4,5)$ in (\ref{A12345final}). The
factorization comes when a particular limit is considered. In that
limit the subamplitude diverges with simple poles and its residues
factorize in terms of the 4-point subamplitude.\\
\noindent Here we follow very closely section 6 of \cite{Mangano1}.
\begin{itemize}
\item{\underline{Soft boson factorization}:}\\
This corresponds to the case when one of the bosons, say the fifth
one, becomes soft ($k_5 \rightarrow 0$). It is very well known (see
\cite{Mangano1}, for example) that the $n$-point Yang-Mills
subamplitude satisfies the factorization mentioned here. In the case
$n=5$ it has the following form:
\begin{eqnarray}
A_{YM}(1,2,3,4,5)_{k_5 \rightarrow 0} \sim \ g \left(\frac{}{}
\frac{\zeta_5 \cdot k_1}{k_5 \cdot k_1} - \frac{\zeta_5 \cdot
k_4}{k_5 \cdot k_4} \frac{}{} \right) \ A_{YM}(1,2,3,4) \ .
\label{softYM5}
\end{eqnarray}
Also, taking $k_5 \rightarrow 0$ in (\ref{AF412345}) it is not
difficult to see that $A_{F^4}(1,2,3,4,5)$ satisfies exactly the
same factorization.
Therefore, taking the $k_5 \rightarrow 0$ limit in
(\ref{A12345final}) and using that
\begin{eqnarray}
A_{YM}(1,2,3,4) & = & 2 \ g^2 \frac{K(\zeta_1, k_1; \zeta_2, k_2;
\zeta_3, k_3; \zeta_4, k_4)}{\alpha_{12} \alpha_{14}} \ , \
\nonumber \\
A_{F^4}(1,2,3,4) & = & - \ 2 \ g^2 \ K(\zeta_1, k_1; \zeta_2, k_2;
\zeta_3, k_3; \zeta_4, k_4) \ , \label{YMF4K}
\end{eqnarray}
we have that
\begin{eqnarray}
A(1,2,3,4,5)_{k_5 \rightarrow 0} & \sim & 2 g^3  \left(\frac{}{}
\frac{\zeta_5 \cdot k_1}{k_5 \cdot k_1} - \frac{\zeta_5 \cdot
k_4}{k_5 \cdot k_4} \frac{}{} \right) \left(\frac{}{}
\frac{1}{\alpha_{12} \alpha_{14}} T \ - \ (2 \alpha')^2 K_3
\frac{}{}
\right)\biggl|_{k_5=0} \times \nonumber \\
&& \times \ K(\zeta_1, k_1; \zeta_2, k_2; \zeta_3, k_3; \zeta_4,
k_4) \ .
\label{almostsoft}
\end{eqnarray}
Now, if $k_5=0$ this implies that $\alpha_{45}=\alpha_{51}=0$ and
$\alpha_{14}=\alpha_{23}$ (see eq. (\ref{alpha14})), so we have that
\begin{eqnarray}
A(1,2,3,4,5)_{k_5 \rightarrow 0} & \sim & 2 g^3  \left(\frac{}{}
\frac{\zeta_5 \cdot k_1}{k_5 \cdot k_1} - \frac{\zeta_5 \cdot
k_4}{k_5 \cdot k_4} \frac{}{} \right) \left(\frac{}{}
\frac{1}{\alpha_{12} \alpha_{23}} T \ - \ (2 \alpha')^2 K_3
\frac{}{}
\right)\biggl|_{\alpha_{45}=\alpha_{51}=0} \times \nonumber \\
&& \times \ K(\zeta_1, k_1; \zeta_2, k_2; \zeta_3, k_3; \zeta_4,
k_4) \ .
\label{almostsoft2}
\end{eqnarray}
In this last relation, the expression evaluated in
$\alpha_{45}=\alpha_{51}=0$ is basically the Gamma factor of the
4-point amplitude evaluated in $\alpha_{12}$ and $\alpha_{23}$ (see
eq. (\ref{relate3})), which finally allows us to write
\begin{eqnarray}
A(1,2,3,4,5)_{k_5 \rightarrow 0} \sim \ g \left(\frac{}{}
\frac{\zeta_5 \cdot k_1}{k_5 \cdot k_1} - \frac{\zeta_5 \cdot
k_4}{k_5 \cdot k_4} \frac{}{} \right) \ A(1,2,3,4) \ .
\label{softA12345}
\end{eqnarray}
\item{\underline{Factorization of collinear poles}:}\\
\noindent This corresponds to the case of two consecutive bosons, say the first
and the second one, that become parallel. The Yang-Mills subamplitude
factorization, in this case, has the following form \cite{Mangano1}:
\begin{multline}
A_{YM}(\zeta_1, k_1;\zeta_2, k_2;\zeta_3, k_3;\zeta_4, k_4;\zeta_5, k_5)_{k_1 || k_2} \sim \\
\begin{split}
& \frac{1}{2(k_1 \cdot k_2)} \ V^{\mu} \frac{\partial}{\partial \zeta^{\mu}}
A_{YM}(\zeta, k_1+k_2;\zeta_3, k_3;\zeta_4, k_4;\zeta_5, k_5) \ ,
\end{split}
\label{collinearYM}
\end{multline}
where
\begin{eqnarray}
V^{\mu} = - g \biggl[(\zeta_1 \cdot \zeta_2) (k_1-k_2)^{\mu}-
2(\zeta_2 \cdot k_1) {\zeta_1}^{\mu}+2(\zeta_1 \cdot k_2)
{\zeta_2}^{\mu} \biggr] \ \label{Vmu}
\end{eqnarray}
comes from the Yang-Mills 3-point vertex given in
(\ref{YMvertex3}):
\begin{eqnarray}
V_{\mu}= -i \ g \ V^{(3)}_{YM \ \mu_1 \mu_2 \mu}(k_1, k_2, -k_1
-k_2) \ \zeta_1^{\mu_1}\zeta_2^{\mu_2} \ . \label{vertex3}
\end{eqnarray}
Notice that the subamplitude on the left handside of
(\ref{collinearYM}) is a 5-point subamplitude while the one on the
right handside
is a 4-point one.\\
\noindent Now, when considering the $k_1 || k_2$ limit in
(\ref{AF412345}) the dominant term is precisely the one which has
the denominator $\alpha_{12}$ (recall that $k_1 \cdot k_2 = 0$ in
this limit, due to the on-shell condition). The three terms in the
residue of $\alpha_{12}$ in (\ref{AF412345}) can be seen to match
with the corresponding ones of $V^{\mu}$, when substituted in
(\ref{Vmu}), so $A_{F^4}(1,2,3,4,5)$ also
satisfies the factorization in (\ref{collinearYM}). \\
\noindent So, in the same way as was done in the previous item, taking the
$k_1 || k_2$ limit in (\ref{A12345final}) leads to
\begin{multline}
A(\zeta_1, k_1;\zeta_2, k_2;\zeta_3, k_3;\zeta_4, k_4;\zeta_5, k_5)_{k_1 || k_2} \sim \\
\begin{split}
& \frac{1}{2(k_1 \cdot k_2)} \ V^{\mu} \frac{\partial}{\partial \zeta^{\mu}} \biggl[
2 \ g^2 \ K(\zeta, k_1+k_2;\zeta_3, k_3;\zeta_4, k_4;\zeta_5, k_5) \biggr] \ \times \\
& \times \left(\frac{}{} \frac{T}{[(k_1+k_2) \cdot k_3] \
[(k_1+k_2) \cdot k_5]} \ - \ (2 \alpha')^2 K_3 \frac{}{}
\right)\biggl|_{\alpha_{12}=0} \ .
\end{split}
\label{collinearA1}
\end{multline}
Using relations (\ref{alpha13}) and (\ref{alpha25}) it can easily be seen that
$(k_1+k_2) \cdot k_3 = \alpha_{45}$ and $(k_1+k_2) \cdot k_5=\alpha_{34}$ when
$\alpha_{12}=0$, so the factor in the third line in (\ref{collinearA1}) becomes,
once more, the Gamma factor of the 4-point subamplitude (see (\ref{relate3})).
So we finally have that:
\begin{multline}
A(\zeta_1, k_1;\zeta_2, k_2;\zeta_3, k_3;\zeta_4, k_4;\zeta_5, k_5)_{k_1 || k_2} \sim \\
\begin{split}
& \frac{1}{2(k_1 \cdot k_2)} \ V^{\mu} \frac{\partial}{\partial \zeta^{\mu}}
 A(\zeta, k_1+k_2;\zeta_3, k_3;\zeta_4, k_4;\zeta_5, k_5) \ .
\end{split}
\label{collinearA3}
\end{multline}
\end{itemize}
\end{enumerate}

\section{5-field terms in the low energy effective lagrangian}
\label{lagrangian}
In the case of the $D^{2n}F^4$ terms
lagrangian, ${\cal L}_{D^{2n}F^4}$, given in
(\ref{Seff-nonabelian}), its determination was quite direct from
the corresponding scattering subamplitude (\ref{A1234}).
Unfortunately, the same procedure does not work in the case of
${\cal L}_{D^{2n}F^5}$: for each of the two terms in the 5-point
subamplitude in (\ref{A12345final}), there does not exist a local
lagrangian, at each order in $\alpha'$, which reproduces the
corresponding subamplitude\footnote{Only for the first term in
(\ref{A12345final}), at order $0$ in $\alpha'$, it is possible to
find a local lagrangian, namely, the Yang-Mills lagrangian.}.
There only exists an effective lagrangian which reproduces the
\underline{sum} of the two terms in (\ref{A12345final}).  This can
be argued by noticing that the $\alpha'$ expansion for each of
those terms does not satisfy the factorizability property but, as
 it was seen in (\ref{softA12345}) and (\ref{collinearA3}), this
 property is indeed satisfied if \underline{both} terms are
 included.\\
\noindent In this section we deal with the main result of this
work, namely, the 5-field terms in the effective lagrangian. By
this we mean the explicit determination of the $D^{2n}F^5$
terms\footnote{The $D^{2n}F^4$ terms also contain 5-field terms,
but in the present context we refer to those which explicitly
contain 5 field strengths and covariant derivatives of them.} by
means of the 5-point subamplitude (\ref{A12345final}). These terms
do not contribute to the \underline{abelian} effective lagrangian:
in the abelian limit all of them become zero\footnote{In the case
of an odd number of strings, the abelian limit of the scattering
amplitude (\ref{general-amplitude}) gives a null result and,
therefore, no interaction term for them goes in the effective
lagrangian. This happens due to the world-sheet parity
antisymmetry, (\ref{twisting}),
of the string subamplitudes.}.\\
\noindent The determination of the $D^{2n}F^5$ terms, by means of
the complete 5-point subamplitude, is by far more complicated than
the corresponding one of the $D^{2n}F^4$ terms \cite{Chandia1}
given in (\ref{Seff-nonabelian}). This happens because, at every
order in $\alpha'$, say ${\alpha'}^{k+3}$ ($k=0, 1, 2, \ldots $),
the 5-point subamplitude receives contributions not only from the
$D^{2k}F^5$ terms, but also from the $D^{2k+2}F^4$ ones. The
contribution of the first type of terms has no poles, but the
contribution of the second type of
terms contains poles and also regular terms. \\
\noindent In the next subsection we first write the final
expression for the $D^{2n}F^5$ terms scattering subamplitude (and
leave the details of its derivation to appendix \ref{derivation}).
In the second subsection we write the desired lagrangian terms in
short notation (using tensors and $\alpha'$ dependent functions)
and in the last one we give explicit examples of the $D^{2n}F^5$
and the $D^{2n}F^4$ terms up to ${\cal O}({\alpha'}^4)$ order.

\subsection{The $D^{2n}F^5$ terms 5-point subamplitude}
\label{D2nF5}

From (\ref{effective1}) we have that the
contribution of the $D^{2n}F^5$ terms to the open superstring
5-point subamplitude is obtained as
\begin{eqnarray}
A_{D^{2n}F^5}(1,2,3,4,5) = A(1,2,3,4,5) - A_{YM}(1,2,3,4,5) -
A_{D^{2n}F^4}(1,2,3,4,5) \ ,
\label{AD2nF5}
\end{eqnarray}
where the $D^{2n}F^4$ terms 5-point subamplitude,
$A_{D^{2n}F^4}(1,2,3,4,5)$, is given in (\ref{AD2nF4}). In appendix
\ref{derivation} we have worked with the expression in
(\ref{AD2nF5}), checking out that all poles cancel (as expected),
arriving to the following expression:
\begin{eqnarray}
A_{D^{2n}F^5}(1,2,3,4,5) &=& g^3 \ \biggl\{ \frac{}{} \biggl[
H^{(1)} \cdot h^{(1)}(\zeta,k) + P^{(1)}
\cdot p^{(1)}(\zeta,k)+\nonumber \\
&&\hphantom{\biggl\{ \frac{}{} \biggl[} + U^{(1)} \cdot
u^{(1)}(\zeta,k) + W^{(1)} \cdot w^{(1)}(\zeta,k)+  Z^{(1)} \cdot
z^{(1)}(\zeta,k) \biggr] +
\nonumber \\
&&\hphantom{\biggl\{ \frac{}{} \biggl[} + (\frac{}{} \mbox{cyclic
permutations} \frac{}{}) \frac{}{} \biggr\} \ + g^3 \ \Delta \cdot
\delta(\zeta,k) , \ \
\label{AD2nF5nopoles}
\end{eqnarray}
where $H^{(1)}$, $P^{(1)}$, $U^{(1)}$, $W^{(1)}$, $Z^{(1)}$ and
$\Delta$ are $\alpha'$ dependent factors (obtained from the Gamma,
the $T$ and the $K_3$ factors) which are given in appendix
\ref{otherfactors}, while $h^{(1)}(\zeta,k)$, $p^{(1)}(\zeta,k)$,
$u^{(1)}(\zeta,k)$, $w^{(1)}(\zeta,k)$, $z^{(1)}(\zeta,k)$ and
$\delta(\zeta,k)$ are kinematical expressions (with no poles) which
depend on the polarizations $\zeta_i$ and momenta $k_i$
($i=1,2,3,4,5$). There is no summation over cyclic permutations of
the term `$g^3 \ \Delta \cdot \delta(\zeta,k)$' because $\Delta$ and
$\delta(\zeta,k)$ are
already cyclic invariant expressions.\\
\noindent The kinematical expressions in (\ref{AD2nF5nopoles}) are
given by
\begin{eqnarray}
\label{h23} h^{(1)}(\zeta,k) &=& t_{(10)}^{\mu_1 \nu_1 \mu_2 \nu_2
\mu_3 \nu_3 \mu_4 \nu_4 \mu_5
\nu_5}\zeta^1_{\mu_1}k^1_{\nu_1}\zeta^2_{\mu_2}k^2_{\nu_2}
\zeta^3_{\mu_3}k^3_{\nu_3}\zeta^4_{\mu_4}k^4_{\nu_4}
\zeta^5_{\mu_5}k^5_{\nu_5} \ ,
\\
\label{p34} p^{(1)}(\zeta,k) &=& 2 \ {(\eta \cdot
t_{(8)})_1}^{\mu_1 \nu_1 \mu_2 \nu_2 \mu_3 \nu_3 \mu_4 \nu_4 \mu_5
\nu_5}\zeta^1_{\mu_1}k^1_{\nu_1}\zeta^2_{\mu_2}k^2_{\nu_2}
\zeta^3_{\mu_3}k^3_{\nu_3}\zeta^4_{\mu_4}k^4_{\nu_4}
\zeta^5_{\mu_5}k^5_{\nu_5} \ ,
\\
\label{u34} u^{(1)}(\zeta ,k) &=&\biggl[ \frac{1}{2} \
t_{(10)}^{\mu _{1}\nu _{1}\mu _{2}\nu _{2}\mu _{3}\nu _{3}\mu
_{4}\nu _{4}\mu _{5}\nu _{5}}\alpha _{34} + 2 \ t_{(8)}^{\mu
_{4}\nu _{4}\mu _{5}\nu _{5}\mu _{1}\nu _{1}\mu _{2}\nu _{2}}
k_{1}^{[\mu _{3}}k_{2}^{\nu _{3}]} + 2 \ t_{(8)}^{\mu _{3}\nu
_{3}\mu _{4}\nu _{4}\mu _{5}\nu _{5}\mu
_{1}\nu _{1}} k_{1}^{[\mu _{2}}k_{3}^{\nu _{2}]} - \notag \\
&&- 2 \ t_{(8)}^{\mu _{1}\nu _{1}\mu _{2}\nu _{2}\mu _{3}\nu
_{3}\mu _{4}\nu _{4}} k_{1}^{[\mu _{5}}k_{4}^{\nu _{5}]} - 2 \
t_{(8)}^{\mu _{5}\nu _{5}\mu _{1}\nu _{1}\mu _{2}\nu _{2}\mu
_{3}\nu _{3}}
k_{1}^{[\mu _{4}}k_{5}^{\nu _{4}]} \biggr] \times \nonumber \\
&& \times \ \zeta _{\mu _{1}}^{1}k_{\nu _{1}}^{1}\zeta _{\mu
_{2}}^{2}k_{\nu _{2}}^{2}\zeta _{\mu _{3}}^{3}k_{\nu _{3}}^{3}\zeta
_{\mu _{4}}^{4}k_{\nu _{4}}^{4}\zeta _{\mu _{5}}^{5}k_{\nu _{5}}^{5}
\ ,
\\
\label{w51} w^{(1)}(\zeta ,k) &=&4 \ t_{(8)}^{\mu _{2}\nu _{2}\mu
_{3}\nu _{3}\mu _{4}\nu _{4}\mu _{5}\nu _{5}} \zeta _{\mu
_{1}}^{1}k_{\nu _{1}}^{1} \zeta _{\mu _{2}}^{2}k_{\nu
_{2}}^{2}\zeta _{\mu _{3}}^{3}k_{\nu _{3}}^{3}\zeta _{\mu
_{4}}^{4}k_{\nu _{4}}^{4}\zeta _{\mu _{5}}^{5}k_{\nu _{5}}^{5}
 k_{2}^{[\mu _{1}}k_{5}^{\nu
_{1}]}  \ ,
\\
\label{z23} z^{(1)}(\zeta,k) &=& 4 \ t_{(8)}^{\mu_2 \nu_2 \mu_3
\nu_3 \mu_4 \nu_4 \mu_5 \nu_5}\zeta^1_{\mu_1}k^1_{\nu_1}
\zeta^2_{\mu_2}k^2_{\nu_2}\zeta^3_{\mu_3}k^3_{\nu_3}
\zeta^4_{\mu_4}k^4_{\nu_4}\zeta^5_{\mu_5}k^5_{\nu_5} \biggl(
k_{3}^{[\mu _{1}}k_{4}^{\nu _{1}]}-k_{2}^{[\mu_{1}}k_{5}^{\nu
_{1}]} \biggr) \ ,
\\
\label{delta}
\delta(\zeta,k) &=& \frac{1}{5}\biggl[ t_{(10)}^{\mu
_{1}\nu _{1}\mu _{2}\nu _{2}\mu _{3}\nu _{3}\mu _{4}\nu _{4}\mu
_{5}\nu _{5}}\alpha _{51}\alpha _{12}
 + 4 {(\eta \cdot t_{(8)})_1}^{\mu _{1}\nu _{1}\mu _{2}\nu
_{2}\mu _{3}\nu _{3}\mu _{4}\nu _{4}\mu _{5}\nu _{5}}\alpha
_{23}\alpha _{45} \biggr] \times \nonumber \\
&& \times \ \zeta _{\mu _{1}}^{1}k_{\nu _{1}}^{1}\zeta _{\mu
_{2}}^{2}k_{\nu _{2}}^{2}\zeta _{\mu _{3}}^{3}k_{\nu
_{3}}^{3}\zeta _{\mu _{4}}^{4}k_{\nu _{4}}^{4}\zeta _{\mu
_{5}}^{5}k_{\nu _{5}}^{5} + (\frac{}{} \mbox{cyclic permutations}
\frac{}{}) \ .
\end{eqnarray}
The expression for the $(\eta \cdot t_{(8)})_1$ tensor of
(\ref{p34}) and (\ref{delta}) is given in equation (\ref{etat8})
and the expression for the $t_{(10)}$ tensor is
given in appendix \ref{t10tensor} . \\
\noindent The expression for $A_{D^{2n}F^5}(1,2,3,4,5)$ in
(\ref{AD2nF5nopoles}) is such that, besides containing no poles and
being cyclic invariant, satisfies (on-shell) gauge invariance and
world-sheet parity\footnote{The upper index $(1)$ in all factors
$H^{(1)}$, $P^{(1)}$, $\ldots$ , denotes that they are invariant
under a twisting transformation with respect to index $1$.
Similarly, the upper index $(1)$ in all kinematical expressions
$h^{(1)}(\zeta,k)$, $p^{(1)}(\zeta,k)$, $\ldots$ , denotes that they
change their sign under a twisting transformation with respect to
index $1$.} on $each$ group of terms, that is, on $[ H^{(1)} \cdot
h^{(1)}(\zeta,k) + (\mbox{cyclic permutations}) ]$, $\ldots$, $[
Z^{(1)} \cdot z^{(1)}(\zeta,k) + (\mbox{cyclic permutations}) ]$ and
$\Delta \cdot \delta(\zeta,k)$, separately. These conditions are
enough to find a local lagrangian for each of those terms, at each
order in $\alpha'$.

\subsection{The main formula}
\label{main}

From formula (\ref{AD2nF5nopoles}), for
$A_{D^{2n}F^5}(1,2,3,4,5)$, and the kinematical expressions
(\ref{h23})-(\ref{delta}), the following effective lagrangian can
be obtained for the $D^{2n}F^5$ terms, up to terms which are
sensible to 6 or higher-point amplitudes\footnote{When considering
the $\alpha'$ expansion of eq. (\ref{LD2nF5final}), whenever two
covariant derivatives $D_{\alpha}$ and $D_{\beta}$ (with $\alpha
\neq \beta$) operate on a $same$ field strength $F_{\mu \nu}(x)$,
the order in which they operate does not matter since the
difference between the two possibilities will be sensible only to
6 or higher-point amplitudes.}:
\begin{multline*}
{\cal L}_{D^{2n}F^5}  =  \ i \ g^3 \ \int \int \int \int \int
\left\{ \frac{}{}\prod_{j=1}^{5} d^{10} x_j \ \delta^{(10)}(x-x_j)
\frac{}{} \right\} \times \\
\begin{split}
& \times  \ \biggl[ \ \biggl\{ \frac{1}{32} H^{(1)} ( - D_{2} \cdot
D_{3}, - D_{3} \cdot D_{4} \ , - D_{4} \cdot D_{5} ) \
t_{(10)}^{\mu_1 \nu_1 \mu_2 \nu_2
\mu_3 \nu_3 \mu_4 \nu_4 \mu_5 \nu_5} + \\
&+ \frac{1}{16} P^{(1)} ( - D_{1} \cdot D_{2}, - D_{2} \cdot D_{3},
- D_{3} \cdot D_{4} \ , - D_{4} \cdot D_{5}, - D_{5} \cdot D_{1} ) \
{(\eta \cdot t_{(8)})_1}^{\mu_1 \nu_1 \mu_2 \nu_2
\mu_3 \nu_3 \mu_4 \nu_4 \mu_5 \nu_5} \biggr\} \times \\
&\times \tr\biggl( F_{\mu_1 \nu_1}(x_1)F_{\mu_2 \nu_2}(x_2)F_{\mu_3
\nu_3}(x_3)F_{\mu_4 \nu_4}(x_4)F_{\mu_5 \nu_5}(x_5) \biggr) \ \ -
\\
& - \ \ U^{(1)} ( - D_{1} \cdot D_{2}, - D_{2} \cdot D_{3}, - D_{4}
\cdot
D_{5}, - D_{5} \cdot D_{1} ) \times \\
& \times \biggl\{\frac{1}{64} t_{(10)}^{\mu_1 \nu_1 \mu_2 \nu_2
\mu_3 \nu_3 \mu_4 \nu_4 \mu_5 \nu_5} \tr\biggl( F_{\mu_1
\nu_1}(x_1)F_{\mu_2 \nu_2}(x_2)D^{\alpha}F_{\mu_3
\nu_3}(x_3)D_{\alpha} F_{\mu_4 \nu_4}(x_4)F_{\mu_5 \nu_5}(x_5)
\biggr) + \\
& + \frac{1}{16} t_{(8)}^{\mu_4 \nu_4 \mu_5 \nu_5 \mu_1 \nu_1 \mu_2
\nu_2} \tr\biggl( D^{\mu_3}F_{\mu_1 \nu_1}(x_1)D^{\nu_3}F_{\mu_2
\nu_2}(x_2)F_{\mu_3 \nu_3}(x_3)F_{\mu_4 \nu_4}(x_4)F_{\mu_5
\nu_5}(x_5) \biggr)+\\
& + \frac{1}{16} t_{(8)}^{\mu_3 \nu_3 \mu_4 \nu_4 \mu_5 \nu_5 \mu_1
\nu_1} \tr\biggl( D^{\mu_2}F_{\mu_1 \nu_1}(x_1)F_{\mu_2
\nu_2}(x_2)D^{\nu_2}F_{\mu_3 \nu_3}(x_3)F_{\mu_4 \nu_4}(x_4)F_{\mu_5
\nu_5}(x_5) \biggr)- \\
& - \frac{1}{16} t_{(8)}^{\mu_1 \nu_1 \mu_2 \nu_2 \mu_3 \nu_3 \mu_4
\nu_4} \tr\biggl( D^{\mu_5}F_{\mu_1 \nu_1}(x_1)F_{\mu_2
\nu_2}(x_2)F_{\mu_3 \nu_3}(x_3)D^{\nu_5}F_{\mu_4 \nu_4}(x_4)F_{\mu_5
\nu_5}(x_5) \biggr)- \\
& - \frac{1}{16} t_{(8)}^{\mu_5 \nu_5 \mu_1 \nu_1 \mu_2 \nu_2
\mu_3 \nu_3} \tr\biggl( D^{\mu_4}F_{\mu_1 \nu_1}(x_1)F_{\mu_2
\nu_2}(x_2)F_{\mu_3 \nu_3}(x_3)F_{\mu_4
\nu_4}(x_4)D^{\nu_4}F_{\mu_5 \nu_5}(x_5) \biggr) \ \biggr\} \ \
-\\
 & - \ \ \frac{1}{8} W^{(1)} ( - D_{1} \cdot D_{2}, - D_{2} \cdot
D_{3}, - D_{3} \cdot D_{4} \ , - D_{4} \cdot D_{5}, - D_{5} \cdot
D_{1} )
\times \\
& \times t_{(8)}^{\mu_2 \nu_2 \mu_3 \nu_3 \mu_4 \nu_4 \mu_5 \nu_5}
\tr\biggl( F_{\mu_1 \nu_1}(x_1)D^{\mu_1}F_{\mu_2
\nu_2}(x_2)F_{\mu_3 \nu_3}(x_3)F_{\mu_4
\nu_4}(x_4)D^{\nu_1}F_{\mu_5 \nu_5}(x_5) \biggr) \ \ -
\end{split}
\end{multline*}
\begin{multline}
\begin{split}
 & - \ \
\frac{1}{8} Z^{(1)} ( - D_{1} \cdot D_{2}, - D_{2} \cdot D_{3}, -
D_{3} \cdot D_{4} \ , - D_{4} \cdot D_{5}, - D_{5} \cdot D_{1} )
\times \\
& \times t_{(8)}^{\mu_2 \nu_2 \mu_3 \nu_3 \mu_4 \nu_4 \mu_5 \nu_5}
\biggl\{ \tr\biggl( F_{\mu_1 \nu_1}(x_1)F_{\mu_2
\nu_2}(x_2)D^{\mu_1}F_{\mu_3 \nu_3}(x_3)D^{\nu_1}F_{\mu_4
\nu_4}(x_4)F_{\mu_5 \nu_5}(x_5) \biggr)
- \\
& \hphantom{\times t_{(8)}^{\mu_2 \nu_2 \mu_3 \nu_3 \mu_4 \nu_4
\mu_5 \nu_5} } - \tr\biggl( F_{\mu_1 \nu_1}(x_1)D^{\mu_1}F_{\mu_2
\nu_2}(x_2)F_{\mu_3 \nu_3}(x_3)F_{\mu_4 \nu_4}(x_4)D^{\nu_1}F_{\mu_5
\nu_5}(x_5) \biggr) \biggr\} \ \ + \\
& + \ \ \frac{1}{160} \Delta( - D_{1} \cdot D_{2}, - D_{2} \cdot
D_{3}, - D_{3}
\cdot D_{4} \ , - D_{4} \cdot D_{5}, - D_{5} \cdot D_{1} ) \times \\
& \times \biggl\{ t_{(10)}^{\mu_1 \nu_1 \mu_2 \nu_2 \mu_3 \nu_3
\mu_4 \nu_4 \mu_5 \nu_5}  \tr\biggl( D^{\alpha} D^{\beta} F_{\mu_1
\nu_1}(x_1)D_{\alpha}F_{\mu_2 \nu_2}(x_2)F_{\mu_3
\nu_3}(x_3)F_{\mu_4
\nu_4}(x_4)D_{\beta}F_{\mu_5 \nu_5}(x_5) \biggr) + \\
& + 4 {(\eta \cdot t_{(8)})_1}^{\mu_1 \nu_1 \mu_2 \nu_2 \mu_3 \nu_3
\mu_4 \nu_4 \mu_5 \nu_5}  \times \\
& \hphantom{ + 4 }  \times \tr\biggl( F_{\mu_1
\nu_1}(x_1)D_{\alpha}F_{\mu_2 \nu_2}(x_2)D^{\alpha} F_{\mu_3
\nu_3}(x_3)D_{\beta} F_{\mu_4 \nu_4}(x_4)D^{\beta}F_{\mu_5
\nu_5}(x_5) \biggr) \ \biggr\} \ \ \ \biggr] \ .
\end{split}
\label{LD2nF5final}
\end{multline}
Notice that in this result we have used that the factors $P^{(1)}$,
$W^{(1)}$, $Z^{(1)}$ and $\Delta$ depend in $all$ five (independent)
$\alpha_{ij}$ variables, while $H^{(1)}$ and $U^{(1)}$ depend on a
$lower$ number of them:
\begin{eqnarray}
\label{explicitH1} H^{(1)} & = & H^{(1)}(\alpha_{23}, \alpha_{34},
\alpha_{45}) \ ,
\\
\label{explicitU1} U^{(1)} & = & U^{(1)}(\alpha_{12}, \alpha_{23},
\alpha_{45}, \alpha_{51}) \ ,
\end{eqnarray}
as may be seen directly in the formulas of appendix
\ref{otherfactors}.\\
\noindent That formula (\ref{LD2nF5final}) is indeed correct may
be confirmed by calculating the 1 PI 5-point function of it,
$\Gamma_{\mu_1 \mu_2 \mu_3 \mu_4 \mu_5}^{(5) \ a_1 a_2 a_3 a_4
a_5}(x_1, x_2, x_3, x_4, x_5)$, and then calculating its Fourier
Transform\footnote{In the case of the 1 PI 4-point function and
the corresponding subamplitude, for example, this was done in the
appendices of \cite{DeRoo1} and \cite{Chandia1}.}. From this last
one it may be checked that the corresponding subamplitude is the
one given in (\ref{AD2nF5nopoles}).

\subsection{Some $\alpha'$ terms}
\label{some}

In this subsection we will see two applications of formula
(\ref{LD2nF5final}), in order to see how it works. Before doing
so, notice that the $\alpha'$ dependent factors of
(\ref{AD2nF5nopoles}) and (\ref{LD2nF5final}) do not begin their
power expansions at the same order in $\alpha'$. As may be seen
from the explicit expansions of them in appendix
\ref{otherfactors}: $H^{(1)}$ and $P^{(1)}$ begin at ${\cal
O}({\alpha'}^3)$ order; $U^{(1)}$, $W^{(1)}$ and $Z^{(1)}$ begin
at ${\cal O}({\alpha'}^4)$ order and $\Delta$ begins at ${\cal
O}({\alpha'}^5)$ order.\\
\noindent As a first application in which the 5-point amplitude is
important, let us see the case of the non-abelian effective
lagrangian of the open superstring at order ${\cal
O}({\alpha'}^3)$ (which was first completely and correctly
calculated in \cite{Koerber1}). At this order, as may be seen in
(\ref{LD2nF4}) and (\ref{LD2nF5}), the lagrangian contains $F^5$
and $D^2F^4$ terms. The first ones can be taken from the
lagrangian in (\ref{LD2nF5final}) and the second ones from the
lagrangian in (\ref{Seff-nonabelian}), giving:
\begin{eqnarray}
{{\cal L}_{\rm ef}}^{(3)}& = &\zeta(3) \ {\alpha'}^3 \biggl[ \ - i
\frac{ \ g^3}{4} \biggl \{ t_{(10)}^{\mu_1 \nu_1 \mu_2 \nu_2 \mu_3
\nu_3 \mu_4 \nu_4 \mu_5 \nu_5} + 2 {(\eta \cdot t_{(8)})_1}^{\mu_1
\nu_1 \mu_2 \nu_2 \mu_3 \nu_3 \mu_4 \nu_4 \mu_5
\nu_5} \biggr\} \times \nonumber \\
&& \hphantom{\zeta(3) \ {\alpha'}^3 [ \ } \times \tr\biggl( F_{\mu_1
\nu_1}F_{\mu_2 \nu_2}F_{\mu_3 \nu_3}F_{\mu_4 \nu_4}F_{\mu_5 \nu_5}
\biggr) \ + \frac{ \ g^2}{2}
t_{(8)}^{\mu_1 \nu_1 \mu_2 \nu_2 \mu_3 \nu_3 \mu_4 \nu_4} \times \nonumber \\
&& \hphantom{\zeta(3) \ {\alpha'}^3 [ \ } \times \biggl\{
\tr\biggl( D^{\alpha}F_{\mu_1 \nu_1}D_{\alpha} F_{\mu_2
\nu_2}F_{\mu_3 \nu_3}F_{\mu_4 \nu_4} \biggr) + \tr\biggl(
D^2F_{\mu_1 \nu_1}F_{\mu_2 \nu_2}F_{\mu_3 \nu_3}F_{\mu_4 \nu_4}
\biggr)
\biggr\} \ \biggr] \ . \nonumber \\
\label{L3}
\end{eqnarray}
In order to compare (\ref{L3}) with one of the expressions of the
known result we use that
\begin{eqnarray}
D^2F_{\mu \nu} =  D^{\alpha}(D_{\alpha} F_{\mu \nu}) =
D_{\mu}(D_{\alpha} {F^{\alpha}}_{\nu}) - D_{\nu}(D_{\alpha}
{F^{\alpha}}_{\mu}) + 2 \ i \ g \ [ F_{\alpha \mu},
{F_{\nu}}^{\alpha} ] \ , \label{D2}
\end{eqnarray}
which may be derived using the Bianchi identity and the $[D,D]F =
-ig[F,F]$ relation (see (\ref{Bianchi}) and (\ref{group4}),
respectively).\\
Substituting (\ref{D2}) in (\ref{L3}), and dropping out on-shell
terms (i.e. the ones which contain $D_{\alpha}
{F^{\alpha}}_{\mu}$)\footnote{In \cite{Tseytlin1} it was seen, at
least up to ${\cal O}({\alpha'}^2)$ order, that the on-shell terms
in the effective lagrangian are not sensible to on-shell
scattering amplitudes.} leads to
\begin{eqnarray}
{{\cal L}_{\rm ef}}^{(3)}& = &\zeta(3) \ {\alpha'}^3 \biggl[ \ - i
\frac{ \ g^3}{4}  ( t_{(10)}+4 (\eta \cdot t_{(8)})_1)^{\mu_1
\nu_1 \mu_2 \nu_2 \mu_3 \nu_3 \mu_4 \nu_4 \mu_5 \nu_5} \tr\biggl(
F_{\mu_1 \nu_1}F_{\mu_2 \nu_2}F_{\mu_3
\nu_3}F_{\mu_4 \nu_4}F_{\mu_5 \nu_5} \biggr) \ +  \nonumber \\
&& \hphantom{\zeta(3) \ {\alpha'}^3 [ \ } + \frac{ \ g^2}{2}
t_{(8)}^{\mu_1 \nu_1 \mu_2 \nu_2 \mu_3 \nu_3 \mu_4 \nu_4}
\tr\biggl( D^{\alpha}F_{\mu_1 \nu_1}D_{\alpha} F_{\mu_2
\nu_2}F_{\mu_3 \nu_3}F_{\mu_4 \nu_4} \biggr) \ \biggr] \ .
\label{newL3}
\end{eqnarray}
\noindent Using the explicit expressions of the $t_{(8)}$ and the
$t_{(10)}$ tensors, and a basis of $F^5$ and $D^2F^4$ terms taken
from \cite{Grasso2}\footnote{See appendix B of this reference.},
it may be verified that the $D^2F^4$ terms of (\ref{newL3}) agree
with the ones in eq. (4.17) of \cite{Grasso1} and that the $F^5$
terms of (\ref{newL3}) agree, on-shell, with the ones of the
mentioned equation after going to $D=4$\footnote{In $D=4$, due to
the Cayley-Hamilton theorem, there is an identity involving the
$F^5$ terms which reduces the number of them which are
independent. That is the reason of why our $F^5$ terms in
(\ref{newL3}) only agree with the ones which go in the effective
action of $N=4$ SYM in \cite{Grasso1} after going to $4$
dimensions. We thank D. Grasso for explaining this to us.}. In
this last reference it was seen that the lagrangian in equation
(4.17) of it agrees (in $D=4$) with the one in \cite{Koerber1},
which was also confirmed in \cite{DeRoo2} and
\cite{Brandt1}.\\
\noindent The next interesting example consists in the ${\cal
O}({\alpha'}^4)$ terms of the effective lagrangian which are
sensible to 5-point amplitudes. We take the $D^2F^5$ terms from
(\ref{LD2nF5final}) and the $D^4F^4$ ones from the lagrangian in
(\ref{Seff-nonabelian}), giving:
\begin{eqnarray}
{\cal L'}^{(4)}& = & {\cal L}_{\rm D^2F^5} + {\cal L}_{\rm D^4F^4}
\ , \label{L4}
\end{eqnarray}
where
\begin{eqnarray}
{\cal L}_{\rm D^2F^5} & = &  i \ \frac{\ \pi^4}{90} \ g^3 \
{\alpha'}^4 \biggl[ \ \ - \frac{1}{8} t_{(10)}^{\mu_1 \nu_1 \mu_2
\nu_2 \mu_3 \nu_3 \mu_4 \nu_4 \mu_5 \nu_5} \biggl \{ 4 \tr\biggl(
F_{\mu_1 \nu_1}F_{\mu_2 \nu_2}F_{\mu_3 \nu_3}D^{\alpha}F_{\mu_4
\nu_4}D_{\alpha}F_{\mu_5 \nu_5} \biggr) +
\nonumber \\
&& + 4 \tr\biggl( F_{\mu_1 \nu_1}D^{\alpha}F_{\mu_2
\nu_2}D_{\alpha}F_{\mu_3 \nu_3}F_{\mu_4 \nu_4}F_{\mu_5 \nu_5}
\biggr) + \tr\biggl( F_{\mu_1 \nu_1}F_{\mu_2
\nu_2}D^{\alpha}F_{\mu_3 \nu_3}D_{\alpha}F_{\mu_4 \nu_4}F_{\mu_5
\nu_5} \biggr) \biggr\} -
\nonumber \\
&& - \frac{1}{4} {(\eta \cdot t_{(8)})_1}^{\mu_1 \nu_1 \mu_2 \nu_2
\mu_3 \nu_3 \mu_4 \nu_4 \mu_5 \nu_5} \biggl\{ 4 \tr\biggl(
F_{\mu_1 \nu_1}F_{\mu_2 \nu_2}D^{\alpha}F_{\mu_3
\nu_3}D_{\alpha}F_{\mu_4 \nu_4}F_{\mu_5 \nu_5} \biggr) +  \nonumber \\
&& + \tr\biggl( F_{\mu_1 \nu_1}F_{\mu_2 \nu_2}F_{\mu_3
\nu_3}D^{\alpha}F_{\mu_4 \nu_4}D_{\alpha}F_{\mu_5 \nu_5} \biggr) +
\tr\biggl( F_{\mu_1 \nu_1}D^{\alpha}F_{\mu_2
\nu_2}D_{\alpha}F_{\mu_3 \nu_3}F_{\mu_4 \nu_4}F_{\mu_5 \nu_5}
\biggr) + \nonumber \\
&& + 4 \tr\biggl( D^{\alpha}F_{\mu_1 \nu_1}F_{\mu_2 \nu_2}F_{\mu_3
\nu_3}F_{\mu_4 \nu_4}D_{\alpha}F_{\mu_5 \nu_5} \biggr) + 4
\tr\biggl( D^{\alpha}F_{\mu_1 \nu_1}D_{\alpha}F_{\mu_2
\nu_2}F_{\mu_3
\nu_3}F_{\mu_4 \nu_4}F_{\mu_5 \nu_5} \biggr) \biggr\} \ -   \nonumber \\
&& -  \frac{1}{4} t_{(10)}^{\mu_1 \nu_1 \mu_2 \nu_2 \mu_3 \nu_3
\mu_4 \nu_4 \mu_5 \nu_5} \tr\biggl( F_{\mu_1 \nu_1}F_{\mu_2
\nu_2}D^{\alpha}F_{\mu_3 \nu_3}D_{\alpha} F_{\mu_4 \nu_4}F_{\mu_5
\nu_5}
\biggr) - \nonumber \\
&& -  t_{(8)}^{\mu_4 \nu_4 \mu_5 \nu_5 \mu_1 \nu_1 \mu_2 \nu_2}
\tr\biggl( D^{\mu_3}F_{\mu_1 \nu_1}D^{\nu_3}F_{\mu_2
\nu_2}F_{\mu_3 \nu_3}F_{\mu_4 \nu_4}F_{\mu_5
\nu_5} \biggr)- \nonumber \\
&& -  t_{(8)}^{\mu_3 \nu_3 \mu_4 \nu_4 \mu_5 \nu_5 \mu_1 \nu_1}
\tr\biggl( D^{\mu_2}F_{\mu_1 \nu_1}F_{\mu_2
\nu_2}D^{\nu_2}F_{\mu_3 \nu_3}F_{\mu_4 \nu_4}F_{\mu_5
\nu_5} \biggr)+ \nonumber \\
&& +  t_{(8)}^{\mu_1 \nu_1 \mu_2 \nu_2 \mu_3 \nu_3 \mu_4 \nu_4}
\tr\biggl( D^{\mu_5}F_{\mu_1 \nu_1}F_{\mu_2 \nu_2}F_{\mu_3
\nu_3}D^{\nu_5}F_{\mu_4 \nu_4}F_{\mu_5
\nu_5} \biggr)+ \nonumber \\
&& +  t_{(8)}^{\mu_5 \nu_5 \mu_1 \nu_1 \mu_2 \nu_2 \mu_3 \nu_3}
\tr\biggl( D^{\mu_4}F_{\mu_1 \nu_1}F_{\mu_2 \nu_2}F_{\mu_3
\nu_3}F_{\mu_4 \nu_4}D^{\nu_4}F_{\mu_5
\nu_5} \biggr) \  -  \nonumber \\
&& -  \frac{1}{2} t_{(8)}^{\mu_2 \nu_2 \mu_3 \nu_3 \mu_4 \nu_4
\mu_5 \nu_5} \tr\biggl( F_{\mu_1 \nu_1}D^{\mu_1}F_{\mu_2
\nu_2}F_{\mu_3 \nu_3}F_{\mu_4 \nu_4}D^{\nu_1}F_{\mu_5 \nu_5}
\biggr) +  \nonumber
\end{eqnarray}
\begin{eqnarray}
 && +  \frac{1}{8} t_{(8)}^{\mu_2 \nu_2 \mu_3
\nu_3 \mu_4 \nu_4 \mu_5 \nu_5} \biggl\{ \tr\biggl( F_{\mu_1
\nu_1}F_{\mu_2 \nu_2}D^{\mu_1}F_{\mu_3 \nu_3}D^{\nu_1}F_{\mu_4
\nu_4}F_{\mu_5
\nu_5} \biggr)- \nonumber\\
 && \hphantom{  +  \frac{1}{8}
t_{(8)}^{\mu_2 \nu_2 \mu_3 \nu_3 \mu_4 \nu_4 \mu_5 \nu_5} \biggl\{
} - \tr\biggl( F_{\mu_1 \nu_1}D^{\mu_1}F_{\mu_2 \nu_2}F_{\mu_3
\nu_3}F_{\mu_4 \nu_4}D^{\nu_1}F_{\mu_5 \nu_5} \biggr) \biggr\} \ \
\ \biggr] \label{LD2F5}
\end{eqnarray}
and
\begin{eqnarray}
{\cal L}_{\rm D^4F^4} & = & \frac{\ \pi^4}{2880} \ g^2 \
{\alpha'}^4 \ t_{(8)}^{\mu_1 \nu_1 \mu_2 \nu_2 \mu_3 \nu_3 \mu_4
\nu_4} \ \biggl[ \ \ \ 9 \tr\biggl( D^2 D^2F_{\mu_1 \nu_1}F_{\mu_2
\nu_2}F_{\mu_3 \nu_3}F_{\mu_4 \nu_4} \biggr) + \nonumber \\
&+&  16  \tr\biggl( D^{\alpha}F_{\mu_1 \nu_1}D_{\alpha}F_{\mu_2
\nu_2}D^{\beta}F_{\mu_3 \nu_3}D_{\beta}F_{\mu_4 \nu_4} \biggr) + 4
\tr\biggl( D^{\alpha}F_{\mu_1 \nu_1}D_{(\beta} D_{\alpha)}F_{\mu_2
\nu_2}D^{\beta}F_{\mu_3 \nu_3}F_{\mu_4
\nu_4} \biggr) + \nonumber \\
&+&  18 \tr\biggl( D^{2}F_{\mu_1 \nu_1}D^{2}F_{\mu_2
\nu_2}F_{\mu_3 \nu_3}F_{\mu_4 \nu_4} \biggr) + 9 \tr\biggl( D^{2}
F_{\mu_1 \nu_1}F_{\mu_2\nu_2}D^{2}F_{\mu_3 \nu_3}
F_{\mu_4 \nu_4} \biggr) + \nonumber \\
&+& 16 \tr\biggl( D^{\alpha}D^{\beta}F_{\mu_1
\nu_1}D_{\alpha}D_{\beta}F_{\mu_2 \nu_2}F_{\mu_3 \nu_3}F_{\mu_4
\nu_4} \biggr) + 34 \tr\biggl( D^{(2} D^{\beta)}F_{\mu_1
\nu_1}D_{\beta}F_{\mu_2 \nu_2}F_{\mu_3 \nu_3}F_{\mu_4
\nu_4} \biggr) + \nonumber \\
&+&  34 \tr\biggl( D^{2}F_{\mu_1 \nu_1}F_{\mu_2
\nu_2}D^{\beta}F_{\mu_3 \nu_3}D_{\beta}F_{\mu_4 \nu_4} \biggr) \ \
\ \biggr] \ . \label{LD4F4}
\end{eqnarray}
At this order the symmetrized prescription mentioned in
(\ref{fsym}) begins to apply: that is why some terms in ${\cal
L}_{\rm D^4F^4}$ contain symmetrized products. Using the explicit
expressions of the $t_{(8)}$ and the $t_{(10)}$ tensors,
integration by parts, the Bianchi and the $[D,D]F=-ig[F,F]$
identities, and dropping out the on-shell terms, the expressions
in (\ref{LD2F5}) and (\ref{LD4F4}) could be made explicit and
reduced to a minimum length,  but we will leave that work to be
done somewhere else and just content ourselves with those
expressions as they are, in order to see how formulas
(\ref{LD2nF5final}) and (\ref{Seff-nonabelian}) have operated at
this order in $\alpha'$. In any case, we have checked that the
abelian limit of (\ref{LD4F4}) agrees
completely with the $\partial^4F^4$ terms of \cite{Andreev1}.\\
\noindent The lagrangian in (\ref{L4}) (with the expressions in
(\ref{LD2F5}) and (\ref{LD4F4})) is not the complete lagrangian at
${\cal O}({\alpha'}^4)$ because the $F^6$ terms should also be
present (that is why we have called it ${\cal L'}^{(4)}$ instead
of ${\cal L}^{(4)}$). A complete lagrangian at this order has been
determined in \cite{Koerber2} by
other method and confirmed in \cite{Sevrin1} and \cite{Nagaoka1}.\\
The procedure to continue writing the $D^{2n}F^5$ terms from
(\ref{LD2nF5final}) and the $D^{2n}F^4$ ones from
(\ref{Seff-nonabelian}) is quite direct using the expansions of
all the $\alpha'$ dependent factors. In the case of the Gamma
factor, the coefficients of the $\alpha'$ expansion are known up
to infinity, while in this work we have expanded the $T$ and
$(2\alpha')^2 K_3$ factors (from which the $P^{(1)}$, $U^{(1)}$,
$W^{(1)}$ and $\Delta$ factors have been obtained) only up to
${\cal O}({\alpha'}^6)$ order. So in this section we could, in
principle, go on with the $D^{2n}F^5$ terms up
to that order, but to save space we will not do it here.\\
In spite of what was mentioned in the previous paragraph, the
structure of the $\alpha'$ expansions of $T$ and $(2\alpha')^2 K_3$
is \underline{completely} known since, as it is proved in detail in
appendix \ref{expansion-K3K}, at each order in $\alpha'$ only cyclic
polynomial invariants (in the $\alpha_{ij}$ variables)
appear\footnote{Further more, the polynomials which appear at each
order in $\alpha'$ should also be invariant under a twisting
transformation, so this diminishes a little the number of them which
appear in the expansion. In appendix \ref{polynomial} we give a
complete list of them up to sixth degree \cite{Medina1}.}. So the
only unknowns are the coefficients which go with each of those
cyclic polynomial invariants: this is a considerably much reduced
number of unknowns than the ones which in principle should be
determined at every $\alpha'$ order (see appendix \ref{polynomial}
for further details).

\section{Summary and final remarks}
\label{final} \noindent We have succeeded in finding the
$D^{2n}F^5$ terms of the open superstring effective lagrangian, to
all order in $\alpha'$. This lagrangian is of importance in the
corresponding sector of the $SO(32)$ Type I theory and in the
description of the low energy interaction of D-branes. To our
knowledge, this is the second $non$-$abelian$ result, reported in
the literature, which is complete (in the sense that all the terms
in the lagrangian and its coefficients are given in a closed form)
and which has been determined to all order in $\alpha'$; the first
one being the determination of the
$D^{2n}F^4$ in \cite{Chandia1}.\\
\noindent This has been a very much more complicated problem than
the determination of the $D^{2n}F^4$ terms, which consisted in a
non-abelian generalization of a previous result in \cite{DeRoo1}.
We have found the general formula and we have given expansions
which allow us to compute explicitly the $D^{2n}F^5$ terms up to
${\cal O}({\alpha'}^6)$ order. We have checked that the $F^5$
terms of our lagrangian, together with the $D^2F^4$ ones found
previously (see eq. (\ref{Seff-nonabelian})), agree with the known
result at ${\cal O}({\alpha'}^3)$ order. At ${\cal
O}({\alpha'}^4)$ order, it would be interesting to see if our
$D^2F^5$ and $D^4F^4$ terms agree with the ones which are sensible
to 4 and 5-point
amplitudes in eqs. (1.5) and (1.6) of \cite{Sevrin1}.\\
\noindent The starting point of all this huge labor has been the
determination of the 5-point subamplitude of massless bosons in
Open Superstring Theory, which we have found in terms of two
kinematical expressions and two $\alpha'$ dependent factors
(see eq. (\ref{A12345final})).\\
\noindent Another important step that we have done, in order to be
able to find all the $D^{2n}F^5$ terms, has been the determination
of the scattering subamplitude $A_{D^{2n}F^5}(1,2,3,4,5)$ as a sum
of terms which have no poles and have manifest cyclic and
(on-shell) gauge invariance, as well as world-sheet parity
symmetry (see eq. (\ref{AD2nF5nopoles})). We have done this by
using the known $t_{(8)}$ and a new $t_{(10)}$ tensor in all the
kinematical
expressions involved in that subamplitude.\\
\noindent The explicit expression of ${\cal L}_{D^{2n}F^5}$
depends directly on the previously found expression of ${\cal
L}_{D^{2n}F^4}$, because the scattering subamplitude
$A_{D^{2n}F^5}(1,2,3,4,5)$ is calculated in terms of
$A_{D^{2n}F^4}(1,2,3,4,5)$ (see eq. (\ref{AD2nF5})). Now, given
that there is not a unique way in choosing the terms of ${\cal
L}_{D^{2n}F^4}$\footnote{Since the only requirement for this
lagrangian is to reproduce superstring 4-point amplitudes, there
is some freedom in the way its covariant derivatives are present
on it.}, and the fact that the $\alpha'$ factors in
(\ref{AD2nF5nopoles}) are not all independent (see, for example,
eq. (\ref{almostthere}) ), it may happen that a final lagrangian
${\cal L}_{D^{2n}F^5}$ could eventually be found in terms of only
the $t_{(10)}$ tensor, as mentioned in \cite{Machado1}. In the
present work we have not look
further in this direction.\\
\noindent The calculations of this work have been extremely long
and, based on the experience we have had in solving them out, we
think it is impossible to have done them without any computer
assistance. In spite of this fact, and contrary to what it is
generally believed, the main result of our paper suggests that it
is indeed possible to take to the level of the effective
lagrangian, to all order in $\alpha'$, the information of (tree
level) superstring scattering amplitudes: at least we have
succeeded on this subject,
up to 5-field terms, in the case of Open Superstring Theory.\\
\noindent On a future paper \cite{Barreiro1} we will use the
results of the present work to determine the tree level
${\alpha'}^4R^5$ terms in the effective lagrangians of the type II
theories, by means of the KLT relations \cite{Kawai1}, and we will
consider the possibility of finding and $all$ $\alpha'$ result for
those actions, as it was done in \cite{Chandia1} in the case of
the 4-Riemann tensor terms.

\section*{Acknowledgements}

R. M. would like to acknowledge useful conversations with F.
Machado, with whom the first steps of this work were given. He
would also like to thank A. Sevrin, N. Berkovits, F. Brandt, P.
Koerber, S. Stieberger and especially D. Grasso and A. Tseytlin,
for useful conversations. This research has been partially
supported by the Brazilian agencies CNPq and FAPEMIG.

\appendix

\section{Conventions and identities}
\label{conventions}

\begin{enumerate}
\item Metric, symmetrization and antisymmetrization over spacetime
indexes:

We use the following convention for the Minkowski metric:
\begin{eqnarray}
\label{metric} \eta_{\mu \nu} = \mbox{diag}(-, +, \ldots , +) \ .
\end{eqnarray}
The symmetrization and antisymmetrization convention that we use,
on the spacetime indexes of a product of two  vectors $A$ and $B$,
is the following:
\begin{eqnarray}
\label{symmetrization} A^{(\mu}B^{\nu)}=\frac{1}{2} (
A^{\mu}B^{\nu}+A^{\nu}B^{\mu}) \ , \\
\label{antisymmetrization} A^{[\mu}B^{\nu]}=\frac{1}{2} (
A^{\mu}B^{\nu}-A^{\nu}B^{\mu} ) \ .
\end{eqnarray}

\item Gauge group generators, field strength and covariant derivative:\\

Gauge fields are matrices in the Lie group internal space, so that
$A_{\mu} = A^{\mu}_{\ a} \lambda^a$, where the $\lambda^a$ are the
generators (in a matrix representation) which satisfy the usual
relation
\begin{eqnarray}
\mbox{tr}(\lambda^a \lambda^b) = \delta^{ab} \ .\label{group1}
\end{eqnarray}
The field strength and the covariant derivative are defined by
\begin{eqnarray}
\label{group2} F_{\mu \nu} & = & \partial_{\mu} A_{\nu} -
\partial_{\nu} A_{\mu}
- ig [ A_{\mu}, A_{\nu}] \ , \\
\label{group3} D_{\mu} \phi & = & \partial_{\mu} \phi - ig [
A_{\mu}, \phi ] \ ,
\end{eqnarray}
and they are related by the identity
\begin{eqnarray}
[D_{\mu}, D_{\nu}] \phi & = & -ig \ [ F_{\mu \nu}, \phi ] \ .
\label{group4}
\end{eqnarray}
Covariant derivatives of field strengths satisfy the Bianchi
identity:
\begin{eqnarray}
\label{Bianchi} D^{\mu} F^{\nu \rho} + D^{\rho} F^{\mu \nu} +
D^{\nu} F^{\rho \mu} = 0 \ .
\end{eqnarray}

\end{enumerate}

\section{Tensors}
\label{tensors}

\subsection{$t_{(8)}$ and $(\eta \cdot t_{(8)})_1$ tensors}
\label{t8tensor}

\noindent The $t_{(8)}$ tensor\footnote{An explicit expression for
it may be found in equation (4.A.21) of \cite{Schwarz1}.},
characteristic of the 4 boson scattering amplitude, is
antisymmetric on each pair $(\mu_j ,\nu_j)$ ($j=1,2,3,4$) and is
symmetric under any exchange of such of pairs.  It satisfies the
identity\footnote{Formula (\ref{t8}) has been taken from appendix
A of \cite{DeRoo1}.}:
\begin{multline}
t^{(8)}_{\mu_1 \nu_1 \mu_2 \nu_2 \mu_3 \nu_3 \mu_4 \nu_4}
A_1^{\mu_1 \nu_1} A_2^{\mu_2 \nu_2} A_3^{\mu_3 \nu_3} A_4^{\mu_4
\nu_4}  = \\
\begin{split}
&  -2 \biggl( \mbox{Tr}(A_1 A_2)\mbox{Tr}(A_3 A_4) + \mbox{Tr}(A_1
A_3)\mbox{Tr}(A_2 A_4) +
\mbox{Tr}(A_1 A_4)\mbox{Tr}(A_2 A_3) \biggr)+ \\
& + 8 \biggl( \mbox{Tr}(A_1 A_2 A_3 A_4) + \mbox{Tr}(A_1 A_3 A_2
A_4) + \mbox{Tr}(A_1 A_3 A_4 A_2)  \biggr) \ ,
\end{split}
\label{t8}
\end{multline}
where the $A_j$ tensors are antisymmetric and where `Tr' means
the trace over the spacetime indexes.\\
\noindent A ten index tensor, which is also antisymmetric on each
pair $(\mu_j ,\nu_j)$, can be constructed from the Minkowski
metric tensor and the $t_{(8)}$ one, as follows:
\begin{eqnarray}
{(\eta \cdot t_{(8)})_1}^{\mu_1 \nu_1 \mu_2 \nu_2 \mu_3 \nu_3
\mu_4 \nu_4 \mu_5 \nu_5} & = & \eta^{\nu_3 \nu_4} t_{(8)}^{\mu_3
\mu_4 \mu_5 \nu_5 \mu_1 \nu_1 \mu_2 \nu_2} + \eta^{\mu_3 \mu_4}
t_{(8)}^{\nu_3 \nu_4 \mu_5 \nu_5 \mu_1 \nu_1 \mu_2 \nu_2} -
\nonumber \\
&&- \ \eta^{\mu_3 \nu_4} t_{(8)}^{\nu_3 \mu_4 \mu_5 \nu_5 \mu_1
\nu_1 \mu_2 \nu_2} - \eta^{\nu_3 \mu_4} t_{(8)}^{\mu_3 \nu_4 \mu_5
\nu_5 \mu_1 \nu_1 \mu_2 \nu_2} \ . \label{etat8}
\end{eqnarray}
This tensor appears in the 5-point amplitude of the open
superstring. It also changes sign under a twisting
transformation\footnote{See the third item of subsection
\ref{properties} for further details about a twisting transformation
on the disk.} with respect to index 1, that is,
\begin{eqnarray}
{(\eta \cdot t_{(8)})_1}^{\mu_1 \nu_1 \mu_5 \nu_5 \mu_4 \nu_4
\mu_3 \nu_3 \mu_2 \nu_2} = - {(\eta \cdot t_{(8)})_1}^{\mu_1 \nu_1
\mu_2 \nu_2 \mu_3 \nu_3 \mu_4 \nu_4 \mu_5 \nu_5} \ .
\label{etat8twisting}
\end{eqnarray}
We have used the subindex $1$ in the $(\eta \cdot t_{(8)})$ tensor
as a reminder of this relation.

\subsection{$t_{(10)}$ tensor}
\label{t10tensor}

The $t_{(10)}$ tensor is another ten index tensor that appears in
the 5-point amplitude of the open superstring. It is linearly
independent to the $(\eta \cdot t_{(8)})_1$ one. It can be
constructed by the following procedure:

\begin{enumerate}
\item Using the Minkowski metric tensor, we make all possible ten
index
tensorial structures:%
\begin{equation*}
\eta ^{\mu _{i}\nu _{n}}\eta ^{\mu _{j}\nu _{o}}\eta ^{\mu _{k}\nu
_{p}}\eta ^{\mu _{l}\nu _{q}}\eta ^{\mu _{m}\nu _{r}},\text{ }\eta
^{\mu _{i}\nu _{n}}\eta ^{\mu _{j}\nu _{o}}\eta ^{\mu _{k}\nu
_{p}}\eta ^{\mu _{l}\mu _{m}}\eta ^{\nu _{r}\nu _{q}}\text{ and
}\eta ^{\mu _{i}\nu _{n}}\eta ^{\mu _{j}\mu _{k}}\eta ^{\nu _{p}\nu
_{o}}\eta ^{\mu _{l}\mu _{m}}\eta ^{\nu _{r}\nu _{q}},
\end{equation*}%
with $i\neq j\neq k\neq l\neq m$ and $n\neq o\neq p\neq q\neq r$.
These
indexes are in the range $1, \ldots, 5$. When applied to the general expression $%
\zeta _{\mu _{1}}^{1}k_{\nu _{1}}^{1}\zeta _{\mu _{2}}^{2}k_{\nu
_{2}}^{2}\zeta _{\mu _{3}}^{3}k_{\nu _{3}}^{3}\zeta _{\mu
_{4}}^{4}k_{\nu _{4}}^{4}\zeta _{\mu _{5}}^{5}k_{\nu _{5}}^{5}$,
many terms will be null due to the physical state condition
$\left( \zeta ^{i}\cdot k^{i}=0\right) $. Despising these terms,
544 structures remain.

\item Making a linear combination of these 544 terms, we demand
antisymmetry on each pair $(\mu _{j},\nu _{j})$ and also a change
of sign under a twisting transformation with respect to index $1$
(the same as in eq. (\ref{etat8twisting}) ). This procedure
results in a ten index tensor with only $14$ free parameters,
which will be our \textit{ansatz} for the $t_{(10)}$ tensor.\\
Unlike the $t_{(8)}$ tensor, the $t_{(10)}$ one \underline{is not
cyclic invariant} in the pair of indexes
$(\mu_{j},\nu_{j})$\footnote{Due to the symmetries of the 4-point
subamplitude, the $t_{(8)}$ tensor is expected to be cyclic
invariant in the pair of indexes $(\mu_{j},\nu_{j})$, but in fact
it happens to be $completely$ symmetric with respect to those
indexes\cite{Schwarz1}.}.

\item Then, the last $14$ parameters are determined by
substituting $t_{(10)}$ in the kinematical expression
$h^{(1)}(\zeta ,k)$, in (\ref{h23}), and  comparing it with the
corresponding kinematical structure in eq. (\ref{AD2nF5nopolesII})
of appendix \ref{derivation}. This comparison is not immediate
since both expressions only agree after using on-shell and
physical state conditions, together with momentum conservation. At
the end we obtain an expression which satisfies an identity
similar to that in (\ref{t8}):
\begin{multline}
t^{(10)}_{\mu_1 \nu_1 \mu_2 \nu_2 \mu_3 \nu_3 \mu_4 \nu_4 \mu_5
\nu_5} A_1^{\mu_1 \nu_1} A_2^{\mu_2 \nu_2} A_3^{\mu_3 \nu_3}
A_4^{\mu_4 \nu_4} A_5^{\mu_5 \nu_5} = \\
\begin{split}
& \hphantom{+ } - 8 \left[  \ \ \mbox{Tr}(A_1 A_2)\mbox{Tr}(A_3
A_4 A_5) + \mbox{Tr}(A_1 A_3)\mbox{Tr}(A_2 A_4 A_5) +
\mbox{Tr}(A_1 A_4)\mbox{Tr}(A_2 A_3 A_5) + \right. \\
& \hphantom{+ \ \ \ 8} + \mbox{Tr}(A_1 A_5)\mbox{Tr}(A_2 A_3 A_4)
+ \mbox{Tr}(A_2 A_3)\mbox{Tr}(A_1 A_4 A_5) +
\mbox{Tr}(A_2 A_4)\mbox{Tr}(A_1 A_3 A_5) + \\
& \hphantom{+ \ \ \ 8} + \mbox{Tr}(A_2 A_5)\mbox{Tr}(A_1 A_3 A_4)
+ \mbox{Tr}(A_3 A_4)\mbox{Tr}(A_1 A_2 A_5) +
\mbox{Tr}(A_3 A_5)\mbox{Tr}(A_1 A_2 A_4) + \\
& \hphantom{+ \ \ \ 8} \left. + \mbox{Tr}(A_4 A_5)\mbox{Tr}(A_1
A_2 A_3) \ \ \right] + \ 48 \ \mbox{Tr}(A_1 A_2 A_3 A_4 A_5) \ +
\\
&+ 16 \left[ \ \ \mbox{Tr}(A_1 A_2 A_3 A_5 A_4) + \mbox{Tr}(A_1
A_2 A_4 A_3 A_5) + \mbox{Tr}(A_1 A_2 A_5 A_3 A_4) + \right. \\
& \hphantom{+ \ 16}+ \mbox{Tr}(A_1 A_2 A_4 A_5 A_3) -
\mbox{Tr}(A_1 A_2 A_5 A_4 A_3) + \mbox{Tr}(A_1 A_3 A_2 A_4 A_5) -
\\
& \hphantom{+ \ 16}- \mbox{Tr}(A_1 A_3 A_2 A_5 A_4) +
\mbox{Tr}(A_1 A_4 A_2 A_3 A_5) + \mbox{Tr}(A_1 A_5 A_2 A_3 A_4) -
\\
& \hphantom{+ \ 16} \left. - \mbox{Tr}(A_1 A_4 A_2 A_5 A_3) -
\mbox{Tr}(A_1 A_5 A_2 A_4 A_3) \ \ \right] \ ,
\end{split}
\label{t10}
\end{multline}
where the $A_j$ fields are antisymmetric. From (\ref{t10}) an
explicit expression of the $t_{(10)}$ tensor may be obtained, once
its symmetry properties are considered.
\end{enumerate}

\section{$\alpha'$ dependent factors}
\label{expansions}

\subsection{Gamma factor}
\label{expansion-Gamma} As remarked in \cite{DeRoo2}, using the
Taylor expansion for $\mbox{ln} \ \Gamma(1+z)$ \footnote{See
formula (10.44c) of \cite{Arfken}, for example.},
\begin{eqnarray}
\mbox{ln} \ \Gamma(1+z) = -\gamma z + \sum_{k=2}^{\infty} (-1)^k
\frac{\zeta(k)}{k} z^k \ \ \ \ \ \ \ (-1 < z \leq 1) \ ,
\label{Taylor}
\end{eqnarray}
it may be proved that the explicit $\alpha'$ expansion for the
Gamma factor in eq. (\ref{A1234}) is given by
\begin{eqnarray}
{\alpha'}^2 \frac{\Gamma(- \alpha' s) \Gamma(- \alpha' t)}
{\Gamma(1- \alpha' s - \alpha' t)} = \frac{1}{st} \cdot
\mbox{exp} \left\{ \sum_{k=2}^{\infty} \frac{\zeta(k)}{k}
{\alpha'}^k (s^k+t^k-(s+t)^k) \right \}   \ . \label{formula1}
\end{eqnarray}
\noindent Up to ${\cal O}({\alpha'}^6)$ terms this gives:
\begin{eqnarray}
{\alpha'}^2 \frac{\Gamma(- \alpha' s) \Gamma(- \alpha' t)}
{\Gamma(1- \alpha' s - \alpha' t)} & = & \frac{1}{st} -\frac{\
\pi^2}{6} {\alpha'}^2 - \zeta(3) (s+t) \ {\alpha'}^3  -
\frac{ \ \pi^4}{360}(4 s^2 + st + 4 t^2) \ {\alpha'}^4 \nonumber \\
 & & + \left[ \frac{\ \pi^2}{6} \zeta(3) \ st (s+t) - \zeta(5)
(s^3 + 2 s^2 t + 2 s t^2+ t^3)  \right]  {\alpha'}^5 \nonumber \\
 & & + \left[ \frac{}{} \frac{1}{2} \zeta(3)^2 st(s+t)^2 - \frac{\
 \pi^6}{15120}(16 s^4 + 12 s^3 t + 23 s^2 t^2 + 12 s t^3 + 16 t^4)
 \right] {\alpha'}^6
\nonumber \\
 & &   + \ {\cal O}({\alpha'}^7) \ .
\label{expansionGamma}
\end{eqnarray}
The function $f(s,t)$, defined in (\ref{f}), consists in this
Gamma factor (divided by $\alpha'^2$) with the pole subtracted, so
it has a well defined power series:
\begin{eqnarray}
f(s,t) = \sum_{m=0}^{\infty} \sum_{n=0}^{\infty} a_{mn} \ s^m t^n
\ {\alpha'}^{m+n} \ ,
\label{fpower}
\end{eqnarray}
where $a_{mn}=a_{nm}$, with $a_{00}= -\pi^2/6$,
$a_{01}=-\zeta(3)$, $a_{11}=-\pi^4/360$, $a_{02}=-\pi^4/90$, etc.

\subsection{$K_2$, $K_3$ and $T$ factors}
\label{expansion-K3K}

In this section we briefly review the definitions of the factors
$K_2$ and $K_3$, given in appendix A.1 of \cite{Brandt1} and we deal
with the factor $T$ introduced in eq. (\ref{K2-K}) of this work. We
study the twisting and the cyclic symmetry and see how they are
present in the $\alpha'$ expansion of $K_3$ and $T$.

\subsubsection{The definitions and the $\alpha_{ij}$ variables}
\label{K2K3Kdefinitions}

The factors $K_2$ and $K_3$ were defined in \cite{Brandt1} as
\begin{eqnarray}
\label{K2-int}
K_2 & = & \int_0^1 d x_3 \int_0^{x_3} d x_2 \
x_3^{2 \alpha' \alpha_{13}} (1-x_3)^{2 \alpha'
  \alpha_{34}-1} {x_2}^{2 \alpha' \alpha_{12}-1} (1-x_2)^{2 \alpha'
  \alpha_{24}} (x_3-x_2)^{2 \alpha' \alpha_{23}} \ , \ \ \ \ \ \ \ \ \\
\label{K3-int}
K_3 & = & \int_0^1 d x_3 \int_0^{x_3} d x_2 \
x_3^{2 \alpha' \alpha_{13}-1} (1-x_3)^{2 \alpha'
  \alpha_{34}} {x_2}^{2 \alpha' \alpha_{12}} (1-x_2)^{2 \alpha'
  \alpha_{24}-1} (x_3-x_2)^{2 \alpha' \alpha_{23}} \ .
\end{eqnarray}

\noindent After making the substitution $x_2 = u \cdot x_3$ in the
inner integral (keeping $x_3$ constant) they become
\begin{eqnarray}
\label{K2-int2} K_2 & = & \int_0^1 d x_3 \int_0^{1} d u \ x_3^{2
\alpha' \rho} (1-x_3)^{2 \alpha'
  \alpha_{34}-1} {u}^{2 \alpha' \alpha_{12}-1} (1-u x_3)^{2 \alpha'
  \alpha_{24}} (1-u)^{2 \alpha' \alpha_{23}} \ , \\
\label{K3-int2}
K_3 & = & \int_0^1 d x_3 \int_0^{1} d u \ x_3^{2
\alpha' \rho} (1-x_3)^{2 \alpha'
  \alpha_{34}} {u}^{2 \alpha' \alpha_{12}} (1-u x_3)^{2 \alpha'
  \alpha_{24}-1} (1-u)^{2 \alpha' \alpha_{23}} \ .
\end{eqnarray}
Here,
\begin{eqnarray}
\rho = \alpha_{12} + \alpha_{13} + \alpha_{23} \ .
\label{rho}
\end{eqnarray}
As was seen in \cite{Kitazawa1}, these factors may be written in
terms of Euler Beta functions and a generalized Hypergeometric
function, as
\begin{eqnarray}
K_2 & = & B(2 \alpha' \alpha_{12}, 1 + 2 \alpha' \alpha_{23})
\cdot B(2 \alpha' \alpha_{34}, 1 + 2 \alpha' \rho) \cdot \nonumber
\\
& & \cdot \ {}_3F_2(1 + 2 \alpha' \rho, 2 \alpha' \alpha_{12}, -2
\alpha' \alpha_{24}; \ 1 + 2 \alpha' \rho + 2 \alpha' \alpha_{34},
1 + 2 \alpha' \alpha_{12} + 2 \alpha' \alpha_{23}; \ 1) \ ,
\label{K2hyperg} \\
K_3 & = & B(1+2 \alpha' \alpha_{12}, 1 + 2 \alpha' \alpha_{23})
\cdot B(1 + 2 \alpha' \alpha_{34}, 1 + 2 \alpha' \rho) \cdot
\nonumber
\\ & & \cdot \ {}_3F_2(1 + 2 \alpha' \rho,1+ 2 \alpha' \alpha_{12},1
-2 \alpha' \alpha_{24}; \ 2 + 2 \alpha' \rho + 2 \alpha'
\alpha_{34}, 2 + 2 \alpha' \alpha_{12} + 2 \alpha' \alpha_{23}; \
1) \ . \ \ \ \ \ \ \label{K3hyperg}
\end{eqnarray}
\noindent We will not use (\ref{K2hyperg}) and (\ref{K3hyperg}) to
find any $\alpha'$ expansion in the present work. In
\cite{Brandt1}, using (\ref{K2-int2}) and (\ref{K3-int2}), the
$\alpha'$ expansions of $K_2$ and $K_3$ were found up to ${\cal
O}(\alpha')$ terms. The double integrals that appear at each power
of $\alpha'$ were calculated using Harmonic Polylogarithms
\cite{Remiddi1}.

\noindent It is important to note that the ten $\alpha_{ij}$
variables defined in (\ref{alphaij}) are not all independent once
the on-shell (${k_i}^2=0$) and the momentum conservation
conditions are taken into account. In fact, only five of them are
independent. When finding the $\alpha'$ expansions of the factors
$K_2$ and $K_3$ we have chosen $\alpha_{12}$, $\alpha_{23}$,
$\alpha_{34}$, $\alpha_{45}$ and $\alpha_{51}$, as independent
variables. The rest of them are given by
\begin{eqnarray}
\label{alpha13}
\alpha_{13} = \alpha_{45} - \alpha_{12} - \alpha_{23} \\
\label{alpha14}
\alpha_{14} = \alpha_{23} - \alpha_{51} - \alpha_{45} \\
\label{alpha24}
\alpha_{24} = \alpha_{51} - \alpha_{23} - \alpha_{34} \\
\label{alpha25}
\alpha_{25} = \alpha_{34} - \alpha_{12} - \alpha_{51} \\
\label{alpha35}
\alpha_{35} = \alpha_{12} - \alpha_{45}
-\alpha_{34}
\end{eqnarray}
Notice that considering (\ref{alpha13}), the $\rho$ variable in
(\ref{rho}) coincides with $\alpha_{45}$.\\

 The $T$ factor has
already been defined in eq. (\ref{K2-K}) as
\begin{eqnarray*}
T = (2 \alpha')^2 \biggl[ \ \alpha_{12} \ \alpha_{34} \ K_2 + (
\alpha_{51} \ \alpha_{12} - \alpha_{12} \ \alpha_{34} +
\alpha_{34} \ \alpha_{45} ) \ K_3 \ \biggr].
\end{eqnarray*}
$T$ and $K_3$ are related to the Gamma factor of the 4-point
amplitude by
\begin{eqnarray}
(2 \alpha')^2 \frac{\Gamma(2 \alpha' \alpha_{34}) \
 \Gamma(2 \alpha' \alpha_{45})}{\Gamma(1+ 2 \alpha' \alpha_{34} + 2 \alpha'
 \alpha_{45})} \ = \ \frac{1}{\alpha_{34}
 \alpha_{45}} \
 T\biggl|_{\alpha_{12}=0} \ - \ (2 \alpha')^2 K_3\biggl|_{\alpha_{12}=0} \ .
\label{relate3}
\end{eqnarray}
To see this we notice that using the definition of $T$, already
mentioned, the relation in (\ref{relate3}) is equivalent to
\begin{eqnarray}
\{ \alpha_{12} \ K_2 \}\biggl|_{\alpha_{12}=0} = \alpha_{45}
\frac{\Gamma(2 \alpha' \alpha_{34})
 \Gamma(2 \alpha' \alpha_{45})}{\Gamma(1+ 2 \alpha' \alpha_{34} + 2 \alpha'
 \alpha_{45})} \ ,
\label{relate32}
\end{eqnarray}
which can be proved using (\ref{K2hyperg}) and the fact that the
Hypergeometric function appearing there becomes $1$ when
$\alpha_{12}=0$.\\
\noindent The following relations (and the analog ones, obtained
by cyclic permutations of them) can also be proved:
\begin{eqnarray}
\label{K3Gamma}
(2 \alpha')^2
K_3\biggl|_{\alpha_{12}=\alpha_{23}=0} & = & - (2 \alpha')^2
\frac{ \ \alpha_{34}f(-2 \alpha_{34}, -2 \alpha_{45})-
\alpha_{51}f(-2 \alpha_{45}, -2 \alpha_{51}) \
}{\alpha_{34}-\alpha_{51}} \ ,
\\
\label{TGamma}
T\biggl|_{\alpha_{12}=\alpha_{23}=0} \ - \ 1& = & -
(2 \alpha')^2 \alpha_{34}\alpha_{45}\alpha_{51} \frac{ f(-2
\alpha_{34}, -2 \alpha_{45})-f(-2 \alpha_{45}, -2 \alpha_{51}) \
}{\alpha_{34}-\alpha_{51}} \ ,
\\
\label{T0}
T\biggl|_{\alpha_{12}=\alpha_{34}=0} \ - \ 1& = & 0 \ .
\end{eqnarray}

\subsubsection{Twisting symmetry}
\label{K2K3Ktwisting}

The twisting transformation was already commented (in the third item
of subsection \ref{properties}) to be equivalent to a parity
transformation in the string world-sheet. In this section we prove
that $K_2$, $K_3$ and $T$ are invariant under a twisting
transformation with respect to the fifth insertion on the disk, that
is, they are invariant under the transformation of indexes given in
(\ref{twisting5}). For this purpose we use that
\begin{eqnarray}
\int_0^1 d x_3 \int_0^{x_3} d x_2 \ g(x_2, x_3) = \int_0^1 d x_3
\int_0^{x_3} d x_2 \ g(1-x_3, 1-x_2) \ . \label{int-twisting}
\end{eqnarray}
This can be easily proved by first changing the order of
integration on the left integral and then making the substitution
\begin{eqnarray}
\left\{ \begin{array}{ccc}
       x_2 & = & 1 - {x_3}'\\
       x_3 & = & 1 - {x_2}'
       \end{array}
       \right.
\label{substitution1}
\end{eqnarray}
Applying the result in (\ref{int-twisting}) to the integrals in
(\ref{K2-int}) and (\ref{K3-int}) we have the desired relation
\begin{eqnarray}
K_i(\alpha_{12}; \alpha_{13}; \alpha_{23}; \alpha_{24};
\alpha_{34}) = K_i(\alpha_{34}; \alpha_{24}; \alpha_{23};
\alpha_{13}; \alpha_{12}) \ \ \ (i=2,3) \ . \label{Ki}
\end{eqnarray}

\noindent Once $K_2$ and $K_3$ are invariant under the twisting
transformation in (\ref{twisting5}), from its definition in
(\ref{K2-K}), it is immediate that $T$ is also invariant under the
same transformation.

\subsubsection{Cyclic symmetry}
\label{K3Kcyclic}

\noindent It is easy to see that $K_2$ is not cyclic invariant.
For this purpose it is enough to look at the first term in the
$\alpha'$ expansion of it, in (\ref{K2}), which clearly does not
respect the cyclic invariance. In this section we will prove that
$K_3$ and $T$ remain invariant under the following cyclic
permutation of indexes:
\begin{eqnarray}
(1,2,3,4,5) \rightarrow (2,3,4,5,1) \ .
\label{permutation}
\end{eqnarray}
So we need to prove that
\begin{eqnarray}
\label{K3cyclic}
K_3(\alpha_{12}; \alpha_{13}; \alpha_{23};
\alpha_{24}; \alpha_{34}) & = & K_3(\alpha_{23}; \alpha_{24};
\alpha_{34};
\alpha_{35}; \alpha_{45}) \ , \\
\label{Kcyclic} T(\alpha_{12}; \alpha_{13}; \alpha_{23};
\alpha_{24}; \alpha_{34}) & = & T(\alpha_{23}; \alpha_{24};
\alpha_{34}; \alpha_{35}; \alpha_{45}) \ .
\end{eqnarray}

\begin{itemize}
\item \underline{Cyclic invariance of $K_3$}:\\

\noindent Making the substitution
\begin{eqnarray}
\left\{ \begin{array}{ccc}
       x_2 & = & 1 - {x_3}'\\
       x_3 & = & (1 - {x_3}')/(1-{x_2}')
       \end{array}
       \right.
\label{substitution2}
\end{eqnarray}
in the integral expression of $K_3$, in (\ref{K3-int}), it is not
difficult to arrive to
\begin{eqnarray}
\label{K3cyclicproof} K_3( \alpha_{12};  \alpha_{13};
\alpha_{23};  \alpha_{24};  \alpha_{34}) & = & K_3( \alpha_{23};
\alpha_{24};  \alpha_{34};  -\alpha_{13}-\alpha_{23}-\alpha_{34};
 \alpha_{12}+\alpha_{13}+\alpha_{23}) \ . \ \ \ \ \ \
\end{eqnarray}
Now, after considering the expressions for $\alpha_{13}$ and
$\alpha_{35}$, given in (\ref{alpha13}) and (\ref{alpha35}), this
last relation becomes the one in (\ref{K3cyclic}).

\item \underline{Cyclic invariance of $T$}:\\

The proof of the cyclic invariance in this case is very much more
involved. We begin noticing that the desired condition
(\ref{Kcyclic}) is equivalent to
\begin{eqnarray}
\alpha_{23} \ \alpha_{45} \ K_2(\alpha_{23}; \alpha_{24};
\alpha_{34};
\alpha_{35}; \alpha_{45}) \ & = & \  \alpha_{12} \ \alpha_{34} \
K_2(\alpha_{12}; \alpha_{13};
\alpha_{23};
\alpha_{24}; \alpha_{34}) \ - \nonumber \\
& & - \ (\alpha_{13}+ \alpha_{23}) \ \alpha_{24} \ K_3(\alpha_{12}; \alpha_{13};
\alpha_{23};
\alpha_{24}; \alpha_{34}) \ , \ \ \ \ \ \ \
\label{aux-relation}
\end{eqnarray}
once the definition of $T(\alpha_{12}; \alpha_{13}; \alpha_{23};
\alpha_{24}; \alpha_{34})$, given in (\ref{K2-K}), together with
the
relations for the $\alpha_{ij}$, given in (\ref{alpha13})-(\ref{alpha35}),
have been considered.\\

\noindent We will prove (\ref{aux-relation}) in five steps:
\begin{enumerate}
\item Using the definition (\ref{K2-int}) we have that
\begin{eqnarray}
K_2(\alpha_{23}; \alpha_{24};
\alpha_{34};
\alpha_{35}; \alpha_{45}) =
\hspace{9.0cm}
\nonumber \\
\int_0^1 d {x_3}' \int_0^{{x_3}'} d {x_2}' \
{x_3}'^{2 \alpha' \alpha_{24}} (1-{x_3}')^{2 \alpha'
  \alpha_{45}-1} {x_2}'^{2 \alpha' \alpha_{23}-1} (1-{x_2}')^{2 \alpha'
  \alpha_{35}} ({x_3}'-{x_2}')^{2 \alpha' \alpha_{34}} \ ,
\nonumber \\
\label{K2prime}
\end{eqnarray}
and making the substitution (\ref{substitution2}) for
$K_2(\alpha_{12}; \alpha_{13}; \alpha_{23};
\alpha_{24}; \alpha_{34})$ and \\
$K_3(\alpha_{12}; \alpha_{13}; \alpha_{23};
\alpha_{24}; \alpha_{34})$, given in (\ref{K2-int}) and (\ref{K3-int}),
we have that
\begin{eqnarray}
\label{K2transformed}
K_2(\alpha_{12}; \alpha_{13}; \alpha_{23};
\alpha_{24}; \alpha_{34}) \ =
\hspace{9.0cm}
\nonumber \\
\int_0^1 d {x_3}' \int_0^{{x_3}'} d {x_2}' \
{x_3}'^{2 \alpha' \alpha_{24}} (1-{x_3}')^{2 \alpha'
  \alpha_{45}} {{x_2}'}^{2 \alpha' \alpha_{23}} (1-{x_2}')^{2 \alpha'
  \alpha_{35}-1} ({x_3}'-{x_2}')^{2 \alpha' \alpha_{34}-1} \ ,
\nonumber \\
\\
K_3(\alpha_{12}; \alpha_{13}; \alpha_{23};
\alpha_{24}; \alpha_{34}) \ =
\hspace{9.0cm}
\nonumber \\
\int_0^1 d {x_3}' \int_0^{{x_3}'} d {x_2}' \
{x_3}'^{2 \alpha' \alpha_{24}-1} (1-{x_3}')^{2 \alpha'
  \alpha_{45}} {x_2}'^{2 \alpha' \alpha_{23}} (1-{x_2}')^{2 \alpha'
  \alpha_{35}-1} ({x_3}'-{x_2}')^{2 \alpha' \alpha_{34}} \ .
\nonumber \\
\label{K3transformed}
\end{eqnarray}

\item Doing integration by parts in (\ref{K2prime}), with respect
to the ${x_2}'$ variable, it may be proved that
\begin{multline}
K_2(\alpha_{23}; \alpha_{24};
\alpha_{34};
\alpha_{35}; \alpha_{45})  = \\
\begin{split}
\frac{\alpha_{35}}{\alpha_{23}}
M(\alpha_{23}; \alpha_{24};
\alpha_{34};
\alpha_{35}; \alpha_{45})
+\frac{\alpha_{34}}{\alpha_{23}}
N(\alpha_{23}; \alpha_{24};
\alpha_{34};
\alpha_{35}; \alpha_{45}) \ ,
\end{split}
\label{int-parts-1}
\end{multline}
where
\begin{eqnarray}
M(\alpha_{23}; \alpha_{24}; \alpha_{34};
\alpha_{35}; \alpha_{45}) \ =
\hspace{9.0cm}
\nonumber \\
\int_0^1 d {x_3}' \int_0^{{x_3}'} d {x_2}' \
{x_3}'^{2 \alpha' \alpha_{24}} (1-{x_3}')^{2 \alpha'
  \alpha_{45}-1} {{x_2}'}^{2 \alpha' \alpha_{23}} (1-{x_2}')^{2 \alpha'
  \alpha_{35}-1} ({x_3}'-{x_2}')^{2 \alpha' \alpha_{34}} \ ,
\nonumber \\
\label{M}
\\
N(\alpha_{23}; \alpha_{24}; \alpha_{34};
\alpha_{35}; \alpha_{45}) \ =
\hspace{9.0cm}
\nonumber \\
\int_0^1 d {x_3}' \int_0^{{x_3}'} d {x_2}' \
{x_3}'^{2 \alpha' \alpha_{24}} (1-{x_3}')^{2 \alpha'
  \alpha_{45}-1} {x_2}'^{2 \alpha' \alpha_{23}} (1-{x_2}')^{2 \alpha'
  \alpha_{35}} ({x_3}'-{x_2}')^{2 \alpha' \alpha_{34}-1} \ .
\nonumber \\
\label{N}
\end{eqnarray}

\item Now, doing integration by parts in (\ref{K3transformed}), with
respect to the ${x_3}'$ variable, it may also be proved that
\begin{multline}
K_3(\alpha_{12}; \alpha_{13};
\alpha_{23};
\alpha_{24}; \alpha_{34})  = \\
\begin{split}
\frac{\alpha_{45}}{\alpha_{24}}
M(\alpha_{23}; \alpha_{24};
\alpha_{34};
\alpha_{35}; \alpha_{45})
+\frac{\alpha_{34}}{\alpha_{24}}
K_2(\alpha_{12}; \alpha_{13};
\alpha_{23};
\alpha_{24}; \alpha_{34}) \ .
\end{split}
\label{int-parts-2}
\end{multline}

\item Noticing that
\begin{eqnarray}
\frac{1}{(1-{x_3}')({x_3}'-{x_2}')} =
\frac{1}{(1-{x_2}')(1-{x_3}')} +
\frac{1}{(1-{x_2}')({x_3}'-{x_2}')} \ ,
\label{partial-fractions}
\end{eqnarray}
and using (\ref{K2transformed}), (\ref{M}) and (\ref{N}), we have that
\begin{eqnarray}
N(\alpha_{23}; \alpha_{24};
\alpha_{34};
\alpha_{35}; \alpha_{45}) =
M(\alpha_{23}; \alpha_{24};
\alpha_{34};
\alpha_{35}; \alpha_{45}) +
K_2(\alpha_{12}; \alpha_{13};
\alpha_{23};
\alpha_{24}; \alpha_{34}) \ .
\nonumber \\
\label{NMK2}
\end{eqnarray}

\item Finally, eliminating $M(\alpha_{23}; \alpha_{24};
\alpha_{34};
\alpha_{35}; \alpha_{45})$ and $N(\alpha_{23}; \alpha_{24};
\alpha_{34};
\alpha_{35}; \alpha_{45})$ from equations (\ref{int-parts-1}),
(\ref{int-parts-2}) and (\ref{NMK2}), leads precisely to (\ref{aux-relation}).
\end{enumerate}
\end{itemize}

\noindent Now that the proof of (\ref{K3cyclic}) and
(\ref{Kcyclic}) has been done, then it may be repeated for the
rest of the cyclic permutations, guaranteeing that $K_3$ and $T$
remain invariant under them. We have also checked this.

\subsubsection{The $\alpha'$ expansions of $(2 \alpha')^2K_3$ and $T$}
\label{K3Kexpansions}

A detailed calculation of the $\alpha'$ expansions of $K_2$ and
$K_3$ was done in \cite{Medina1}, up to ${\cal O}((2 \alpha')^4)$
terms in both cases. This was done using Harmonic Polylogarithms
\cite{Remiddi1} and the $harmpol$ package of FORM \cite{Vermaseren1}
(see appendix A.3. of \cite{Brandt1} for more details about the type
of calculations involved). The $\alpha'$ expansion of $T$ was also
obtained in \cite{Medina1}, up to ${\cal O}((2 \alpha')^6)$ terms,
by directly using its definition (\ref{K2-K}) and the expansions of
$K_2$ and $K_3$. In this section we will only write the $\alpha'$
expansions of $(2 \alpha')^2 K_3$ and $T$.

\noindent It is not very difficult to prove that the $\alpha'$
expansions of $(2 \alpha')^2 K_3$ and $T$ are power series, that
is, they do not have poles in $\alpha'$ (as $K_2$
does\footnote{See eq. (\ref{K2}).}). So, at every order in
$\alpha'$, there only appear polynomial expressions which, after
using the relations (\ref{alpha13})-(\ref{alpha35}), may be
written in terms of the five $\alpha_{ij}$ variables mentioned in
section \ref{K2K3Kdefinitions}. Each of these polynomial
expressions respects the $twisting$ and the $cyclic$ symmetry,
that were seen to be satisfied by $(2 \alpha')^2 K_3$ and $T$ in
sections \ref{K2K3Ktwisting} and \ref{K3Kcyclic}. The result
obtained, written in terms of polynomial cyclic invariants (which
will be specified in the next subsection), is the following
\cite{Medina1}:
\begin{eqnarray}
(2 \alpha')^2 K_3 & = & (2 \alpha')^2 \ \zeta(2) \ I_1^{(0)} - \
(2 \alpha')^3 \ \zeta(3) \ I_1^{(1)} \ + \ (2 \alpha')^4 \
\frac{2}{5} \zeta(2)^2 \ ( \ I_1^{(2)} +
\frac{1}{4}I_2^{(2)} + I_3^{(2)} \ ) + \nonumber \\
 & + &  (2 \alpha')^5 \ \biggl[ \frac{}{} -\zeta(5) \
 ( \ I_1^{(3)}+I_3^{(3)}+I_4^{(3)} \ ) + ( \ -2
 \zeta(5)+\zeta(2)\zeta(3) \ ) \ ( \ I_2^{(3)}+I_5^{(3)}+I_6^{(3)} \ )+
 \nonumber \\
 & &\hphantom{(2 \alpha')^5 [}  + ( \ \frac{7}{2}
 \zeta(5)-2\zeta(2)\zeta(3) \ ) \ I_7^{(3)} \frac{}{} \biggr] +
 \nonumber \\
 & + &  (2 \alpha')^6 \ \biggl[ \frac{}{} \frac{8}{35}\zeta(2)^3 \
 ( \ I_1^{(4)}+I_3^{(4)}+I_4^{(4)}+I_7^{(4)} \ )
 +\nonumber \\
  &  & \hphantom{(2 \alpha')^6 \ \biggl[}
 ( \ \frac{6}{35}\zeta(2)^3-\frac{1}{2}\zeta(3)^2 \ ) \
 ( \ I_2^{(4)}+I_5^{(4)}+I_8^{(4)}-I_9^{(4)}
 -I_{12}^{(4)}+I_{13}^{(4)} \ )+
 \nonumber \\
 & &\hphantom{(2 \alpha')^5 [}  + ( \ \frac{23}{70}
 \zeta(2)^3-\zeta(3)^2 \ )( \ I_6^{(4)}+ I_{10}^{(4)} \ ) + ( \ -\frac{26}{105}
 \zeta(2)^3+\zeta(3)^2 \ ) \ I_{11}^{(4)} + \nonumber \\
 & &\hphantom{(2 \alpha')^5 [}  + ( \ -\frac{67}{105}
 \zeta(2)^3+2 \zeta(3)^2 \ ) \ I_{14}^{(4)} \frac{}{} \biggr] \ + \ {\cal O}( \ (2
 \alpha')^7 \ ) \ ,
 \label{K3final} \\
T & = & I_1^{(0)} \  - \ (2 \alpha')^3 \ \zeta(3) \ I_6^{(3)} \ +
\ (2 \alpha')^4 \ \frac{2}{5} \zeta(2)^2 \ ( \ I_8^{(4)} +
\frac{1}{4}I_{10}^{(4)} +
I_{13}^{(4)} + I_{14}^{(4)} \ ) + \nonumber \\
 & + &  (2 \alpha')^5 \ \biggl[ \frac{}{} -\zeta(5) \ ( \ I_{10}^{(5)}+I_{15}^{(5)}
 +I_{19}^{(5)}+I_{22}^{(5)}+I_{25}^{(5)} \ ) + \nonumber \\
  & & \hphantom{(2 \alpha')^5 [} + ( \ -2
 \zeta(5)+\zeta(2)\zeta(3) \ ) \ ( \ I_{12}^{(5)}+I_{16}^{(5)}+I_{18}^{(5)} \ )+
 \nonumber \\
 & &\hphantom{(2 \alpha')^5 [}  + ( \ \frac{7}{2}
 \zeta(5)-2\zeta(2)\zeta(3) \ ) \ ( \ I_{23}^{(5)}+I_{24}^{(5)} \ ) \frac{}{} \biggr] +
 \nonumber \\
 & + &  (2 \alpha')^6 \ \biggl[ \frac{}{} \frac{8}{35} \zeta(2)^3 \
 ( \ I_{10}^{(6)}+I_{15}^{(6)}
 +I_{21}^{(6)}+I_{26}^{(6)}+I_{30}^{(6)}+I_{33}^{(6)}+I_{41}^{(6)} \ )
 +\nonumber \\
  &  & \hphantom{(2 \alpha')^6 \ \biggl[}
 ( \ \frac{6}{35}\zeta(2)^3-\frac{1}{2}\zeta(3)^2 \ ) \
 ( \ I_{12}^{(6)}+I_{18}^{(6)}
 +I_{29}^{(6)}-I_{31}^{(6)}-I_{32}^{(6)}+I_{34}^{(6)}-I_{36}^{(6)}-I_{38}^{(6)}
  \ )+
 \nonumber \\
 & &\hphantom{(2 \alpha')^5 [}  + ( \ \frac{23}{70}
 \zeta(2)^3-\zeta(3)^2 \ ) \ (   \ I_{20}^{(6)}+I_{27}^{(6)})
 + \nonumber \\
  & &\hphantom{(2 \alpha')^5 [} + ( \ -\frac{26}{105}
 \zeta(2)^3+ \zeta(3)^2 \ ) \ ( \ I_{39}^{(6)} + I_{40}^{(6)} \ ) +
 ( \ -\frac{67}{105}
 \zeta(2)^3+2 \zeta(3)^2 \ ) \ I_{37}^{(6)} \frac{}{} + \nonumber \\
  & &\hphantom{(2 \alpha')^5 [} + ( \ -\frac{109}{210} \zeta(2)^3
-\frac{3}{2} \zeta(3)^2 \ ) \ I_{42}^{(6)} \biggr] \ + \ {\cal O}(
\ (2 \alpha')^7 \ ) \ .
 \label{Kfinal}
\end{eqnarray}
Here $I_{j}^{(i)}$ denotes the $j$-th polynomial cyclic invariant of
degree $i$.

\subsubsection{Polynomial cyclic invariants}
\label{polynomial}

\noindent The polynomial cyclic invariants are uniquely determined
up to a global factor which we have chosen to be $1$. In the next
lines we list them up to sixth degree. At degrees $0$, $1$, $2$,
$3$, $4$, $5$ and $6$ there are, respectively, one, one, three,
seven, fourteen, twenty six and forty two linearly independent
polynomial cyclic invariants.
\begin{itemize}
\item Degree 0:
\begin{eqnarray}
I_{1}^{(0)} = 1 \ .
\label{I10}
\end{eqnarray}
\item Degree 1:
\begin{eqnarray}
I_{1}^{(1)} = \alpha_{12} + \alpha_{23} + \alpha_{34} +
\alpha_{45} + \alpha_{51} \ .
\label{I11}
\end{eqnarray}
\item Degree 2:
\begin{eqnarray}
I_{1}^{(2)} & = & \alpha_{12}^2 + \alpha_{23}^2 + \alpha_{34}^2 +
\alpha_{45}^2 + \alpha_{51}^2 \ , \nonumber \\
I_{2}^{(2)} & = & \alpha_{12}\alpha_{23} + \alpha_{23}\alpha_{34}
+ \alpha_{34}\alpha_{45}  + \alpha_{45}\alpha_{51}  +
\alpha_{51}\alpha_{12}  \ , \label{Icyclic2} \\
I_{3}^{(2)} & = &
\alpha_{12}\alpha_{34} + \alpha_{23}\alpha_{45} +
\alpha_{34}\alpha_{51}  + \alpha_{45}\alpha_{12}  +
\alpha_{51}\alpha_{23}  \ . \nonumber
\end{eqnarray}
\item Degree 3:
\begin{eqnarray}
I_{1}^{(3)} & = & \alpha_{12}^3 + \alpha_{23}^3 + \alpha_{34}^3 +
\alpha_{45}^3 + \alpha_{51}^3 \ , \nonumber \\
I_{2}^{(3)} & = & \alpha_{12}^2 \alpha_{23} + \alpha_{23}^2
\alpha_{34}  + \alpha_{34}^2 \alpha_{45}  + \alpha_{45}^2
\alpha_{51}  + \alpha_{51}^2 \alpha_{12}  \ , \nonumber \\
I_{3}^{(3)} & = & \alpha_{12}^2 \alpha_{34} + \alpha_{23}^2
\alpha_{45}  + \alpha_{34}^2 \alpha_{51}  + \alpha_{45}^2
\alpha_{12}  + \alpha_{51}^2\alpha_{23}  \ , \nonumber \\
I_{4}^{(3)} & = & \alpha_{12}^2 \alpha_{45} + \alpha_{23}^2
\alpha_{51}  + \alpha_{34}^2 \alpha_{12}  + \alpha_{45}^2
\alpha_{23}  + \alpha_{51}^2\alpha_{34}  \ , \label{Icyclic3} \\
I_{5}^{(3)} & = & \alpha_{12}^2 \alpha_{51} + \alpha_{23}^2
\alpha_{12}  + \alpha_{34}^2 \alpha_{23}  + \alpha_{45}^2
\alpha_{34}  + \alpha_{51}^2\alpha_{45}  \ , \nonumber \\
I_{6}^{(3)} & = & \alpha_{12}\alpha_{23}\alpha_{34} +
\alpha_{23}\alpha_{34}\alpha_{45}   +
\alpha_{34}\alpha_{45}\alpha_{51}  +
\alpha_{45}\alpha_{51}\alpha_{12}  +
\alpha_{51}\alpha_{12}\alpha_{23}   \ , \nonumber \\
I_{7}^{(3)} & = & \alpha_{12}\alpha_{34}\alpha_{45} +
\alpha_{23}\alpha_{45}\alpha_{51}   +
\alpha_{34}\alpha_{51}\alpha_{12}  +
\alpha_{45}\alpha_{12}\alpha_{23}  +
\alpha_{51}\alpha_{23}\alpha_{34}   \ . \nonumber
\end{eqnarray}
\item Degree 4:
\begin{eqnarray}
I_{1}^{(4)} & = & \alpha_{12}^4 + \alpha_{23}^4 + \alpha_{34}^4 +
\alpha_{45}^4 + \alpha_{51}^4 \ , \nonumber \\
I_{2}^{(4)} & = & \alpha_{12}^3 \alpha_{23} + \alpha_{23}^3
\alpha_{34}  + \alpha_{34}^3 \alpha_{45}  + \alpha_{45}^3
\alpha_{51}  + \alpha_{51}^3 \alpha_{12}  \ ,
\nonumber \\
I_{3}^{(4)} & = & \alpha_{12}^3 \alpha_{34} + \alpha_{23}^3
\alpha_{45}  + \alpha_{34}^3 \alpha_{51}  + \alpha_{45}^3
\alpha_{12}  + \alpha_{51}^3\alpha_{23}  \ .
\nonumber \\
I_{4}^{(4)} & = & \alpha_{12}^3 \alpha_{45} + \alpha_{23}^3
\alpha_{51}  + \alpha_{34}^3 \alpha_{12}  + \alpha_{45}^3
\alpha_{23}  + \alpha_{51}^3\alpha_{34}  \ .
\nonumber \\
I_{5}^{(4)} & = & \alpha_{12}^3 \alpha_{51} + \alpha_{23}^3
\alpha_{12}  + \alpha_{34}^3 \alpha_{23}  + \alpha_{45}^3
\alpha_{34}  + \alpha_{51}^3 \alpha_{45}  \ .
\nonumber \\
I_{6}^{(4)} & = & \alpha_{12}^2 \alpha_{23}^2 + \alpha_{23}^2
\alpha_{34}^2  + \alpha_{34}^2 \alpha_{45}^2  + \alpha_{45}^2
\alpha_{51}^2  + \alpha_{51}^2 \alpha_{12}^2  \ ,
\nonumber \\
I_{7}^{(4)} & = & \alpha_{12}^2 \alpha_{34}^2 + \alpha_{23}^2
\alpha_{45}^2  + \alpha_{34}^2 \alpha_{51}^2  + \alpha_{45}^2
\alpha_{12}^2  + \alpha_{51}^2 \alpha_{23}^2  \ ,
\nonumber \\
I_{8}^{(4)} & = & \alpha_{12}^2\alpha_{23}\alpha_{34} +
\alpha_{23}^2\alpha_{34}\alpha_{45}   +
\alpha_{34}^2\alpha_{45}\alpha_{51}  +
\alpha_{45}^2\alpha_{51}\alpha_{12}  +
\alpha_{51}^2\alpha_{12}\alpha_{23}   \ ,
\label{Icyclic4} \\
I_{9}^{(4)} & = & \alpha_{12}^2\alpha_{23}\alpha_{45} +
\alpha_{23}^2\alpha_{34}\alpha_{51}   +
\alpha_{34}^2\alpha_{45}\alpha_{12}  +
\alpha_{45}^2\alpha_{51}\alpha_{23}  +
\alpha_{51}^2\alpha_{12}\alpha_{34}   \ ,
\nonumber \\
I_{10}^{(4)} & = & \alpha_{12}^2\alpha_{23}\alpha_{51} +
\alpha_{23}^2\alpha_{34}\alpha_{12}   +
\alpha_{34}^2\alpha_{45}\alpha_{23}  +
\alpha_{45}^2\alpha_{51}\alpha_{34}  +
\alpha_{51}^2\alpha_{12}\alpha_{45}   \ ,
\nonumber \\
I_{11}^{(4)} & = & \alpha_{12}^2\alpha_{34}\alpha_{45} +
\alpha_{23}^2\alpha_{45}\alpha_{51}   +
\alpha_{34}^2\alpha_{51}\alpha_{12}  +
\alpha_{45}^2\alpha_{12}\alpha_{23}  +
\alpha_{51}^2\alpha_{23}\alpha_{34}   \ ,
\nonumber \\
I_{12}^{(4)} & = & \alpha_{12}^2\alpha_{34}\alpha_{51} +
\alpha_{23}^2\alpha_{45}\alpha_{12}   +
\alpha_{34}^2\alpha_{51}\alpha_{23}  +
\alpha_{45}^2\alpha_{12}\alpha_{34}  +
\alpha_{51}^2\alpha_{23}\alpha_{45}   \ ,
\nonumber \\
I_{13}^{(4)} & = & \alpha_{12}^2\alpha_{45}\alpha_{51} +
\alpha_{23}^2\alpha_{51}\alpha_{12}   +
\alpha_{34}^2\alpha_{12}\alpha_{23}  +
\alpha_{45}^2\alpha_{23}\alpha_{34}  +
\alpha_{51}^2\alpha_{34}\alpha_{45}   \ ,
\nonumber \\
I_{14}^{(4)} & = & \alpha_{12}\alpha_{23}\alpha_{34}\alpha_{45} +
\alpha_{23}\alpha_{34}\alpha_{45}\alpha_{51} +
\alpha_{34}\alpha_{45}\alpha_{51}\alpha_{12} +
\alpha_{45}\alpha_{51}\alpha_{12}\alpha_{23} +
\alpha_{51}\alpha_{12}\alpha_{23}\alpha_{34}  \ . \nonumber
\end{eqnarray}
\item Degree 5:
\begin{eqnarray}
I_{1}^{(5)} & = & \alpha_{12}^5 + \alpha_{23}^5 + \alpha_{34}^5 +
\alpha_{45}^5 + \alpha_{51}^5 \ ,
\nonumber \\
I_{2}^{(5)} & = & \alpha_{12}^4 \alpha_{23} + \alpha_{23}^4
\alpha_{34}  + \alpha_{34}^4 \alpha_{45}  + \alpha_{45}^4
\alpha_{51}  + \alpha_{51}^4 \alpha_{12}  \ ,
\nonumber \\
I_{3}^{(5)} & = & \alpha_{12}^4 \alpha_{34} + \alpha_{23}^4
\alpha_{45}  + \alpha_{34}^4 \alpha_{51}  + \alpha_{45}^4
\alpha_{12}  + \alpha_{51}^4\alpha_{23}  \ ,
\nonumber \\
I_{4}^{(5)} & = & \alpha_{12}^4 \alpha_{45} + \alpha_{23}^4
\alpha_{51}  + \alpha_{34}^4 \alpha_{12}  + \alpha_{45}^4
\alpha_{23}  + \alpha_{51}^4\alpha_{34}  \ ,
\nonumber \\
I_{5}^{(5)} & = & \alpha_{12}^4 \alpha_{51} + \alpha_{23}^4
\alpha_{12}  + \alpha_{34}^4 \alpha_{23}  + \alpha_{45}^4
\alpha_{34}  + \alpha_{51}^4 \alpha_{45}  \ ,
\nonumber \\
I_{6}^{(5)} & = &  \alpha_{12}^3 \alpha_{23}^2 + \alpha_{23}^3
\alpha_{34}^2  + \alpha_{34}^3 \alpha_{45}^2  + \alpha_{45}^3
\alpha_{51}^2  + \alpha_{51}^3 \alpha_{12}^2  \ ,
\nonumber \\
I_{7}^{(5)} & = &  \alpha_{12}^3 \alpha_{34}^2 + \alpha_{23}^3
\alpha_{45}^2  + \alpha_{34}^3 \alpha_{51}^2  + \alpha_{45}^3
\alpha_{12}^2  + \alpha_{51}^3 \alpha_{23}^2  \ ,
\nonumber \\
I_{8}^{(5)} & = &  \alpha_{12}^3 \alpha_{45}^2 + \alpha_{23}^3
\alpha_{51}^2  + \alpha_{34}^3 \alpha_{12}^2  + \alpha_{45}^3
\alpha_{23}^2  + \alpha_{51}^3 \alpha_{34}^2  \ ,
\nonumber \\
I_{9}^{(5)} & = &  \alpha_{12}^3 \alpha_{51}^2 + \alpha_{23}^3
\alpha_{12}^2  + \alpha_{34}^3 \alpha_{23}^2  + \alpha_{45}^3
\alpha_{34}^2  + \alpha_{51}^3 \alpha_{45}^2  \ ,
\nonumber \\
I_{10}^{(5)} & = &  \alpha_{12}^3\alpha_{23}\alpha_{34} +
\alpha_{23}^3\alpha_{34}\alpha_{45}   +
\alpha_{34}^3\alpha_{45}\alpha_{51}  +
\alpha_{45}^3\alpha_{51}\alpha_{12}  +
\alpha_{51}^3\alpha_{12}\alpha_{23}   \ ,
\nonumber \\
I_{11}^{(5)} & = &  \alpha_{12}^3\alpha_{23}\alpha_{45} +
\alpha_{23}^3\alpha_{34}\alpha_{51}   +
\alpha_{34}^3\alpha_{45}\alpha_{12}  +
\alpha_{45}^3\alpha_{51}\alpha_{23}  +
\alpha_{51}^3\alpha_{12}\alpha_{34}   \ ,
\nonumber \\
I_{12}^{5)} & = &  \alpha_{12}^3\alpha_{23}\alpha_{51} +
\alpha_{23}^3\alpha_{34}\alpha_{12}   +
\alpha_{34}^3\alpha_{45}\alpha_{23}  +
\alpha_{45}^3\alpha_{51}\alpha_{34}  +
\alpha_{51}^3\alpha_{12}\alpha_{45}   \ ,
\nonumber \\
I_{13}^{(5)} & = &  \alpha_{12}^3\alpha_{34}\alpha_{45} +
\alpha_{23}^3\alpha_{45}\alpha_{51}  +
\alpha_{34}^3\alpha_{51}\alpha_{12}  +
\alpha_{45}^3\alpha_{12}\alpha_{23}  +
\alpha_{51}^3\alpha_{23}\alpha_{34}   \ ,
\nonumber \\
I_{14}^{(5)} & = &  \alpha_{12}^3\alpha_{34}\alpha_{51} +
\alpha_{23}^3\alpha_{45}\alpha_{12}   +
\alpha_{34}^3\alpha_{51}\alpha_{23}  +
\alpha_{45}^3\alpha_{12}\alpha_{34}  +
\alpha_{51}^3\alpha_{23}\alpha_{45}   \ ,
\label{Icyclic5} \\
I_{15}^{(5)} & = &  \alpha_{12}^3\alpha_{45}\alpha_{51} +
\alpha_{23}^3\alpha_{51}\alpha_{12}   +
\alpha_{34}^3\alpha_{12}\alpha_{23}  +
\alpha_{45}^3\alpha_{23}\alpha_{34}  +
\alpha_{51}^3\alpha_{34}\alpha_{45}   \ ,
\nonumber \\
I_{16}^{(5)} & = &  \alpha_{12}^2\alpha_{23}^2\alpha_{34} +
\alpha_{23}^2\alpha_{34}^2\alpha_{45}  +
\alpha_{34}^2\alpha_{45}^2\alpha_{51}  +
\alpha_{45}^2\alpha_{51}^2\alpha_{12}  +
\alpha_{51}^2\alpha_{12}^2\alpha_{23}   \ ,
\nonumber \\
I_{17}^{(5)} & = &  \alpha_{12}^2\alpha_{23}^2\alpha_{45} +
\alpha_{23}^2\alpha_{34}^2\alpha_{51}  +
\alpha_{34}^2\alpha_{45}^2\alpha_{12}  +
\alpha_{45}^2\alpha_{51}^2\alpha_{23}  +
\alpha_{51}^2\alpha_{12}^2\alpha_{34}   \ ,
\nonumber \\
I_{18}^{(5)} & = &  \alpha_{12}^2\alpha_{23}^2\alpha_{51} +
\alpha_{23}^2\alpha_{34}^2\alpha_{12}  +
\alpha_{34}^2\alpha_{45}^2\alpha_{23}  +
\alpha_{45}^2\alpha_{51}^2\alpha_{34}  +
\alpha_{51}^2\alpha_{12}^2\alpha_{45}   \ ,
\nonumber \\
I_{19}^{(5)} & = &  \alpha_{12}^2\alpha_{34}^2\alpha_{23} +
\alpha_{23}^2\alpha_{45}^2\alpha_{34}  +
\alpha_{34}^2\alpha_{51}^2\alpha_{45}  +
\alpha_{45}^2\alpha_{12}^2\alpha_{51}  +
\alpha_{51}^2\alpha_{23}^2\alpha_{12}   \ ,
\nonumber \\
I_{20}^{(5)} & = &  \alpha_{12}^2\alpha_{34}^2\alpha_{45} +
\alpha_{23}^2\alpha_{45}^2\alpha_{51}  +
\alpha_{34}^2\alpha_{51}^2\alpha_{12}  +
\alpha_{45}^2\alpha_{12}^2\alpha_{23}  +
\alpha_{51}^2\alpha_{23}^2\alpha_{34}   \ ,
\nonumber \\
I_{21}^{(5)} & = &  \alpha_{12}^2\alpha_{34}^2\alpha_{51} +
\alpha_{23}^2\alpha_{45}^2\alpha_{12}  +
\alpha_{34}^2\alpha_{51}^2\alpha_{23}  +
\alpha_{45}^2\alpha_{12}^2\alpha_{34}  +
\alpha_{51}^2\alpha_{23}^2\alpha_{45}   \ ,
\nonumber \\
I_{22}^{(5)} & = &  \alpha_{12}^2\alpha_{23}\alpha_{34}\alpha_{45}
+ \alpha_{23}^2\alpha_{34}\alpha_{45}\alpha_{51} +
\alpha_{34}^2\alpha_{45}\alpha_{51}\alpha_{12} +
\alpha_{45}^2\alpha_{51}\alpha_{12}\alpha_{23} +
\alpha_{51}^2\alpha_{12}\alpha_{23}\alpha_{34}  \ ,
\nonumber \\
I_{23}^{(5)} & = &  \alpha_{12}^2\alpha_{23}\alpha_{34}\alpha_{51}
+ \alpha_{23}^2\alpha_{34}\alpha_{45}\alpha_{12} +
\alpha_{34}^2\alpha_{45}\alpha_{51}\alpha_{23} +
\alpha_{45}^2\alpha_{51}\alpha_{12}\alpha_{34} +
\alpha_{51}^2\alpha_{12}\alpha_{23}\alpha_{45}  \ ,
\nonumber \\
I_{24}^{(5)} & = &  \alpha_{12}^2\alpha_{23}\alpha_{45}\alpha_{51}
+ \alpha_{23}^2\alpha_{34}\alpha_{51}\alpha_{12} +
\alpha_{34}^2\alpha_{45}\alpha_{12}\alpha_{23} +
\alpha_{45}^2\alpha_{51}\alpha_{23}\alpha_{34} +
\alpha_{51}^2\alpha_{12}\alpha_{34}\alpha_{45}  \ ,
\nonumber \\
I_{25}^{(5)} & = &  \alpha_{12}^2\alpha_{34}\alpha_{45}\alpha_{51}
+ \alpha_{23}^2\alpha_{45}\alpha_{51}\alpha_{12} +
\alpha_{34}^2\alpha_{51}\alpha_{12}\alpha_{23} +
\alpha_{45}^2\alpha_{12}\alpha_{23}\alpha_{34} +
\alpha_{51}^2\alpha_{23}\alpha_{34}\alpha_{45}  \ ,
\nonumber \\
I_{26}^{(5)} & = &
\alpha_{12}\alpha_{23}\alpha_{34}\alpha_{45}\alpha_{51} \ .
\nonumber
\end{eqnarray}
\item Degree 6:
\begin{eqnarray}
I_{1}^{(6)} & = &  \alpha_{12}^6 + \alpha_{23}^6 + \alpha_{34}^6 +
\alpha_{45}^6 + \alpha_{51}^6 \ ,
\nonumber \\
I_{2}^{(6)} & = &  \alpha_{12}^5 \alpha_{23} + \alpha_{23}^5
\alpha_{34}  + \alpha_{34}^5 \alpha_{45}  + \alpha_{45}^5
\alpha_{51}  + \alpha_{51}^5 \alpha_{12}  \ ,
\nonumber \\
I_{3}^{(6)} & = &  \alpha_{12}^5 \alpha_{34} + \alpha_{23}^5
\alpha_{45}  + \alpha_{34}^5 \alpha_{51}  + \alpha_{45}^5
\alpha_{12}  + \alpha_{51}^5\alpha_{23}  \ ,
\nonumber \\
I_{4}^{(6)} & = &  \alpha_{12}^5 \alpha_{45} + \alpha_{23}^5
\alpha_{51}  + \alpha_{34}^5 \alpha_{12}  + \alpha_{45}^5
\alpha_{23}  + \alpha_{51}^5\alpha_{34}  \ ,
\nonumber \\
I_{5}^{(6)} & = &  \alpha_{12}^5 \alpha_{51} + \alpha_{23}^5
\alpha_{12}  + \alpha_{34}^5 \alpha_{23}  + \alpha_{45}^5
\alpha_{34}  + \alpha_{51}^5 \alpha_{45}  \ .
\nonumber \\
I_{6}^{(6)} & = &  \alpha_{12}^4 \alpha_{23}^2 + \alpha_{23}^4
\alpha_{34}^2  + \alpha_{34}^4 \alpha_{45}^2  + \alpha_{45}^4
\alpha_{51}^2  + \alpha_{51}^4 \alpha_{12}^2  \ ,
\nonumber \\
I_{7}^{(6)} & = &  \alpha_{12}^4 \alpha_{34}^2 + \alpha_{23}^4
\alpha_{45}^2  + \alpha_{34}^4 \alpha_{51}^2  + \alpha_{45}^4
\alpha_{12}^2  + \alpha_{51}^4 \alpha_{23}^2  \ ,
\nonumber \\
I_{8}^{(6)} & = &  \alpha_{12}^4 \alpha_{45}^2 + \alpha_{23}^4
\alpha_{51}^2  + \alpha_{34}^4 \alpha_{12}^2  + \alpha_{45}^4
\alpha_{23}^2  + \alpha_{51}^4 \alpha_{34}^2  \ ,
\nonumber \\
I_{9}^{(6)} & = &  \alpha_{12}^4 \alpha_{51}^2 + \alpha_{23}^4
\alpha_{12}^2  + \alpha_{34}^4 \alpha_{23}^2  + \alpha_{45}^4
\alpha_{34}^2  + \alpha_{51}^4 \alpha_{45}^2  \ ,
\nonumber \\
I_{10}^{(6)} & = &  \alpha_{12}^4\alpha_{23}\alpha_{34} +
\alpha_{23}^4\alpha_{34}\alpha_{45}   +
\alpha_{34}^4\alpha_{45}\alpha_{51}  +
\alpha_{45}^4\alpha_{51}\alpha_{12}  +
\alpha_{51}^4\alpha_{12}\alpha_{23}   \ , \nonumber
\end{eqnarray}
\begin{eqnarray}
I_{11}^{(6)} & = &  \alpha_{12}^4\alpha_{23}\alpha_{45} +
\alpha_{23}^4\alpha_{34}\alpha_{51}  +
\alpha_{34}^4\alpha_{45}\alpha_{12}  +
\alpha_{45}^4\alpha_{51}\alpha_{23}  +
\alpha_{51}^4\alpha_{12}\alpha_{34}   \ ,
\nonumber \\
I_{12}^{(6)} & = &  \alpha_{12}^4\alpha_{23}\alpha_{51} +
\alpha_{23}^4\alpha_{34}\alpha_{12}  +
\alpha_{34}^4\alpha_{45}\alpha_{23}  +
\alpha_{45}^4\alpha_{51}\alpha_{34}  +
\alpha_{51}^4\alpha_{12}\alpha_{45}   \ ,
\nonumber \\
I_{13}^{(6)} & = &  \alpha_{12}^4\alpha_{34}\alpha_{45} +
\alpha_{23}^4\alpha_{45}\alpha_{51}  +
\alpha_{34}^4\alpha_{51}\alpha_{12}  +
\alpha_{45}^4\alpha_{12}\alpha_{23}  +
\alpha_{51}^4\alpha_{23}\alpha_{34}   \ ,
\nonumber \\
I_{14}^{(6)} & = &  \alpha_{12}^4\alpha_{34}\alpha_{51} +
\alpha_{23}^4\alpha_{45}\alpha_{12}  +
\alpha_{34}^4\alpha_{51}\alpha_{23}  +
\alpha_{45}^4\alpha_{12}\alpha_{34}  +
\alpha_{51}^4\alpha_{23}\alpha_{45}   \ ,
\nonumber \\
I_{15}^{(6)} & = &  \alpha_{12}^4\alpha_{45}\alpha_{51} +
\alpha_{23}^4\alpha_{51}\alpha_{12}  +
\alpha_{34}^4\alpha_{12}\alpha_{23}  +
\alpha_{45}^4\alpha_{23}\alpha_{34}  +
\alpha_{51}^4\alpha_{34}\alpha_{45}   \ ,
\nonumber \\
I_{16}^{(6)} & = &  \alpha_{12}^3 \alpha_{23}^3 + \alpha_{23}^3
\alpha_{34}^3  + \alpha_{34}^3 \alpha_{45}^3  + \alpha_{45}^3
\alpha_{51}^3  + \alpha_{51}^3 \alpha_{12}^3  \ ,
\nonumber \\
I_{17}^{(6)} & = &  \alpha_{12}^3 \alpha_{34}^3 + \alpha_{23}^3
\alpha_{45}^3  + \alpha_{34}^3 \alpha_{51}^3  + \alpha_{45}^3
\alpha_{12}^3  + \alpha_{51}^3 \alpha_{23}^3  \ ,
\nonumber \\
I_{18}^{(6)} & = &  \alpha_{12}^3\alpha_{23}^2\alpha_{34} +
\alpha_{23}^3\alpha_{34}^2\alpha_{45} +
\alpha_{34}^3\alpha_{45}^2\alpha_{51}  +
\alpha_{45}^3\alpha_{51}^2\alpha_{12}  +
\alpha_{51}^3\alpha_{12}^2\alpha_{23}   \ ,
\nonumber \\
I_{19}^{(6)} & = &  \alpha_{12}^3\alpha_{23}^2\alpha_{45} +
\alpha_{23}^3\alpha_{34}^2\alpha_{51} +
\alpha_{34}^3\alpha_{45}^2\alpha_{12}  +
\alpha_{45}^3\alpha_{51}^2\alpha_{23} +
\alpha_{51}^3\alpha_{12}^2\alpha_{34}   \ ,
\nonumber \\
I_{20}^{(6)} & = &  \alpha_{12}^3\alpha_{23}^2\alpha_{51} +
\alpha_{23}^3\alpha_{34}^2\alpha_{12} +
\alpha_{34}^3\alpha_{45}^2\alpha_{23}  +
\alpha_{45}^3\alpha_{51}^2\alpha_{34} +
\alpha_{51}^3\alpha_{12}^2\alpha_{45}   \ ,
\nonumber \\
I_{21}^{(6)} & = &  \alpha_{12}^3\alpha_{34}^2\alpha_{23} +
\alpha_{23}^3\alpha_{45}^2\alpha_{34} +
\alpha_{34}^3\alpha_{51}^2\alpha_{45}  +
\alpha_{45}^3\alpha_{12}^2\alpha_{51} +
\alpha_{51}^3\alpha_{23}^2\alpha_{12}   \ ,
\nonumber \\
I_{22}^{(6)} & = &  \alpha_{12}^3\alpha_{34}^2\alpha_{45} +
\alpha_{23}^3\alpha_{45}^2\alpha_{51} +
\alpha_{34}^3\alpha_{51}^2\alpha_{12}  +
\alpha_{45}^3\alpha_{12}^2\alpha_{23} +
\alpha_{51}^3\alpha_{23}^2\alpha_{34}   \ ,
\label{Icyclic6} \\
I_{23}^{(6)} & = &  \alpha_{12}^3\alpha_{34}^2\alpha_{51} +
\alpha_{23}^3\alpha_{45}^2\alpha_{12} +
\alpha_{34}^3\alpha_{51}^2\alpha_{23}  +
\alpha_{45}^3\alpha_{12}^2\alpha_{34} +
\alpha_{51}^3\alpha_{23}^2\alpha_{45}   \ ,
\nonumber \\
I_{24}^{(6)} & = &  \alpha_{12}^3\alpha_{45}^2\alpha_{23} +
\alpha_{23}^3\alpha_{51}^2\alpha_{34} +
\alpha_{34}^3\alpha_{12}^2\alpha_{45}  +
\alpha_{45}^3\alpha_{23}^2\alpha_{51} +
\alpha_{51}^3\alpha_{34}^2\alpha_{12}   \ ,
\nonumber \\
I_{25}^{(6)} & = &  \alpha_{12}^3\alpha_{45}^2\alpha_{34} +
\alpha_{23}^3\alpha_{51}^2\alpha_{45} +
\alpha_{34}^3\alpha_{12}^2\alpha_{51}  +
\alpha_{45}^3\alpha_{23}^2\alpha_{12} +
\alpha_{51}^3\alpha_{34}^2\alpha_{23}   \ ,
\nonumber \\
I_{26}^{(6)} & = &  \alpha_{12}^3\alpha_{45}^2\alpha_{51} +
\alpha_{23}^3\alpha_{51}^2\alpha_{12} +
\alpha_{34}^3\alpha_{12}^2\alpha_{23}  +
\alpha_{45}^3\alpha_{23}^2\alpha_{34} +
\alpha_{51}^3\alpha_{34}^2\alpha_{45}   \ ,
\nonumber\\
I_{27}^{(6)} & = &  \alpha_{12}^3\alpha_{51}^2\alpha_{23} +
\alpha_{23}^3\alpha_{12}^2\alpha_{34} +
\alpha_{34}^3\alpha_{23}^2\alpha_{45}  +
\alpha_{45}^3\alpha_{34}^2\alpha_{51} +
\alpha_{51}^3\alpha_{45}^2\alpha_{12}   \ ,
\nonumber\\
I_{28}^{(6)} & = & \alpha_{12}^3\alpha_{51}^2\alpha_{34} +
\alpha_{23}^3\alpha_{12}^2\alpha_{45} +
\alpha_{34}^3\alpha_{23}^2\alpha_{51}  +
\alpha_{45}^3\alpha_{34}^2\alpha_{12} +
\alpha_{51}^3\alpha_{45}^2\alpha_{23}    \ ,
\nonumber\\
I_{29}^{(6)} & = &  \alpha_{12}^3\alpha_{51}^2\alpha_{45} +
\alpha_{23}^3\alpha_{12}^2\alpha_{51} +
\alpha_{34}^3\alpha_{23}^2\alpha_{12}  +
\alpha_{45}^3\alpha_{34}^2\alpha_{23} +
\alpha_{51}^3\alpha_{45}^2\alpha_{34}   \ ,
\nonumber\\
I_{30}^{(6)} & = &  \alpha_{12}^3\alpha_{23}\alpha_{34}\alpha_{45}
+ \alpha_{23}^3\alpha_{34}\alpha_{45}\alpha_{51} +
\alpha_{34}^3\alpha_{45}\alpha_{51}\alpha_{12} +
\alpha_{45}^3\alpha_{51}\alpha_{12}\alpha_{23} +
\alpha_{51}^3\alpha_{12}\alpha_{23}\alpha_{34}   \ ,
\nonumber\\
I_{31}^{(6)} & = & \alpha_{12}^3\alpha_{23}\alpha_{34}\alpha_{51}
+ \alpha_{23}^3\alpha_{34}\alpha_{45}\alpha_{12} +
\alpha_{34}^3\alpha_{45}\alpha_{51}\alpha_{23} +
\alpha_{45}^3\alpha_{51}\alpha_{12}\alpha_{34} +
\alpha_{51}^3\alpha_{12}\alpha_{23}\alpha_{45}   \ ,
\nonumber\\
I_{32}^{(6)} & = & \alpha_{12}^3\alpha_{23}\alpha_{45}\alpha_{51}
+ \alpha_{23}^3\alpha_{34}\alpha_{51}\alpha_{12} +
\alpha_{34}^3\alpha_{45}\alpha_{12}\alpha_{23} +
\alpha_{45}^3\alpha_{51}\alpha_{23}\alpha_{34} +
\alpha_{51}^3\alpha_{12}\alpha_{34}\alpha_{45}    \ ,
\nonumber\\
I_{33}^{(6)} & = & \alpha_{12}^3\alpha_{34}\alpha_{45}\alpha_{51}
+ \alpha_{23}^3\alpha_{45}\alpha_{51}\alpha_{12} +
\alpha_{34}^3\alpha_{51}\alpha_{12}\alpha_{23} +
\alpha_{45}^3\alpha_{12}\alpha_{23}\alpha_{34} +
\alpha_{51}^3\alpha_{23}\alpha_{34}\alpha_{45}     \ ,
\nonumber\\
I_{34}^{(6)} & = &  \alpha_{12}^2\alpha_{23}^2\alpha_{34}^2 +
\alpha_{23}^2\alpha_{34}^2\alpha_{45}^2 +
\alpha_{34}^2\alpha_{45}^2\alpha_{51}^2  +
\alpha_{45}^2\alpha_{51}^2\alpha_{12}^2 +
\alpha_{51}^2\alpha_{12}^2\alpha_{23}^2   \ ,
\nonumber\\
I_{35}^{(6)} & = &  \alpha_{12}^2\alpha_{23}^2\alpha_{45}^2 +
\alpha_{23}^2\alpha_{34}^2\alpha_{51}^2 +
\alpha_{34}^2\alpha_{45}^2\alpha_{12}^2  +
\alpha_{45}^2\alpha_{51}^2\alpha_{23}^2 +
\alpha_{51}^2\alpha_{12}^2\alpha_{34}^2    \ ,
\nonumber\\
I_{36}^{(6)} & = &
\alpha_{12}^2\alpha_{23}^2\alpha_{34}\alpha_{45} +
\alpha_{23}^2\alpha_{34}^2\alpha_{45}\alpha_{51} +
\alpha_{34}^2\alpha_{45}^2\alpha_{51}\alpha_{12} +
\alpha_{45}^2\alpha_{51}^2\alpha_{12}\alpha_{23} +
\alpha_{51}^2\alpha_{12}^2\alpha_{23}\alpha_{34}   \ ,
\nonumber\\
I_{37}^{(6)} & = &
\alpha_{12}^2\alpha_{23}^2\alpha_{34}\alpha_{51} +
\alpha_{23}^2\alpha_{34}^2\alpha_{45}\alpha_{12} +
\alpha_{34}^2\alpha_{45}^2\alpha_{51}\alpha_{23} +
\alpha_{45}^2\alpha_{51}^2\alpha_{12}\alpha_{34} +
\alpha_{51}^2\alpha_{12}^2\alpha_{23}\alpha_{45}   \ ,
\nonumber\\
I_{38}^{(6)} & = &
\alpha_{12}^2\alpha_{23}^2\alpha_{45}\alpha_{51} +
\alpha_{23}^2\alpha_{34}^2\alpha_{51}\alpha_{12} +
\alpha_{34}^2\alpha_{45}^2\alpha_{12}\alpha_{23} +
\alpha_{45}^2\alpha_{51}^2\alpha_{23}\alpha_{34} +
\alpha_{51}^2\alpha_{12}^2\alpha_{34}\alpha_{45}   \ ,
\nonumber\\
I_{39}^{(6)} & = &
\alpha_{12}^2\alpha_{34}^2\alpha_{23}\alpha_{45} +
\alpha_{23}^2\alpha_{45}^2\alpha_{34}\alpha_{51} +
\alpha_{34}^2\alpha_{51}^2\alpha_{45}\alpha_{12} +
\alpha_{45}^2\alpha_{12}^2\alpha_{51}\alpha_{23} +
\alpha_{51}^2\alpha_{23}^2\alpha_{12}\alpha_{34}  \ ,
\nonumber \\
I_{40}^{(6)} & = &
\alpha_{12}^2\alpha_{34}^2\alpha_{23}\alpha_{51} +
\alpha_{23}^2\alpha_{45}^2\alpha_{34}\alpha_{12} +
\alpha_{34}^2\alpha_{51}^2\alpha_{45}\alpha_{23} +
\alpha_{45}^2\alpha_{12}^2\alpha_{51}\alpha_{34} +
\alpha_{51}^2\alpha_{23}^2\alpha_{12}\alpha_{45}  \ ,
\nonumber \\
I_{41}^{(6)} & = &
\alpha_{12}^2\alpha_{34}^2\alpha_{45}\alpha_{51} +
\alpha_{23}^2\alpha_{45}^2\alpha_{51}\alpha_{12} +
\alpha_{34}^2\alpha_{51}^2\alpha_{12}\alpha_{23} +
\alpha_{45}^2\alpha_{12}^2\alpha_{23}\alpha_{34} +
\alpha_{51}^2\alpha_{23}^2\alpha_{34}\alpha_{45}  \ ,
\nonumber \\
I_{42}^{(6)} & = &
\alpha_{12}^2\alpha_{23}\alpha_{34}\alpha_{45}\alpha_{51} +
\alpha_{23}^2\alpha_{34}\alpha_{45}\alpha_{51}\alpha_{12} +
\alpha_{34}^2\alpha_{45}\alpha_{51}\alpha_{12}\alpha_{23} +
\alpha_{45}^2\alpha_{51}\alpha_{12}\alpha_{23}\alpha_{34} +
\nonumber \\
& & + \alpha_{51}^2\alpha_{12}\alpha_{23}\alpha_{34}\alpha_{45} \
.
\nonumber
\end{eqnarray}
\end{itemize}

\subsection{$H^{(1)}$, $P^{(1)}$, $U^{(1)}$, $W^{(1)}$, $Z^{(1)}$ and
$\Delta$ factors}
\label{otherfactors}

The $\alpha'$ dependent factors of formula (\ref{AD2nF5nopoles}),
except for $Z^{(1)}$, are all given explicitly in terms of the
function $f$ of (\ref{f}), the factor $K_3$ of (\ref{K3-int}) and
the factor $T$ of (\ref{K2-K})\footnote{In this formula, the factors
$H^{(j)}$ (with $j=2,3,4,5$) are constructed by means of cyclic
permutations of the factor $H^{(1)}$. For example, to obtain the
expression for $H^{(2)}$ we make the following change of indexes in
the $\alpha_{ij}$ variables: $(1,2,3,4,5) \rightarrow
(2,3,4,5,1)$.}:
\begin{eqnarray}
\label{H23} H^{(1)} &=& - (2 \alpha')^2 \frac{f(-2 \alpha_{23}, -2
\alpha_{34})-f(-2 \alpha_{34}, -2
\alpha_{45})}{\alpha_{23}-\alpha_{45}} \ ,
\\
\label{P34} P^{(1)} &=&  (2 \alpha')^2 \frac{ \ \ K_3 -
K_3\biggl|_{\alpha_{34}=0}}{\alpha_{34}}  \ ,
\\
\label{U34} U^{(1)} &=& \frac{(2 \alpha')^2\biggl\{f(-2 \alpha_{51},
-2 \alpha_{12}) + \ K_3\biggl|_{\alpha_{34}=0} \biggr\}  - \
\alpha_{45}H^{(3)} \ - \ \alpha_{23}H^{(4)}}
{\alpha_{23}\alpha_{45}} \ ,
\\
\label{W51} W^{(1)} &=& \frac{(2 \alpha')^2 \biggl[K_3 -
K_3\biggl|_{\alpha_{51}=0} - K_3\biggl|_{\alpha_{12}=0}
\biggr]+\alpha_{23}H^{(1)}- (2 \alpha')^2 f(-2 \alpha_{34}, -2
\alpha_{45}) \ }{\alpha_{51}\alpha_{12}} \ ,
\\
\label{Z23} Z^{(1)} &=& (2 \alpha')^2  \times \nonumber
\\
&& \times \frac{G(-2 \alpha_{34}, -2
\alpha_{45},-\alpha_{25}-\alpha_{34},-\alpha_{23}-\alpha_{45})-G(-2
\alpha_{34}, -2
\alpha_{23},-\alpha_{25}-\alpha_{34},-\alpha_{23}-\alpha_{45})}{\alpha_{23}-\alpha_{45}}
\ , \nonumber
\\
\\
\label{Delta} \Delta &=&  \frac{ \ T -1 - \biggl[\frac{}{}\biggl\{
T\biggl|_{\alpha_{23}=0}- \ 1 \ - \
\alpha_{12}\alpha_{23}\alpha_{34}H^{(5)} \biggr\}+(\mbox{cyclic
permutations})\frac{}{}\biggr]
}{\alpha_{12}\alpha_{23}\alpha_{34}\alpha_{45}\alpha_{51}} \ ,
\end{eqnarray}
where the function $G(a,b,c,d)$ that appears in (\ref{Z23}) is
defined by the $\alpha'$ series
\begin{eqnarray}
G(a,b,c,d) = \sum_{m=1}^{\infty} \sum_{n=0}^{\infty} \frac{ \
a_{mn} \ {\alpha'}^{m+n}}{\left(\begin{array}{c}
                          m+n \\
                          m
                          \end{array} \right)}
\sum_{p=0}^{m-1} \sum_{q=0}^{n} \left(\begin{array}{c}
                          m-1-p+n-q \\
                          m-1-p
                          \end{array} \right)
                          \left(\begin{array}{c}
                          p+q \\
                          p
                          \end{array} \right)
a^{m-1-p} \ b^{n-q} \ c^p \ d^q \ . \nonumber \\
\label{G}
\end{eqnarray}
Here, the (completely known) coefficients $a_{mn}$ are the ones that
appear in the expansion of the function $f(s,t)$, in (\ref{fpower}).
May be it is possible to write the function $G(a,b,c,d)$ in a closed
form (in terms of Hypergeometric functions, for
example), but we have not succeeded in doing so.\\
\noindent $G(a,b,c,d)$ begins like
\begin{eqnarray}
G(a,b,c,d) & = & -\zeta(3) \ \alpha' - \frac{ \ \pi^4}{720}
\biggl( \ 8a + 8c + b + d  \biggr) \
{\alpha'}^2 + \nonumber \\
&& + \biggl( -\frac{1}{3} \zeta(5) \
(3a^2+3ac+3c^2+2b^2+2bd+2d^2+4ab+2ad+2bc+4cd) + \nonumber \\
&& \hphantom{ + \biggl( }+ \frac{ \ \pi^2}{18}\zeta(3)
(b^2+bd+d^2+2ab+ad+bc+2cd) \biggr) \ {\alpha'}^3 \ +{\cal
O}({\alpha'}^4). \label{Gpower}
\end{eqnarray}
\noindent Using eqs. (\ref{K3hyperg}), (\ref{K2-K}) and
(\ref{K2hyperg}), the $\alpha'$ factors defined in equations
(\ref{H23})-(\ref{Z23}) can all be written in terms of the Gamma
factor, Beta and Hypergeometric functions. It is also not
difficult to see that these factors are all invariant under a
twisting transformation with respect to index $1$. The factor
$\Delta$ in (\ref{Delta}) is invariant under a twisting
transformation with respect to any of the five
indexes and it is also cyclic invariant.\\
\noindent Formulas
(\ref{H23}), (\ref{P34}) and (\ref{Z23}) can be understood to have
no poles since the numerator is a power series of the $\alpha_{ij}$
and, whenever the denominator is zero, so happens with the
numerator. This means that, in each of those
cases, the numerator is divisible by the corresponding denominator.\\
\noindent In the case of formulas (\ref{U34}), (\ref{W51}) and
(\ref{Delta}) it is not straight forward to see that those
expressions contain no poles, but the argument is similar: in all
cases the numerator involves a power series of the $\alpha_{ij}$
and, whenever the denominator is zero, so happens with the
numerator. For example, in the case of (\ref{U34}) and (\ref{W51}),
using (\ref{K3Gamma}) (and one of the cyclic permutations of that
relation) it can be proved that the numerator becomes zero if any of
the $\alpha_{ij}$ of the denominator are zero. In the case of
(\ref{Delta}), using (\ref{TGamma}) and (\ref{T0}) (and the cyclic
permutations of them) it can be proved that if any of the five
$\alpha_{ij}$ of the denominator is zero,
so happens with the numerator.\\
\noindent Up to ${\cal O}({\alpha'}^6)$ order, the expansions of the
factors defined in (\ref{H23})-(\ref{Delta}) are given by
\begin{eqnarray}
\label{H23exp} H^{(1)} &=&  \ -\zeta(3)(2
\alpha')^3+\frac{\pi^4}{360}\biggl[4\alpha_{45}+4\alpha_{23}+\alpha_{34}\biggr](2
\alpha')^4+ \biggl[\frac{\
\pi^2}{6}\zeta(3)\alpha_{34}\biggl(\alpha_{34}+\alpha_{45}+\alpha_{23}\biggr)-
\nonumber
\\
&& \ \ \ \
-\zeta(5)\biggl(2\alpha_{34}\alpha_{23}+\alpha_{23}^2+2\alpha_{45}\alpha_{34}+2\alpha_{34}^2
+\alpha_{45}^2+\alpha_{45}\alpha_{23}\biggr)\biggr](2 \alpha')^5 +
\nonumber \\
&& + \biggl[ -\frac{1}{2}\zeta(3)^2
\alpha_{34}\biggl(2\alpha_{23}\alpha_{34}+\alpha_{45}\alpha_{23}+2\alpha_{45}\alpha_{34}
+\alpha_{45}^2+\alpha_{23}^2+\alpha_{34}^2\biggr) + \frac{ \
\pi^6}{15120}\biggl(16\alpha_{45}\alpha_{23}^2+\nonumber
\\
&&\hphantom{+ \biggl[ } +12\alpha_{34}\alpha_{23}^2
+12\alpha_{45}\alpha_{34}\alpha_{23}+16\alpha_{23}\alpha_{45}^2+23\alpha_{23}\alpha_{34}^2
+23\alpha_{45}\alpha_{34}^2+16\alpha_{45}^3+12\alpha_{34}^3+
\nonumber \\
&&\hphantom{+ \biggl[ } +16\alpha_{23}^3
+12\alpha_{34}\alpha_{45}^2\biggr) \biggr](2 \alpha')^6 + \ {\cal
O}( \ (2 \alpha')^7 \ )  \ ,
\\
\label{P34exp} P^{(1)} &=& -\zeta(3)(2
\alpha')^3+\frac{\pi^4}{360}\biggl[4\alpha_{34}+\alpha_{45}+\alpha_{23}
+4\alpha_{51}+4\alpha_{12}\biggr] (2 \alpha')^4+  \biggl[\frac{\
\pi^2}{6}\zeta(3)\biggl(\alpha_{23}\alpha_{45}-  \nonumber
\\
&& \ \ \ \ -2\alpha_{12}\alpha_{45}+\alpha_{45}^2
+\alpha_{23}^2+\alpha_{34}\alpha_{45}+\alpha_{12}\alpha_{23}-2\alpha_{51}\alpha_{23}
+\alpha_{23}\alpha_{34}+\alpha_{45}\alpha_{51}-2\alpha_{51}\alpha_{12}\biggr)-
\nonumber
\\
&& \ \ \ \ -\frac{1}{2}\zeta(5)\biggl(2\alpha_{34}^2+
2\alpha_{12}^2+4\alpha_{23}^2+2\alpha_{12}\alpha_{34}+2\alpha_{51}\alpha_{34}
+2\alpha_{51}^2 +4\alpha_{45}^2+4\alpha_{23}\alpha_{34}+\nonumber
\\
&& \ \ \ \
+4\alpha_{12}\alpha_{23}-7\alpha_{12}\alpha_{45}-7\alpha_{51}\alpha_{23}
+4\alpha_{23}\alpha_{45}+4\alpha_{45}\alpha_{51}-7\alpha_{51}\alpha_{12}+\nonumber
\\
&&\ \ \ \ +4\alpha_{34}\alpha_{45}\biggr)\biggr](2 \alpha')^5
+\biggl[
\frac{1}{2}\zeta(3)^2\biggl(2\alpha_{51}^2\alpha_{23}+2\alpha_{12}^2\alpha_{45}
-\alpha_{23}^2\alpha_{45}+\alpha_{23}^2\alpha_{51}-\alpha_{51}^2\alpha_{45}
-\nonumber \\
&& \ \ \
-\alpha_{34}^2\alpha_{45}-\alpha_{34}^2\alpha_{23}+\alpha_{12}^2\alpha_{51}
-2\alpha_{12}\alpha_{23}^2-2\alpha_{45}^2\alpha_{51}-2\alpha_{23}^2\alpha_{34}
-2\alpha_{45}^2\alpha_{34}+\alpha_{45}^2\alpha_{12}+\nonumber \\
&& \ \ \ +\alpha_{51}^2\alpha_{12}
-\alpha_{45}^2\alpha_{23}-\alpha_{12}^2\alpha_{23}-\alpha_{45}^3-\alpha_{23}^3
+2\alpha_{12}\alpha_{34}\alpha_{51}-2\alpha_{23}\alpha_{34}\alpha_{45}
+\alpha_{12}\alpha_{45}\alpha_{34}+\nonumber
\\
&& \ \ \ +4\alpha_{51}\alpha_{12}\alpha_{23}
+\alpha_{51}\alpha_{23}\alpha_{34}+4\alpha_{12}\alpha_{23}\alpha_{45}
+4\alpha_{45}\alpha_{51}\alpha_{23}-\alpha_{12}\alpha_{23}\alpha_{34}
-\alpha_{34}\alpha_{45}\alpha_{51}+\nonumber
\\
&& \ \ \ +4\alpha_{45}\alpha_{51}\alpha_{12}\biggr) + \frac{ \
\pi^6}{45360}\biggl(-52\alpha_{51}^2\alpha_{23}-52\alpha_{12}^2\alpha_{45}
+36\alpha_{23}^2\alpha_{45}-28\alpha_{23}^2\alpha_{51}+36\alpha_{51}^2\alpha_{45}
+\nonumber \\
&& \ \ \
+36\alpha_{34}^2\alpha_{45}+36\alpha_{34}^2\alpha_{23}-28\alpha_{12}^2\alpha_{51}
+69\alpha_{12}\alpha_{23}^2+69\alpha_{45}^2\alpha_{51}+69\alpha_{23}^2\alpha_{34}
+69\alpha_{45}^2\alpha_{34}-\nonumber
\\
&& \ \ \ -28\alpha_{45}^2\alpha_{12}-28\alpha_{51}^2\alpha_{12}
+36\alpha_{45}^2\alpha_{23}+36\alpha_{12}^2\alpha_{23}+48\alpha_{34}^3+36\alpha_{45}^3
+36\alpha_{23}^3+48\alpha_{51}^3+\nonumber
\\
&& \ \ \ +48\alpha_{12}^3+48\alpha_{34}\alpha_{51}^2
-52\alpha_{12}\alpha_{34}\alpha_{51}+69\alpha_{23}\alpha_{34}\alpha_{45}
-28\alpha_{12}\alpha_{45}\alpha_{34}-134\alpha_{51}\alpha_{12}\alpha_{23}
-\nonumber \\
&& \ \ \
-28\alpha_{51}\alpha_{23}\alpha_{34}-134\alpha_{12}\alpha_{23}\alpha_{45}
-134\alpha_{45}\alpha_{51}\alpha_{23}+36\alpha_{12}\alpha_{23}\alpha_{34}
+36\alpha_{34}\alpha_{45}\alpha_{51}-\nonumber
\\
&& \ \ \ -134\alpha_{45}\alpha_{51}\alpha_{12}
+48\alpha_{51}\alpha_{34}^2+48\alpha_{12}\alpha_{34}^2+48\alpha_{34}\alpha_{12}^2\biggr)\biggr](2
\alpha')^6+{\cal O}(  \ (2 \alpha')^7 \ )  \ ,
\end{eqnarray}
\begin{eqnarray}
\label{U34exp} U^{(1)} &=& \frac{\pi^4}{90}(2
\alpha')^4+\biggl[-\frac{\ \pi^2}{3}
\zeta(3)\biggl(\alpha_{51}+\alpha_{12}\biggr)
+\frac{1}{2}\zeta(5)\biggl(7\alpha_{51}+7\alpha_{12}-2\alpha_{45}
-2\alpha_{23}\biggr) \biggr](2 \alpha')^5 + \nonumber
\\
&&+ \biggl[
\frac{1}{2}\zeta(3)^2\biggl(4\alpha_{51}\alpha_{12}+\alpha_{51}^2+\alpha_{12}^2
+2\alpha_{51}\alpha_{23}+2\alpha_{45}\alpha_{12}+\alpha_{45}\alpha_{51}
+\alpha_{12}\alpha_{23}\biggr) + \nonumber \\
&& \ \ \ + \frac{ \
\pi^6}{22680}\biggl(-14\alpha_{45}\alpha_{51}+24\alpha_{23}^2-26\alpha_{45}\alpha_{12}
-26\alpha_{51}\alpha_{23}+24\alpha_{45}^2-67\alpha_{51}\alpha_{12}
-\nonumber \\
&& \ \ \ -14\alpha_{12}^2
+24\alpha_{45}\alpha_{23}-14\alpha_{51}^2-14\alpha_{12}\alpha_{23}\biggr)\biggr](2
\alpha')^6 + \ {\cal O}( \ (2 \alpha')^7 \ ) \ ,
\\
\label{W51exp} W^{(1)} &=& \frac{\pi^4}{360}(2
\alpha')^4+\biggl[\frac{ \
\pi^2}{6}\zeta(3)\biggl(\alpha_{45}+\alpha_{23}-2\alpha_{34}
+\alpha_{51}+\alpha_{12}\biggr)
+\frac{1}{2}\zeta(5)\biggl(7\alpha_{34}-4\alpha_{45}-\nonumber
\\
&& \ \ \ -4\alpha_{23} -4\alpha_{51}-4\alpha_{12}\biggr)\biggr](2
\alpha')^5 + \biggl[
-\frac{1}{2}\zeta(3)^2\biggl(\alpha_{45}^2+2\alpha_{45}\alpha_{51}
+\alpha_{51}^2+\alpha_{12}^2+\alpha_{23}^2-\nonumber\\
&& \ \ \
-4\alpha_{23}\alpha_{45}+2\alpha_{12}\alpha_{23}+\alpha_{45}\alpha_{12}
-2\alpha_{34}^2
-\alpha_{34}\alpha_{51}+\alpha_{51}\alpha_{23}-4\alpha_{23}\alpha_{34}
-\alpha_{12}\alpha_{34}-\nonumber \\
&& \ \ \ -4\alpha_{34}\alpha_{45}+2\alpha_{51}\alpha_{12}\biggr) +
\frac{ \
\pi^6}{45360}\biggl(36\alpha_{51}\alpha_{23}-28\alpha_{12}\alpha_{34}+69\alpha_{51}\alpha_{12}
+36\alpha_{23}^2+36\alpha_{45}^2+\nonumber
\\
&& \ \ \ +69\alpha_{12}\alpha_{23}+36\alpha_{51}^2+36\alpha_{12}^2
-134\alpha_{23}\alpha_{34}-52\alpha_{34}^2+69\alpha_{45}\alpha_{51}
-134\alpha_{34}\alpha_{45}+\nonumber \\
&& \ \ \
+36\alpha_{45}\alpha_{12}-28\alpha_{34}\alpha_{51}-134\alpha_{23}\alpha_{45}\biggr)\biggr](2
\alpha')^6 + \ {\cal O}( \ (2 \alpha')^7 \ )  \ ,
\\
\label{Z23exp} Z^{(1)} &=&  -\frac{ \ \pi^4}{1440}(2
\alpha')^4+\biggl(-\frac{ \
\pi^2}{72}\zeta(3)+\frac{1}{6}\zeta(5)\biggr)\biggl(-\alpha_{12}+3\alpha_{45}-\alpha_{51}
+6\alpha_{34}+3\alpha_{23}\biggr)(2 \alpha')^5 + \nonumber
\\
&& +
\biggl[\frac{1}{192}\zeta(3)^2\biggl(21\alpha_{45}^2+6\alpha_{12}\alpha_{51}
-16\alpha_{23}\alpha_{51}
+3\alpha_{12}^2+3\alpha_{51}^2+21\alpha_{23}^2+96\alpha_{34}\alpha_{45}-
\nonumber \\
&&\hphantom{ +
\biggl[}-16\alpha_{45}\alpha_{12}+72\alpha_{34}^2-16\alpha_{23}\alpha_{12}
-24\alpha_{34}\alpha_{51}-24\alpha_{34}\alpha_{12}-16\alpha_{45}\alpha_{51}
+96\alpha_{23}\alpha_{34}+\nonumber
\\
&&\hphantom{ + \biggl[}+30\alpha_{45}\alpha_{23}\biggr) -\frac{ \
\pi^6}{362880}\biggl(276\alpha_{23}\alpha_{34}-72\alpha_{34}\alpha_{51}
-46\alpha_{45}\alpha_{51}
+90\alpha_{45}\alpha_{23}+63\alpha_{23}^2+\nonumber
\\
&&\hphantom{ + \biggl[}+216\alpha_{34}^2 -46\alpha_{23}\alpha_{51}
+18\alpha_{12}\alpha_{51}+9\alpha_{12}^2+9\alpha_{51}^2
+63\alpha_{45}^2-46\alpha_{23}\alpha_{12}
+276\alpha_{34}\alpha_{45}-\nonumber \\
&&\hphantom{ + \biggl[}-46\alpha_{45}\alpha_{12}
-72\alpha_{34}\alpha_{12}\biggr)\biggr](2 \alpha')^6 + \ {\cal O}( \
(2 \alpha')^7 \ ) \ ,
\\
\label{Deltaexp} \Delta &=& \biggl(\frac{3}{2}\pi^2
\zeta(3)-\frac{35}{2}\zeta(5)\biggr)(2 \alpha')^5 + \nonumber
\\
&& \ \ \  +
\biggl(-\frac{3}{2}\zeta(3)^2+\frac{109}{45360}\pi^6\biggr)
\biggl(\alpha_{12}+\alpha_{23}+\alpha_{34}+\alpha_{45}+
\alpha_{51}\biggr)(2 \alpha')^6 + {\cal O}(  \ (2 \alpha')^7 \ ) \
.
\end{eqnarray}

\section{The Yang-Mills 5-point subamplitude}
\label{YM5} The calculation of the Yang-Mills 5-point subamplitude
was already considered in the appendix D of
\cite{Brandt1}\footnote{In \cite{Brandt1} it was called as `5-point
amplitude'.}. Its (off-shell) expression is the following:
\begin{multline}
A_{YM}(1,2,3,4,5)= -i \ g^3 \times \\
\begin{split}
&\times \biggl\{ \frac{ {V_{YM}}_{\mu_1 \mu_2}^{ (3) \ \rho}(k_1,
k_2, -k_1-k_2){V_{YM}}^{(3) \sigma }_{\ \ \ \mu_3
\mu_4}(-k_3-k_4,k_3,k_4) {V_{YM}}^{(3)}_{\ \mu_5 \rho
\sigma}(k_5,k_1+k_2, k_3+k_4)}{(k_1+k_2)^2
(k_3+k_4)^2} \ - \\
&\hphantom{-i \ g^3 \biggl\{ }- \ \frac{{V_{YM}}_{\mu_1 \mu_2}^{
(3) \ \rho}(k_1, k_2, -k_1-k_2){V_{YM}}^{(4)}_ {\rho \mu_3 \mu_4
\mu_5}(k_1+k_2, k_3, k_4, k_5)}{(k_1+k_2)^2} \biggr\}
\zeta_1^{\mu_1}\zeta_2^{\mu_2}\zeta_3^{\mu_3}\zeta_4^{\mu_4}
\zeta_5^{\mu_5} \ + \\
&+ \ \left( \frac{}{}\mbox{cyclic permutations} \frac{}{} \right)
\ ,
\end{split}
\label{AYM5}
\end{multline}
where the Yang-Mills 3 and 4-point vertices (which do not carry
color indices) are given, respectively, by
\begin{eqnarray}
\label{YMvertex3}
V^{(3)}_{YM \ \mu_1 \mu_2 \mu_3}(k_1, k_2, k_3)
&=& -i \ [\frac{}{} \eta_{\mu_1 \mu_2} (k_1-k_2)_{\mu_3} +
\eta_{\mu_2 \mu_3}
(k_2-k_3)_{\mu_1} + \eta_{\mu_3 \mu_1} (k_3-k_1)_{\mu_2} \frac{}{}] \ , \ \ \ \ \ \ \\
\label{YMvertex4} V^{(4)}_{YM \ \mu_1 \mu_2 \mu_3 \mu_4}(k_1, k_2,
k_3, k_4 ) &=& - \ [\frac{}{} \eta_{\mu_1 \mu_2} \eta_{\mu_3
\mu_4} - 2 \eta_{\mu_1 \mu_3} \eta_{\mu_2 \mu_4} + \eta_{\mu_4
\mu_1} \eta_{\mu_2 \mu_3} \frac{}{}] \ .
\end{eqnarray}
We have not worked any further trying to find a tensor notation
which would shorten the expression of $A_{YM}(1,2,3,4,5)$ in
(\ref{AYM5}).

\section{The $D^{2n}F^4$ terms 5-point subamplitude}
\label{D2nF4}

The 5-point subamplitude the comes from the lagrangian ${\cal
L}_{D^{2n}F^4}$, given in (\ref{Seff-nonabelian}), is given by:
\begin{multline}
A_{D^{2n}F^4}(1,2,3,4,5) = -2 \ (2 \alpha')^ 2 \ g^3 \ \biggl\{ \ \
f(-2 \alpha_{34}, -2
\alpha_{45}) \ \frac{T_{12}(\zeta,k)}{\alpha_{12}}  \ +  \\
\begin{split}
& \ \ + \ K(\zeta_1, k_1; \zeta_2, k_2; \zeta_3, k_3; \zeta_4,
k_4) \ S_{5}(\zeta_5; k_1, k_2, k_3, k_4,k_5; \alpha') \ \
\biggr\} \ + \ \left( \frac{}{}\mbox{cyclic permutations}
\frac{}{} \right) \ ,
\end{split}
\label{AD2nF4}
\end{multline}
where
\begin{eqnarray}
T_{12}(\zeta,k) & = & \alpha_{12} \ K(\zeta_1, \zeta_2; \zeta_3,
k_3; \zeta_4, k_4; \zeta_5, k_5)  + (\zeta_1 \cdot \zeta_2) K(k_1,
k_2; \zeta_3,
k_3; \zeta_4, k_4; \zeta_5, k_5) + \nonumber \\
&+& (\zeta_1 \cdot k_2) K(\zeta_2, k_1 + k_2; \zeta_3, k_3; \zeta_4,
k_4; \zeta_5, k_5)  -  (\zeta_2 \cdot k_1) K(\zeta_1, k_1 + k_2;
\zeta_3, k_3; \zeta_4, k_4; \zeta_5, k_5)  \
\nonumber \\
\label{T12}
\end{eqnarray}
and
\begin{multline}
S_{5}(\zeta_5; k_1, k_2, k_3, k_4,k_5; \alpha') =\ \biggl[ \
\left( \frac{}{} G(-2 \alpha_{34},-2 \alpha_{23},- \alpha_{12}-
\alpha_{34},-\alpha_{14}- \alpha_{23}) + \right. \\
\begin{split}
\left.+ \ G(-2 \alpha_{12},-2 \alpha_{23}, - \alpha_{12}-
\alpha_{34},-\alpha_{14}- \alpha_{23}) \frac{}{} \right) \ \zeta_5
\cdot (k_1 + k_2) \ \ +
\\
+ \ \left( \frac{}{} G(-2 \alpha_{23}, -2 \alpha_{34},- \alpha_{14}-
\alpha_{23},-\alpha_{12}- \alpha_{34}) - \right. \\
\left.- \ G(-2 \alpha_{23},-2 \alpha_{12}, - \alpha_{14}-
\alpha_{23},-\alpha_{12}- \alpha_{34}) \frac{}{} \right) \ \zeta_5
\cdot (k_1 + k_4) \ \biggr] \ .
\end{split}
\label{S12}
\end{multline}
\noindent In these formulas, by $K(A, a; B, b; C, c; D, d)$ we mean
the same expression of the 4-point amplitude kinematic factor, in
(\ref{kinematicfactor}), evaluated in the corresponding variables.
The function $f(s,t)$ in the second line of (\ref{AD2nF4}) is the
one defined in (\ref{f}) and the function $G(a,b,c,d)$ that goes in
the second square bracket of (\ref{AD2nF4}) is the one defined in
(\ref{G}).

\begin{figure}[h]
\centerline{\includegraphics*[scale=0.7,angle=0]{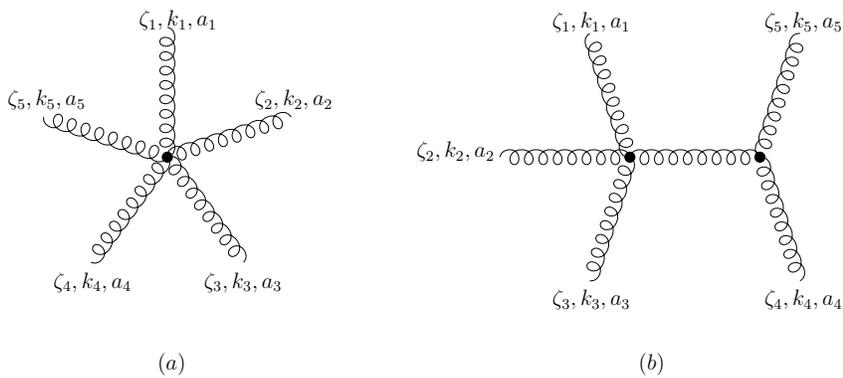}}
\caption{Type of 5-particle diagrams, with $\alpha'$ dependence,
which come from the effective lagrangian. The Feynman diagram in
(a) is one particle irreducible (1 PI), while the one in (b) is
one particle reducible. } \label{twodiagrams}
\end{figure}

\noindent The derivation of our formula in (\ref{AD2nF4}) is quite
nontrivial and very lengthy. We will not give all the details
here, but will comment some aspects about it. It was obtained by
first finding the 4 and 5-point vertices, $V^{(4)}_{\mu_1 \mu_2
\mu_3 \mu_4}(k_1, k_2, k_3, k_4)$ and $V^{(5)}_{\mu_1 \mu_2 \mu_3
\mu_4 \mu_5}(k_1, k_2, k_3, k_4, k_5)$ (which do not carry color
indices), of the lagrangian in (\ref{Seff-nonabelian}). This was
done along the same lines of the Appendices of
\cite{DeRoo1,Chandia1}, that is, by means of the 1-particle
irreducible 4 and 5-point functions in momentum space.\\
In the case of the 1-particle irreducible 4-point function,
$\Gamma_{\mu_1 \mu_2 \mu_3 \mu_4}^{(4)} (k_1, k_2, k_3, k_4)$, the
calculation was straight forward since only the abelian terms of
${\cal L}_{D^{2n}F^4}$ were required. \\
In the case of the 1-particle irreducible 5-point function,
$\Gamma_{\mu_1 \mu_2 \mu_3 \mu_4 \mu_5}^{(5)} (k_1, k_2, k_3, k_4,
k_5)$, the calculation was much more involved than the previous
one. On each $D^{2n}F^4$ term there are two types of 5-field terms
present:
\begin{enumerate}
\item The ones that come from the non-abelian $F^4$ terms
($F^4 \rightarrow A^2 \cdot (\partial A)^3$), with the covariant
derivatives acting as ordinary ones ($D^{2n} \rightarrow
{\partial}^{2n})$.\\
\noindent In the scattering amplitude (\ref{AD2nF4}) their
contribution is the one that goes with the $f$ factor, with no
poles.
\item The ones that pick an $A^{a}_{\mu}$ term from the covariant
derivatives ($D^{2n} \rightarrow A \cdot {\partial}^{2n-1}$) and
the other four from the abelian part of the $F^4$ term
($F^4 \rightarrow (\partial A)^4$).\\
\noindent These are the terms of the scattering amplitude
(\ref{AD2nF4}) which are written in terms of the function
$G(a,b,c,d)$. They begin to contribute only at order ${\alpha'}^3$
(while the rest of the terms begin to contribute at ${\alpha'}^2$
order) .
\end{enumerate}
Once the 4 and 5-point vertices have been found, to all order in
$\alpha'$, the calculation of the 5-point subamplitude in
(\ref{AD2nF4}) can be done using the corresponding Feynman rules.
This subamplitude receives contributions from one particle
irreducible (1 PI) diagrams and one particle reducible ones which
contain only simple poles, as shown in
figure \ref{twodiagrams}.\\
\noindent By construction, $A_{D^{2n}F^4}(1,2,3,4,5)$ remains
invariant under cyclic permutations of indexes \\
$(1,2,3,4,5)$. It is not difficult to see that it also remains
invariant under a
world-sheet parity (or twisting) transformation.\\
Since the lagrangian ${\cal L}_{D^{2n}F^4}$ is gauge invariant,
$A_{D^{2n}F^4}(1,2,3,4,5)$ should also satisfy (on-shell) gauge
invariance. This is a very non trivial test of formula
(\ref{AD2nF4}). In fact, demanding that it should become zero
after doing $\zeta_5 \rightarrow k_5$, it leads to the following
relation to be satisfied:
\begin{eqnarray}
f(-2 \alpha_{12}, -2 \alpha_{23}) - f(-2 \alpha_{23}, -2
\alpha_{34}) & = &  \left[ \frac{}{} G(-2 \alpha_{34},-2
\alpha_{23},- \alpha_{12}-
\alpha_{34},-\alpha_{14}- \alpha_{23}) +  \right. \nonumber \\
& + & \left.   \ G(-2 \alpha_{12},-2 \alpha_{23}, - \alpha_{12}-
\alpha_{34},-\alpha_{14}- \alpha_{23}) \frac{}{} \right]
(\alpha_{51}+ \alpha_{25})  + \nonumber
\\
&&  + \ \left[ \frac{}{} G(-2 \alpha_{23}, -2 \alpha_{34},-
\alpha_{14}- \alpha_{23},-\alpha_{12}-
\alpha_{34}) - \right. \nonumber \\
& - &  \left.  \ G(-2 \alpha_{23},-2 \alpha_{12}, - \alpha_{14}-
\alpha_{23},-\alpha_{12}- \alpha_{34}) \frac{}{} \right]
(\alpha_{51} + \alpha_{45})  . \nonumber \\
\label{gauge-inv-AD2nF4}
\end{eqnarray}
We have checked this relation using the expansions of $f(s,t)$ and
$G(a,b,c,d)$, given in (\ref{fpower}) and (\ref{G}), respectively,
up to ${\cal O}({\alpha'}^{15})$ terms.

\section{Derivation of the $D^{2n}F^5$ terms 5-point subamplitude}
\label{derivation}

Together with the derivation of formulas (\ref{A12345final}) and
(\ref{AD2nF4}), the calculations of the present appendix
constitute an essential part of this work. There are three main
achievements that we arrive to in the expression for
$A_{D^{2n}F^5}(1,2,3,4,5)$: it explicitly has no poles; it is
written as a sum of terms that have manifest cyclic, (on-shell)
gauge invariance and world-sheet parity symmetry; it is written in
terms of tensors in such a way that a local
lagrangian can be found almost directly in terms of them.\\
We have divided the calculations of this appendix in five steps:\\

\noindent {\bf \underline{Step 1}:} \ \ Treatment of the poles of $T
\cdot A_{YM}(1,2,3,4,5)$ in
(\ref{A12345final}).\\

\noindent We begin writing the mentioned term as
\begin{multline}
T \cdot A_{YM}(1,2,3,4,5)= A_{YM}(1,2,3,4,5) - i g^3 \biggl\{ \ \
\biggl(T\biggl|_{\alpha_{12}=0}- \ 1 \ \biggr) \times \\
\begin{split}
&\times \biggl[ \frac{ {V_{YM}}_{\mu_1 \mu_2}^{ (3) \ \rho}(k_1,
k_2, -k_1-k_2){V_{YM}}^{(3) \sigma }_{\ \ \ \mu_3
\mu_4}(-k_3-k_4,k_3,k_4) {V_{YM}}^{(3)}_{\ \mu_5 \rho
\sigma}(k_5,k_1+k_2, k_3+k_4)}{4 \alpha_{12} \alpha_{34}} \ - \\
&- \ \frac{{V_{YM}}_{\mu_1 \mu_2}^{ (3) \ \rho}(k_1, k_2,
-k_1-k_2){V_{YM}}^{(4)}_ {\rho \mu_3 \mu_4 \mu_5}(k_1+k_2, k_3, k_4,
k_5)}{2 \alpha_{12}} \biggr]
\zeta_1^{\mu_1}\zeta_2^{\mu_2}\zeta_3^{\mu_3}\zeta_4^{\mu_4}
\zeta_5^{\mu_5} \ \ + \\
&+ \ \left( \frac{}{}\mbox{cyclic permutations} \frac{}{} \right) \
\ \biggl\} \  \hspace{0.9cm} - \hspace{0.9cm} i g^3 \biggl\{ \ \
\biggl(T \ - \ T\biggl|_{\alpha_{12}=0} \biggr) \times \\
&\times \biggl[ \frac{ {V_{YM}}_{\mu_1 \mu_2}^{ (3) \ \rho}(k_1,
k_2, -k_1-k_2){V_{YM}}^{(3) \sigma }_{\ \ \ \mu_3
\mu_4}(-k_3-k_4,k_3,k_4) {V_{YM}}^{(3)}_{\ \mu_5 \rho
\sigma}(k_5,k_1+k_2, k_3+k_4)}{4 \alpha_{12} \alpha_{34}} \ - \\
&- \ \frac{{V_{YM}}_{\mu_1 \mu_2}^{ (3) \ \rho}(k_1, k_2,
-k_1-k_2){V_{YM}}^{(4)}_ {\rho \mu_3 \mu_4 \mu_5}(k_1+k_2, k_3, k_4,
k_5)}{2 \alpha_{12}} \biggr]
\zeta_1^{\mu_1}\zeta_2^{\mu_2}\zeta_3^{\mu_3}\zeta_4^{\mu_4}
\zeta_5^{\mu_5} \ \ + \\
&+ \ \left( \frac{}{}\mbox{cyclic permutations} \frac{}{} \right) \
\ \biggl\} \ \ ,
\end{split}
\label{term1}
\end{multline}
Here we have split the left-hand member of the equality into three
terms on the right-hand side:  $A_{YM}(1,2,3,4,5)$, the term which
contains a factor $( \ T|_{\alpha_{12}=0} \ - \ 1 \ )$ (and cyclic
permutations) and the term which contains a factor $( \ T \ - \
T|_{\alpha_{12}=0} \ )$ (and cyclic permutations). We have used the
on-shell expression of $A_{YM}(1,2,3,4,5)$ in these last two
terms\footnote{This has been done by substituting
$(k_i+k_j)^2=2\alpha_{ij}$ in the denominators of (\ref{AYM5}).} and
we have also used the fact that $T$ is cyclic invariant.\\
In principle, $T \cdot A_{YM}(1,2,3,4,5)$ should have double poles
because $A_{YM}(1,2,3,4,5)$ does, but it happens that the mentioned
expression has only simple ones. To see this we notice that $( \ T \
- \ T|_{\alpha_{12}=0} \ )$ is factorable by $\alpha_{12}$ and also,
using (\ref{K2-K}) and (\ref{relate32}), it may be proved that $( \
T|_{\alpha_{12}=0} \ - \ 1 \ )$ is factorable by
$\alpha_{34}\alpha_{45}$. Using this information in (\ref{term1}) we
can arrive to the following result:
\begin{multline}
T \cdot A_{YM}(1,2,3,4,5)= A_{YM}(1,2,3,4,5) - i g^3 \biggl\{ \ \
\frac{1}{\alpha_{12}} \ \ \biggl[ \ \ \
\biggl(\frac{T|_{\alpha_{12}=0}- \ 1 \ }
{\alpha_{34}\alpha_{45}}\biggr)
\times \\
\begin{split}
& \times \biggl(\frac{\alpha_{45}}{4} {V_{YM}}_{\mu_1 \mu_2}^{ (3) \
\rho}(k_1, k_2, -k_1-k_2){V_{YM}}^{(3) \sigma }_{\ \ \ \mu_3
\mu_4}(-k_3-k_4,k_3,k_4) {V_{YM}}^{(3)}_{\ \mu_5 \rho
\sigma}(k_5,k_1+k_2, k_3+k_4) + \\
&\hphantom{\times \biggl( \ } + \frac{\alpha_{34}\alpha_{45}}{2} \
{V_{YM}}_{\mu_1 \mu_2}^{ (3) \ \rho}(k_1, k_2,
-k_1-k_2){V_{YM}}^{(4)}_ {\rho \mu_3 \mu_4 \mu_5}(k_1+k_2, k_3, k_4,
k_5) \biggr) \ \ + \\
&\hphantom{T \cdot A_{YM}(1,2,3,4,5)= A_{YM}(1,2,3,4,5) - i g^3
\biggl\{ \ \frac{1}{\alpha_{12}} \ \  }  + \ \biggl(\frac{T \ -
\ T|_{\alpha_{45}=0} \ }{\alpha_{45}}\biggr) \times \\
& \times \frac{1}{4} {V_{YM}}_{\mu_1 \mu_2}^{ (3) \ \rho}(k_1, k_2,
-k_1-k_2){V_{YM}}^{(3) \sigma }_{\ \ \ \mu_3
\mu_4}(-k_3-k_4,k_3,k_4) {V_{YM}}^{(3)}_{\ \mu_5 \rho
\sigma}(k_5,k_1+k_2, k_3+k_4) \ \ \ \biggr] \times \\
& \times \
\zeta_1^{\mu_1}\zeta_2^{\mu_2}\zeta_3^{\mu_3}\zeta_4^{\mu_4}
\zeta_5^{\mu_5} \ \ \ \ + \ \ \ \ \left( \frac{}{}\mbox{cyclic
permutations} \frac{}{} \right) \ \ \biggr\} \ \ \ \ - \\
&- i g^3 \biggl\{ \ \ \biggl(\frac{T \ - \ T|_{\alpha_{12}=0} \
}{\alpha_{12}}\biggr) \frac{1}{2} {V_{YM}}_{\mu_1 \mu_2}^{ (3) \
\rho}(k_1, k_2, -k_1-k_2){V_{YM}}^{(4)}_ {\rho \mu_3 \mu_4
\mu_5}(k_1+k_2, k_3, k_4, k_5) \ \times \\
& \times \
\zeta_1^{\mu_1}\zeta_2^{\mu_2}\zeta_3^{\mu_3}\zeta_4^{\mu_4}
\zeta_5^{\mu_5} \ \ \ \ + \ \ \ \ \left( \frac{}{}\mbox{cyclic
permutations} \frac{}{} \right) \ \ \biggr\} \ ,
\end{split}
\label{term2}
\end{multline}
in which is manifest that $T \cdot A_{YM}(1,2,3,4,5)$ contains only
simple poles: they come in the first curly bracket of this
equation.\\

\noindent {\bf \underline{Step 2}:} \ \ Treatment of the poles of
$(2 \alpha')^2 K_3 \cdot A_{F^4}(1,2,3,4,5)$ in
(\ref{A12345final}).\\

\noindent In a similar way as it was done in Step 1, using the
expression in (\ref{AF412345}) for $A_{F^4}(1,2,3,4,5)$, we arrive
to:
\begin{eqnarray}
(2 \alpha')^2 K_3 \cdot A_{F^4}(1,2,3,4,5) & = & -2 g^3 \biggl\{ (2
\alpha')^2 f(-2 \alpha_{34}, -2 \alpha_{45}) \frac{T_{12}(\zeta,
k)}{\alpha_{12}} + \left( \frac{}{}\mbox{cyclic permutations}
\frac{}{} \right) \ \biggr\} + \nonumber \\
&+&2 g^3 \biggl\{ \biggl(\frac{T|_{\alpha_{12}=0}- \ 1 \ }
{\alpha_{34}\alpha_{45}}\biggr) \frac{T_{12}(\zeta, k)}{\alpha_{12}}
+ \left( \frac{}{}\mbox{cyclic permutations}
\frac{}{} \right) \ \biggr\} + \nonumber \\
&+& 2 g^3 \biggl\{ (2 \alpha')^2 \biggl(\frac{K_3 \ - \
{K_3}|_{\alpha_{12}=0} \ }{\alpha_{12}}\biggr) T_{12}(\zeta, k)+
\left( \frac{}{}\mbox{cyclic
permutations} \frac{}{} \right) \ \biggr\} \ , \nonumber \\
\label{term3}
\end{eqnarray}
where $T_{12}(\zeta, k)$ is given in (\ref{T12}).\\
The simple poles appear in the first and second lines of
(\ref{term3}). In the second line of this equation we used that
\begin{eqnarray}
(2 \alpha')^2 K_3\biggl|_{\alpha_{12}=0} \ +  \ \ (2 \alpha')^2 f(-2
\alpha_{34}, -2 \alpha_{45}) \ = \ \frac{T|_{\alpha_{12}=0}- \ 1 \ }
{\alpha_{34}\alpha_{45}} \ , \label{identity}
\end{eqnarray}
which is equivalent to (\ref{relate3}), once the double pole of that
equation has been subtracted in both sides of it.\\
We notice that the simple poles in the first line of (\ref{term3})
are \underline{exactly the same ones} of\\
$A_{D^{2n}F^4}(1,2,3,4,5)$ (see eq. (\ref{AD2nF4}) ).\\

\noindent {\bf \underline{Step 3}:} \ \ Taking the poles to
$A_{D^{2n}F^4}(1,2,3,4,5)$.\\

\noindent Using the expressions in (\ref{term2}) and (\ref{term3})
we have that
\begin{multline}
T \cdot A_{YM}(1,2,3,4,5)+(2 \alpha')^2 K_3 \cdot A_{F^4}(1,2,3,4,5)
= \\
\begin{split}
&=A_{YM}(1,2,3,4,5) - 2 g^3 \biggl\{ (2 \alpha')^2 f(-2 \alpha_{34},
-2 \alpha_{45}) \frac{T_{12}(\zeta, k)}{\alpha_{12}}   + \left(
\frac{}{}\mbox{cyclic
permutations} \frac{}{} \right) \ \biggr\} \ +\\
&+ g^3 \biggl\{ \ \ \frac{1}{\alpha_{12}} \ \ \biggl[ \ \ \
\biggl(\frac{T|_{\alpha_{12}=0}- \ 1 \ }
{\alpha_{34}\alpha_{45}}\biggr)
\times \biggl( \ 2 \ T_{12}(\zeta, k) \ \ -  \\
& - i \frac{\alpha_{45}}{4} {V_{YM}}_{\mu_1 \mu_2}^{ (3) \
\rho}(k_1, k_2, -k_1-k_2){V_{YM}}^{(3) \sigma }_{\ \ \ \mu_3
\mu_4}(-k_3-k_4,k_3,k_4) {V_{YM}}^{(3)}_{\ \mu_5 \rho
\sigma}(k_5,k_1+k_2, k_3+k_4) + \\
&\hphantom{\times \biggl( \ } +i \frac{\alpha_{34}\alpha_{45}}{2} \
{V_{YM}}_{\mu_1 \mu_2}^{ (3) \ \rho}(k_1, k_2,
-k_1-k_2){V_{YM}}^{(4)}_ {\rho \mu_3 \mu_4 \mu_5}(k_1+k_2, k_3, k_4,
k_5) \biggr) \ \ - \\
&\hphantom{T \cdot A_{YM}(1,2,3,4,5)= A_{YM}(1,2,3,4,5) - i g^3
\biggl\{ \ \frac{1}{\alpha_{12}} \ \  }  - \ \biggl(\frac{T \ -
\ T|_{\alpha_{45}=0} \ }{\alpha_{45}}\biggr) \times \\
& \times \frac{i}{4} {V_{YM}}_{\mu_4 \mu_5}^{ (3) \ \rho}(k_4, k_5,
-k_4-k_5){V_{YM}}^{(3) \sigma }_{\ \ \ \mu_1
\mu_2}(-k_1-k_2,k_1,k_2) {V_{YM}}^{(3)}_{\ \mu_3 \rho
\sigma}(k_3,k_4+k_5, k_1+k_2) \ \ \ \biggr] \times \\
& \times \
\zeta_1^{\mu_1}\zeta_2^{\mu_2}\zeta_3^{\mu_3}\zeta_4^{\mu_4}
\zeta_5^{\mu_5} \ \ \ \ + \ \ \ \ \left( \frac{}{}\mbox{cyclic
permutations} \frac{}{} \right) \ \ \biggr\} \ \ \ \ - \\
&- i g^3 \biggl\{ \ \ \biggl(\frac{T \ - \ T|_{\alpha_{12}=0} \
}{\alpha_{12}}\biggr) \frac{1}{2} {V_{YM}}_{\mu_1 \mu_2}^{ (3) \
\rho}(k_1, k_2, -k_1-k_2){V_{YM}}^{(4)}_ {\rho \mu_3 \mu_4
\mu_5}(k_1+k_2, k_3, k_4, k_5) \ \times \\
& \times \
\zeta_1^{\mu_1}\zeta_2^{\mu_2}\zeta_3^{\mu_3}\zeta_4^{\mu_4}
\zeta_5^{\mu_5} \ \ \ \ + \ \ \ \ \left( \frac{}{}\mbox{cyclic
permutations} \frac{}{} \right) \ \ \biggr\} \ + \\
&+ 2 g^3 \biggl\{ (2 \alpha')^2 \biggl(\frac{K_3 \ - \
{K_3}|_{\alpha_{12}=0} \ }{\alpha_{12}}\biggr) T_{12}(\zeta, k)+
\left( \frac{}{}\mbox{cyclic
permutations} \frac{}{} \right) \ \biggr\} \ .
\end{split}
\label{term4}
\end{multline}
Now, we first notice that
\begin{eqnarray}
\biggl(\frac{T \ - \ T|_{\alpha_{45}=0} \ }{\alpha_{45}}\biggr) -
\alpha_{34}\biggl(\frac{T|_{\alpha_{12}=0}- \ 1 \ }
{\alpha_{34}\alpha_{45}}\biggr) \label{M4512}
\end{eqnarray}
is factorable by $\alpha_{12}$\footnote{The expression in
(\ref{M4512}) has a power series expansion in $\alpha_{12}$ and
using (one of the cyclic permutations of) (\ref{T0}) it can be
proved that it becomes $0$ when $\alpha_{12}=0$.}, so we will call
it $\alpha_{12}M_{12}$, where $M_{12}$ has a well defined $\alpha'$
series expansion with no poles. Eliminating $( \ T \ - \
T|_{\alpha_{45}=0} \ )/\alpha_{45}$ from it and substituting it in
(\ref{term4}), together with the expression of
$A_{D^{2n}F^4}(1,2,3,4,5)$, given in (\ref{AD2nF4}), we arrive to
\begin{multline}
T \cdot A_{YM}(1,2,3,4,5)+(2 \alpha')^2 K_3 \cdot A_{F^4}(1,2,3,4,5)
= \\
\begin{split}
&=A_{YM}(1,2,3,4,5)+A_{D^{2n}F^4}(1,2,3,4,5) + g^3 \biggl\{ \ \
\biggl(\frac{T|_{\alpha_{12}=0}- \ 1 \ }
{\alpha_{34}\alpha_{45}}\biggr) \times \ \biggl( \ \ \
\frac{1}{\alpha_{12}} \
 \biggl[ \ 2 \ T_{12}(\zeta, k) \ -  \\
& - i \frac{\alpha_{45}}{4} {V_{YM}}_{\mu_1 \mu_2}^{ (3) \
\rho}(k_1, k_2, -k_1-k_2){V_{YM}}^{(3) \sigma }_{\ \ \ \mu_3
\mu_4}(-k_3-k_4,k_3,k_4) {V_{YM}}^{(3)}_{\ \mu_5 \rho
\sigma}(k_5,k_1+k_2, k_3+k_4) - \\
& -i \frac{\alpha_{34}}{4} {V_{YM}}_{\mu_4 \mu_5}^{ (3) \ \rho}(k_4,
k_5, -k_4-k_5){V_{YM}}^{(3) \sigma }_{\ \ \ \mu_1
\mu_2}(-k_1-k_2,k_1,k_2) {V_{YM}}^{(3)}_{\ \mu_3 \rho
\sigma}(k_3,k_4+k_5, k_1+k_2) \ + \\
&+i \frac{\alpha_{34}\alpha_{45}}{2} \ {V_{YM}}_{\mu_1 \mu_2}^{ (3)
\ \rho}(k_1, k_2, -k_1-k_2){V_{YM}}^{(4)}_ {\rho \mu_3 \mu_4
\mu_5}(k_1+k_2, k_3, k_4, k_5) \ \biggr] \ + \\
&+i \frac{\alpha_{34}}{2} \ {V_{YM}}_{\mu_4 \mu_5}^{ (3) \
\rho}(k_4, k_5, -k_4-k_5){V_{YM}}^{(4)}_ {\rho \mu_1 \mu_2
\mu_3}(k_4+k_5, k_1, k_2, k_3) \ \ \ \biggr) \
\zeta_1^{\mu_1}\zeta_2^{\mu_2}\zeta_3^{\mu_3}\zeta_4^{\mu_4}
\zeta_5^{\mu_5} \ \ + \\
&+ \ \left( \frac{}{}\mbox{cyclic permutations} \frac{}{} \right) \
\ \ \biggr\} \ \ \ \ - \ \ i g^3 \biggl\{ \ \ M_{12} \ \times
\\
&\times \biggl( \frac{1}{4} {V_{YM}}_{\mu_4 \mu_5}^{ (3) \
\rho}(k_4, k_5, -k_4-k_5){V_{YM}}^{(3) \sigma }_{\ \ \ \mu_1
\mu_2}(-k_1-k_2,k_1,k_2) {V_{YM}}^{(3)}_{\ \mu_3 \rho
\sigma}(k_3,k_4+k_5, k_1+k_2) \ - \\
&- \frac{\alpha_{12}}{2} \ {V_{YM}}_{\mu_4 \mu_5}^{ (3) \ \rho}(k_4,
k_5, -k_4-k_5){V_{YM}}^{(4)}_ {\rho \mu_1 \mu_2 \mu_3}(k_4+k_5, k_1,
k_2, k_3) \ \biggr) \
\zeta_1^{\mu_1}\zeta_2^{\mu_2}\zeta_3^{\mu_3}\zeta_4^{\mu_4}
\zeta_5^{\mu_5} \ \  + \\
& + \  \left( \frac{}{}\mbox{cyclic
permutations} \frac{}{} \right) \ \ \biggr\} \ + \\
&+ 2 g^3 \biggl\{ (2 \alpha')^2 \biggl(\frac{K_3 \ - \
{K_3}|_{\alpha_{12}=0} \ }{\alpha_{12}}\biggr) T_{12}(\zeta, k)+
\left( \frac{}{}\mbox{cyclic permutations} \frac{}{} \right) \
\biggr\} \ + \\
& +2 \ g^3 \ (2 \alpha')^ 2  \ \biggl\{ \ K(\zeta_1, k_1; \zeta_2,
k_2; \zeta_3, k_3; \zeta_4, k_4) \ S_{5}(\zeta_5; k_1, k_2, k_3, k_4,k_5; \alpha')  \ + \\
&\hphantom{+2 \ g^3 \ (2 \alpha')^ 2  \ \biggl\{ \ }  + \left(
\frac{}{}\mbox{cyclic permutations} \frac{}{} \right) \ \biggr\} \ .
\end{split}
\label{term5}
\end{multline}
This is a huge expression. At this moment our main interest lies
in the term in the square bracket (the one that goes multiplying
the factor $1/\alpha_{12}$) and its cyclic permutations because,
besides the poles of $A_{D^{2n}F^4}(1,2,3,4,5)$, those are the
only places where poles could come from. We have verified
computationally that, using on-shell and physical state
conditions, together with momentum conservation, the term in the
square bracket is factorable by $\alpha_{12}$, thus eliminating
all poles which are not contained in $A_{D^{2n}F^4}(1,2,3,4,5)$.\\

\noindent {\bf \underline{Step 4}:} \ \ Derivation of an expression for
$A_{D^{2n}F^5}(1,2,3,4,5)$ as a sum of terms which are gauge invariant
and have no poles.\\

\noindent Since the left hand-side of (\ref{term5}) is precisely
$A(1,2,3,4,5)$, using (\ref{AD2nF5}) we have that the sum of curly
brackets in (\ref{term5}) corresponds to $A_{D^{2n}F^5}(1,2,3,4,5)$.
This expression explicitly has no poles and is (on-shell) gauge
invariant as a whole, but each of the terms in curly brackets is not
individually gauge invariant. This means that, in order to write
$A_{D^{2n}F^5}(1,2,3,4,5)$ as a sum of terms which have both
properties, some redistribution of the terms in (\ref{term5}) should
be done. Only
after succeeding in doing this redistribution it will be possible to
find a local lagrangian for each group of terms.\\
\noindent This is a very non trivial step and in the following lines
we will only summarize the operations which took us to the desired
expression:
\begin{enumerate}
\item In (\ref{term5}) we substitute the following expression (and
the cyclic permutations of it in the corresponding cases) for the
kinematical expression $T_{12}(\zeta,k)$, defined in (\ref{T12}):
\begin{eqnarray}
T_{12}(\zeta,k) & = & {(\eta \cdot t_{(8)})_4}^{\mu_1 \nu_1 \mu_2
\nu_2 \mu_3 \nu_3 \mu_4 \nu_4 \mu_5 \nu_5} \zeta _{\mu
_{1}}^{1}k_{\nu _{1}}^{1}\zeta _{\mu _{2}}^{2}k_{\nu _{2}}^{2}\zeta
_{\mu _{3}}^{3}k_{\nu _{3}}^{3}\zeta _{\mu _{4}}^{4}k_{\nu
_{4}}^{4}\zeta _{\mu _{5}}^{5}k_{\nu _{5}}^{5} + \nonumber
\\
&&+ (\zeta^1 \cdot k^2) K( \zeta^2, k^2; \zeta^3, k^3; \zeta^4, k^4;
\zeta^5, k^5) - (\zeta^2 \cdot k^1) K( \zeta^1, k^1; \zeta^3, k^3;
\zeta^4, k^4; \zeta^5, k^5) \ . \nonumber \\
\label{T12II}
\end{eqnarray}
The term with the $(\eta \cdot t_{(8)})_4$ tensor is the part of
$T_{12}(\zeta,k)$ which is gauge invariant. That term changes sign
under a twisting transformation with respect to index $4$.
\item In the term $S_{5}(\zeta_5; k_1, k_2, k_3, k_4,k_5; \alpha')$
given in (\ref{S12}), which also goes in (\ref{term5}), we
substitute the following relation (and the corresponding cyclic
permutations of it):
\begin{multline}
G(-2 \alpha_{34},-2 \alpha_{23},- \alpha_{12}-
\alpha_{34},-\alpha_{14}- \alpha_{23}) + \ G(-2 \alpha_{12},-2
\alpha_{23}, - \alpha_{12}-
\alpha_{34},-\alpha_{14}- \alpha_{23}) = \\
\begin{split}
&\frac{f(-2 \alpha_{12}, -2 \alpha_{23}) - f(-2 \alpha_{23}, -2
\alpha_{34})}{\alpha_{34}- \alpha_{12}} \ - \  \left[ \frac{}{} G(-2
\alpha_{23}, -2 \alpha_{34},- \alpha_{14}- \alpha_{23},-\alpha_{12}
 - \alpha_{34}) \ - \right. \\
& \left. - \  G(-2 \alpha_{23},-2 \alpha_{12}, - \alpha_{14}-
\alpha_{23},-\alpha_{12}- \alpha_{34}) \frac{}{} \right] \
\left(\frac{\alpha_{51} + \alpha_{45}}{\alpha_{34}-
\alpha_{12}}\right) .
\end{split}
\label{G1G2}
\end{multline}
This relation has been obtained from (\ref{gauge-inv-AD2nF4}) and in
the denominators we have used, from (\ref{alpha25}), that
$\alpha_{51}+\alpha_{25}=\alpha_{34}-\alpha_{12}$.
\item We then substitute in (\ref{term5}) the following
expressions for $M_{12}$ and $(T|_{\alpha_{12}=0}- \ 1 \ )/
(\alpha_{34}\alpha_{45})$ (and the corresponding cyclic
permutations of them):
\begin{eqnarray}
\label{M12} M_{12}&=& \alpha_{51} H^{(3)} +
\alpha_{51}\alpha_{23}U^{(1)}+\alpha_{23}\alpha_{34}U^{(3)}
+\alpha_{34}\alpha_{51}U^{(5)} +
\alpha_{34}\alpha_{51}\alpha_{23}\Delta  \ , \ \ \ \ \ \ \ \ \\
\label{G12} \frac{T|_{\alpha_{12}=0}- \ 1 \ }
{\alpha_{34}\alpha_{45}}&=& \alpha_{23} H^{(1)} + \alpha_{51}
H^{(2)}+\alpha_{23}\alpha_{51}U^{(4)} \ .
\end{eqnarray}
These relations have been obtained from the definition of
$M_{12}$, given in (\ref{M4512}), and using the definitions of the
$H^{(k)}$, $U^{(k)}$ and $\Delta$ factors of appendix
\ref{otherfactors}, together with (\ref{identity}). \\
\noindent As an outcome, the resulting expression of
$A_{D^{2n}F^5}(1,2,3,4,5)$ will depend on kinematical expressions
and on the factors $H^{(k)}$, $P^{(k)}$, $U^{(k)}$, $Z^{(k)}$
($k=1, \ldots , 5.$) and $\Delta$. It will no longer depend
(explicitly) on the $T$ and $K_3$ factors.
\item We next require on-shell gauge invariance in the resulting
expression of $A_{D^{2n}F^5}(1,2,3,4,5)$ by doing, for example,
$\zeta_1 \rightarrow k_1$ and demanding that the expression should
vanish after using on-shell and physical state conditions, together
with momentum conservation. This leads us to the following condition
to be satisfied:
\begin{eqnarray}
\alpha_{51}P^{(3)}-\alpha_{12}P^{(4)}=
\alpha_{51}H^{(2)}-\alpha_{12}H^{(5)}+\alpha_{23}\alpha_{51}U^{(4)}-
\alpha_{12}\alpha_{45}U^{(3)} \ ,
\label{almostthere}
\end{eqnarray}
which can be seen to be valid after using the definitions of the
$H^{(k)}$, $P^{(k)}$ and $U^{(k)}$ factors, given in appendix
\ref{otherfactors}.\\
\noindent In order for the condition (\ref{almostthere}) to be
automatically satisfied, without needing to make any substitutions,
we introduce a new $\alpha'$ dependent factor, $W^{(1)}$, and its
cyclic permutations $W^{(k)}$ ($k=2,3,4,5$). This factor has been
defined in eq. (\ref{W51}) and it may be proved that it satisfies
the following relations:
\begin{eqnarray}
P^{(4)} = H^{(5)}+\alpha_{45}U^{(3)}+\alpha_{51}W^{(1)} \ ,
\\
P^{(3)} = H^{(2)}+\alpha_{23}U^{(4)}+\alpha_{12}W^{(1)} \ ,
\label{relateW}
\end{eqnarray}
which automatically fulfill (\ref{almostthere}).
\end{enumerate}

\noindent {\bf \underline{Step 5}:} \ \ Final expression for
$A_{D^{2n}F^5}(1,2,3,4,5)$.\\

\noindent From the previous step we find the following expression
for $A_{D^{2n}F^5}(1,2,3,4,5)$:
\begin{eqnarray}
A_{D^{2n}F^5}(1,2,3,4,5) &=& g^3 \ \biggl\{ \frac{}{} \biggl[
H^{(1)} \cdot h(\zeta,k) + P^{(1)}
\cdot p^{(1)}(\zeta,k)+\nonumber \\
&&\hphantom{\biggl\{ \frac{}{} \biggl[} + U^{(1)} \cdot u(\zeta,k) +
W^{(1)} \cdot w(\zeta,k)+  Z^{(1)} \cdot z^{(1)}(\zeta,k) \biggr] +
\nonumber \\
&&\hphantom{\biggl\{ \frac{}{} \biggl[} + (\frac{}{} \mbox{cyclic
permutations} \frac{}{}) \frac{}{} \biggr\} \ + g^3 \ \Delta \cdot
\delta'(\zeta,k) \ ,
\label{AD2nF5nopolesII}
\end{eqnarray}
where, by now, only $p^{(1)}(\zeta,k)$ and $z^{(1)}(\zeta,k)$ (and
their cyclic permutations) are known to be the same kinematical
expressions of equations (\ref{p34}) and (\ref{z23}), respectively.\\
\noindent In (\ref{AD2nF5nopolesII}) we have an expression which, on
$each$ group of terms, has no poles and is (on-shell) gauge
invariant. We have also verified that the world-sheet parity
condition (\ref{twisting}) is satisfied by $each$ group of terms in
(\ref{AD2nF5nopolesII}). But unfortunately, for $h(\zeta,k)$,
$u(\zeta,k)$, $w(\zeta,k)$ and $\delta'(\zeta,k)$ we only have long
(computer saved) expressions, whose specific structure is not
explicitly known.\\
\noindent Our final labor has then been to determine the structure
of $h(\zeta,k)$, $u(\zeta,k)$, $w(\zeta,k)$ and $\delta'(\zeta,k)$.
In appendix \ref{t10tensor} we have explained with some detail how
we have obtained an on-shell equivalent expression for $h(\zeta,k)$
(which we have called $h^{(1)}(\zeta,k)$ in equations
(\ref{AD2nF5nopoles}) and (\ref{h23})) in which the gauge symmetry
is manifest. This has been done by introducing a ten index tensor,
$t_{(10)}^{\mu_1 \nu_1 \mu_2 \nu_2 \mu_3 \nu_3 \mu_4 \nu_4 \mu_5
\nu_5}$, which is antisymmetric on each pair $(\mu_j \ \nu_j)$ and
which has the twisting symmetry mentioned in eq. (\ref{etat8twisting}).\\
\noindent For each remaining kinematical expression we have also
used an $ansatz$, in analogy to $h^{(1)}(\zeta,k)$, which consists
in the most general expression that can be constructed and which is
manifestly gauge invariant. It is quite remarkable that using
on-shell, physical state and momentum conservation conditions, we
have been able to determine their structure completely in terms of
only two tensors: the known $t_{(8)}$ and the same $t_{(10)}$ that
we have mentioned in the above lines. The resulting expressions are
the ones that we have named $u^{(1)}(\zeta,k)$, $w^{(1)}(\zeta,k)$
and $\delta(\zeta,k)$, respectively, in subsection \ref{D2nF5}.\\
\noindent The final expression that we have for
$A_{D^{2n}F^5}(1,2,3,4,5)$ is, then, the one in
(\ref{AD2nF5nopoles}), which has no poles, is cyclic and
(on-shell) gauge invariant on $each$ group of terms, as explained
in subsection \ref{D2nF5}, and which also has the world-sheet
parity symmetry manifest. All this final step has been
fundamental, in order to go from the subamplitude in
(\ref{AD2nF5nopoles}) to the effective lagrangian, ${\cal
L}_{D^{2n}F^5}$, in (\ref{LD2nF5final}).

\end{document}